\documentclass{article}
\pdfoutput=1
\usepackage[]{fontenc} 
\usepackage{amsmath}
\usepackage{amssymb}
\usepackage{hyperref}
\usepackage{latexsym}
\usepackage{graphicx}
\usepackage{color}
\usepackage{braket}
\usepackage{amssymb}
\usepackage{bbold}

\usepackage{afterpage}

\usepackage{cite}

\newcommand{\Imag}{\mathop{\mathrm{Im}}}

\newcommand{\mL}{\mathcal{L}}

\newcommand{\mO}{\mathcal{O}}

\newcommand{\be}{\begin{equation}}
\newcommand{\ee}{\end{equation}}
\newcommand{\ba}{\begin{eqnarray}}
\newcommand{\ea}{\end{eqnarray}}
\begin{document}
\title{Collider production of Electroweak resonances from  $\gamma\gamma$ states}

\author
{
Rafael L. Delgado$^\dagger$,  Antonio Dobado, Miguel Espada,\\
Felipe J. Llanes-Estrada, and Iv\'an Le\'on Merino\\
Departamento de F\'isica Te\'orica I, 
Universidad Complutense de Madrid, \\ Plaza de las Ciencias 2,
28040 Madrid, Spain.\\
$^\dagger$ Now at  Physik-Department T30f, Technische Universit\"at M\"unchen\\ 
James-Franck-Str. 1, D-85747 Garching, Germany
}

\date{October 5$^{\rm th}$ 2018}

\def\thefootnote{\arabic{footnote}}
\setcounter{page}{0}
\setcounter{footnote}{0}

\maketitle

\begin{abstract}
We estimate production cross sections for 2-body resonances of the Electroweak Symmetry Breaking sector (in $W_LW_L$ and $Z_LZ_L$ rescattering) from $\gamma\gamma$ scattering. We employ unitarized Higgs Effective Field Theory amplitudes previously computed coupling the two photon channel to the EWSBS. We work in the Effective Photon Approximation and examine both $e^-e^+$ collisions at energies of order 1-2 TeV (as relevant for future lepton machines) and $pp$ collisions at LHC energies. 
Dynamically generating a spin-0 resonance around 1.5 TeV  (by appropriately choosing the parameters of the effective theory) we find that the differential cross section per unit $s$, $p_t^2$ is of order 0.01 fbarn/TeV$^4$ at the LHC. Injecting a spin-2 resonance around 2 TeV we find an additional factor 100 suppression for $p_t$ up to 200 GeV. The very small cross sections put these $\gamma\gamma$ processes, though very clean, out of reach of immediate future searches.
\end{abstract}
\maketitle

\section{Introduction}

Accelerator-based particle physics is making progress in the exploration of the TeV energy range at the LHC. At a minimum, one may make headway in understanding the sector of the Standard Model (SM) responsible for Electroweak Symmetry Breaking (EWSBS), composed of the new Higgs boson $h$ and the  longitudinal components of gauge boson pairs $W_L W_L$ and $Z_L Z_L$. These are equivalent to the $\omega^a$ Goldstone bosons of electroweak symmetry breaking, in the sense of the Equivalence Theorem~\cite{ET}. Under its hypothesis, that the energy of longitudinal gauge boson scattering is large $E_{\omega\omega}^2= s_{\omega\omega}\gg M_W^2,\ M_Z^2,\ m_h^2$, the scattering amplitudes involving the $W_L$ and $Z_L$ (that come to dominate $W$ and $Z$ scattering anyway at high energy) can be exchanged for the scattering amplitudes of the scalar $\omega^a$. Employing the latter is advantageous because of the absence of spin complications and because many of their couplings are related, in a transparent manner, by the pattern of symmetry breaking,
$SU(2)_L\times SU(2)_R\to SU(2)_c$.

Much of the LHC strategy so far has focused on hard collisions, with multiple tracks in the central rapidity region of the detectors, triggering for various high-$p_t$ (transverse momentum) scenarios. To reduce noise produced by hadron remainders, and also to directly access quartic gauge couplings, the isolation of $\gamma\gamma$ initiated events is an interesting additional alley of investigation.

In fact, run-I of the LHC has already found some events corresponding to the reaction $\gamma\gamma\to W^+W^-$, initially with low $p_t$ below $100\,{\rm GeV}$ \cite{Piotrzkowski:2014bia}, and now up to $200$-$300\,{\rm GeV}$~\cite{Khachatryan:2016mud}. This later publication presents marginal ($3.4\sigma$) evidence with approximately 20 inverse femtobarn of integrated luminosity taken at $7$ and $8\,{\rm TeV}$ in pp collisions. They have a total of 15 reconstructed events in both sets of data (with expected backgrounds summing about 5 events). The data is used to constrain coefficients of the linear realization of the Standard Model Effective Theory (SMEFT), following earlier Tevatron studies~\cite{Abazov:2013opa}, but not the nonlinear Higgs EFT (HEFT) that we employ.

Encouraged by this success, CMS and Totem have joined~\cite{Albrow:2015ois} into the CMS-Totem Precision Proton Spectrometer (CTPPS) that will employ the LHC bending magnets to curve the trajectory of slightly deflected protons and detect them off-beam. The ATLAS collaboration is also working in at least two subprojects~\cite{Hamal:2014cia}, AFP and ALFA, that allow to identify one or even the two elastically scattered protons a couple hundred meters down the beampipe from the main detector. Tagging of the outgoing protons with these detectors will allow rather exclusive measurements, among others, of $\gamma\gamma$ initiated reactions, efficiently exploiting the LHC as a photon-photon collider. 

Meanwhile, a new generation of $e^- e^+$ colliders is in very advanced design stages. CLIC~\cite{Abramowicz:2016zbo} and the ILC~\cite{vanderKolk:2016akp} would naturally run in the $350$-$500\,{\rm GeV}$ region (just above the $t\bar{t}$ threshold, but in a second stage they could reach up to 1-5 to $3\,{\rm TeV}$ (CLIC) and $1\,{\rm TeV}$ (ILC) which would allow many interesting new physics studies with $WW$ pairs~\cite{Wang:2016eln}. The lepton colliders can also easily be adapted to perform $\gamma\gamma$ physics, and LEP was indeed used this way~\cite{Brodsky:2005wk}.

Therefore, it is sensible to carry out theoretical studies of the EWSBS in photon-photon collisions since the experimental prospects are reasonably good. Since no clear direction for new physics searches is emerging yet from the LHC~\cite{searches}, there has been a revival of the electroweak chiral Lagrangian --now including an explicit Higgs boson, in what has been called~\cite{Alonso:2015fsp} the Higgs Effective Field Theory (HEFT)-- and other effective theory formulations. 

HEFT is valid to about $4\pi v\simeq 3\,{\rm TeV}$ (or $4\pi f$ in the presence of a new physics scale such as in Composite Higgs Models). Because we use the Equivalence Theorem that requires high energies, we address the $500\,{\rm GeV}$-$3\,{\rm TeV}$ region (other groups have examined the lower--energy $\gamma\gamma$ production of new resonances). In this energy range, $m_h$ is negligible, and we thus consistently neglect the Higgs-potential self-couplings of order $m_h^2$. Except for this small assumption, a feature of many BSM (Beyond the Standard Model) approaches, our setup is rather encompassing, as several BSM theories may be cast, at moderate energy, in HEFT.

Several groups~\cite{Espriu:2013fia,Azatov:2012bz,Brivio:2013pma,Alonso:2012px,Pich:2013fba,Jenkins:2013zja,Degrande:2012wf,Buchalla:2013rka,Buchalla:2012qq} have studied in detail this EFT and its derived scattering amplitudes. Since those EFTs violate unitarity (see subsec.~\ref{subsec:unitarity} below for a summary), we~\cite{Delgado:2013loa,Delgado:2015kxa,Delgado:2014dxa} and others~\cite{Espriu:2013fia,Espriu:2012ih,Corbett:2015lfa,Sekulla:2016yku,Kilian:2014zja,Alboteanu:2008my} have pursued methods of unitarization that are sensible in the resonance region.

In a recent contribution~\cite{Delgado:2016rtd} we have coupled the EWSBS, well studied in HEFT+unitarity in that body of work, to the $\gamma\gamma$ channel. The motivation is clear: now we are prepared to address the production cross section of $\omega\omega$ bosons via $\gamma\gamma$ intermediate states. That is the thrust of the present document.

The electric field of a fast charge is Lorentz contracted in the longitudinal direction and thus practically transverse, appearing as an electromagnetic wave travelling parallel to the particle's momentum, as observed by Fermi~\cite{Fermi:1924tc}; the theory  was further developed by Weizs\"acker and Williams~\cite{Williams:1935dka,vonWeizsacker:1934nji} (at a classical level) while Pomeranchuk and Shmushkevitch~\cite{Pomeranchuk} offered a consistent covariant formulation. The resulting "Equivalent Photon Approximation" whereby the moving charge  is accompanied by a quantized radiation field is reviewed and detailed in~\cite{Budnev:1974de,Terazawa:1973tb}, from which we will draw all needed material.

Because we are working under kinematic conditions that make the Equivalence Theorem a good approximation, throughout the article we will use interchangeably the notations $W_LW_L$ and $\omega\omega$ for the charged, longitudinal gauge bosons and $Z_LZ_L$ or $zz$ for the neutral ones, computing all amplitudes in terms of the Goldstone bosons.

\section{$\gamma\gamma\to \omega\omega$ Differential cross section} \label{sigma:photo}

\subsection{Partial waves in perturbation theory}\label{subsec:partwaves}

The lowest-order  $\gamma\gamma$ partial waves that do not vanish (which we denote by a $(0)$ superindex) are given next in Eq.~(\ref{A:scatter:gammagamma:partialWaves:F:alpha}). They are Next to Leading Order (NLO) for $J=0$ while Leading Order (LO) suffices for $J=2$. 

We obtained them in terms of the fine structure constant $\alpha=e^2/4\pi$ and the parameters of the EWSBS (that the LHC is constraining) in~\cite{Delgado:2016rtd}, from earlier work on the effective Lagrangian and the invariant amplitude involving two photons in~\cite{Delgado:2014jda}. They read
\begin{subequations}\label{A:scatter:gammagamma:partialWaves:F:alpha}
\begin{align}
  P_{00}^{(0)} &= \frac{\alpha s_{\gamma\gamma}}{8\sqrt{6}}(2A_C + A_N) & %
  P_{02}^{(0)} &= \frac{\alpha}{6\sqrt{2}} %
  \label{A:scatter:gammagamma:partialWaves:F:alpha:1}\\
  P_{20}^{(0)} &= \frac{\alpha s_{\gamma\gamma}}{8\sqrt{3}}(A_C - A_N) &
  P_{22}^{(0)} &= \frac{\alpha}{12} %
  \label{A:scatter:gammagamma:partialWaves:F:alpha:2}
\end{align}
\end{subequations}
where the combinations $A_C$ and $A_N$ refer to the charged basis $W^+W^-$ and $ZZ$, which here appear mixed because we employ the custodial isospin basis that characterizes the final state, since the photon coupling is isospin violating and can yield both $I=0$ and $I=2$. $I=1$ is discarded because the $\omega\omega$ state must be Bose symmetric, entailing $J=1$, and the $\gamma\gamma$ state cannot be arranged with one unit of angular momentum as per Landau-Yang's theorem. $A_C$ and $A_N$ can be written as
\ba
A_N &\equiv&  \frac{2ac_\gamma^r}{v^2} + \frac{a^2-1}{4\pi^2 v^2} \\
A_C &\equiv&
\frac{8(a_1^r-a_2^r+a_3^r)}{v^2} +\frac{2ac_\gamma^r}{v^2} +\frac{a^2-1}{8\pi^2 v^2} \ .
\ea

For completeness, let us quote also the scalar partial wave yielding the scalar-isoscalar $hh$ final state, which only couples with positive parity states 
\begin{equation}\label{A:scatter:gammagamma:partialWaves:F:alpha:3}
R_0^{(0)} = \frac{\alpha}{32\sqrt{2}\pi^2 v^2}(a^2-b) \ .
\end{equation}

The scalar partial waves $P_{I0}^{(0)}$ at this order, and all waves at higher orders, grow polynomially with Mandelstam $s$ according to the chiral counting, if there is BSM physics in the EWSBS, until the new scale of that physics is approached. Therefore, chiral perturbation theory (ChPT) eventually breaks down; the amplitudes can still be represented from first principles (unitarity and causality) by a dispersive analysis, with chiral perturbation theory supplying the low-energy behavior (subtraction constants for the dispersion relations) which gives rise to the well-known unitarized EFT. In the next subsection we quickly recall the application of this unitarization to amplitudes involving two photons.

If no new physics is within reach at the LHC, the corresponding SM expressions are $a=1$, $c_\gamma =a_i=0$, $b=a^2$ and thus $R^{(0)}_0=0$, as well as $A_N=A_C=0$, so that $P^{(0)}_{00}=0=P^{(0)}_{20}$ (while $P_{02}$ and $P_{22}$ remain nonvanishing).

\subsection{Unitarity and resonances}\label{subsec:unitarity}
In this article we do not consider the final $hh$ state, and for simplicity we also assume that it is decoupling from $\omega\omega\simeq W_LW_L$ so we set $a^2=b$ (as well as the other parameters coupling both channels, $d=e=0$).

The scattering amplitude linking $\omega\omega$ and $\gamma\gamma$ is then a three by three matrix~\cite{Delgado:2016rtd} due to custodial isospin. The two-photon state can couple to both $I=0,2$ breaking custodial symmetry, though the presumed BSM interactions in the BSM do not connect the two channels. For each of them, angular momentum can be  0 or 2. This matrix is
\begin{equation}\label{chLagr:UnitProceduresWW:Fcoupled1}
  F(s) = %
   \begin{pmatrix}
      A_{0J}(s)   & 0          & P_{0J}(s) \\
      0           & A_{2J}(s)  & P_{2J}(s) \\
      P_{0J}(s)   & P_{2J}(s)  & 0
   \end{pmatrix} + \mO(\alpha^2),
\end{equation}
where the $A_{IJ}(s)$ are the elastic partial waves $\omega\omega\to\omega\omega$ from~\cite{Delgado:2013hxa,Delgado:2015kxa}, and the $P_{IJ}(s)$ photon-photon amplitudes are taken from subsec.~\ref{subsec:partwaves}. The two zeroes in the upper left box encode isospin symmetry in the EWSBS; the zero in the lower right corner arises because we work at LO in $\alpha$, so that $\Braket{\gamma\gamma |F^{(0)}|\gamma\gamma} \simeq 0$.

The unitarity condition for this matrix amplitude 
\begin{equation}\label{unitarity}
\Imag F(s) = F(s) F(s)^\dagger
\end{equation}
is not satisfied by the perturbative amplitude because of the derivative couplings growing with $s$, so unitarization is needed.  But since  $\alpha$ is a small parameter, it can be taken at leading order. Then, Eq.~(\ref{unitarity}) can be satisfied, in very good approximation, to all orders in $s$ but only to LO in $\alpha$. Substituting Eq.~(\ref{chLagr:UnitProceduresWW:Fcoupled1}) in Eq.~(\ref{unitarity}) yields
\begin{subequations}\label{unitarityexpanded}
\begin{align}
   \Imag A_{IJ} &= \lvert A_{IJ}\rvert^2\\
   \Imag P_{IJ} &= P_{IJ}A_{IJ}^*\ .
\end{align}
\end{subequations}
In the second equation, the $\gamma\gamma\to\gamma\gamma$ amplitude has been neglected as it would exceed first order in the $\alpha$ expansion.

The elastic $\omega\omega\to\omega\omega$ amplitude may be expanded in the HEFT (as recounted in~\cite{Delgado:2015kxa}) by
\begin{equation} \label{EFTexpansion}
   A(s) = A^{(0)}(s) + A^{(1)}(s) + \mO(s^3) \ .
\end{equation}
This amplitude violates exact elastic unitarity $|A|^2={\rm Im} A$, satisfying it only in perturbation theory $|A^{(0)}|^2={\rm Im} A^{(1)}$, which is an important handicap of EFTs and leads to large separations from data at mid-energy (few-hundred MeV above threshold) in hadronic physics. However, if it is employed as the low-energy limit of a $\tilde{A}$ satisfying exact unitarity and obtained from dispersion relations, it gives rise to successful methods (such as the IAM, N/D, Improved-K matrix, large-N unitarization, etc.). These methods differ in numerical accuracy but not in substance~\cite{Delgado:2013loa,Delgado:2015kxa}, as they all reproduce the same resonances in each elastic $IJ$ channel for similar values of the chiral parameters.

The $P$ amplitudes, by Watson's theorem, need to have the same phase as $\tilde{A}$ due to strong rescattering. This we guarantee by satisfying Eq.~(\ref{unitarityexpanded}). Observing that at low energies, $P\approx P^{(0)}$, and enforcing the correct analytical structure in the complex $s$ plane, we proposed~\cite{Delgado:2016rtd} the following unitarization method for the $\gamma\gamma\to \omega\omega$ \emph{scalar} amplitudes,
\begin{equation} \label{Omnes}
   \tilde{P} = \frac{P^{(0)}}{1-\frac{A^{(1)}}{A^{(0)}}} = \frac{P^{(0)}}{A^{(0)}}\tilde{A}\ ,
\end{equation}
which implements the IAM philosophy; here, $\tilde{A}(s) = A^{(0)}(s)/(1-\frac{A^{(1)}(s)}{A^{(0)}(s)})$ is the elastic IAM. Now, for $J=2$, the IAM cannot be employed, and then we resort to the well-known $N/D$ method (we have also checked  that employing the N/D for both $J=0$ and $J=2$ leads to little material difference). Then, a formula similar to Eq.~(\ref{Omnes}) can be used
\begin{equation}\label{chLagr:UnitProceduresWW:Fcoupled3:fin}
   \tilde{P}_{I2} = \frac{P_{I2}^{(0)}}{A_{{\rm L},I2}}A_{I2}^{\rm N/D},
   \quad I=0,2 .
\end{equation}
Here, the $N/D$ elastic amplitude has been employed; this is somewhat more complicated than the IAM,
\begin{equation}\label{unitar:ND:elastic}
\tilde{A} = A^{\rm N/D} = \frac{A_L(s)}{1+\frac{1}{2}g(s)A_L(-s)},
\end{equation}
and requires giving further detail on Eq.~(\ref{EFTexpansion}), as the quantities
\begin{subequations}
\begin{align}
   g(s)   &= \frac{1}{\pi}\left(\frac{B(\mu)}{D}+\log\frac{-s}{\mu^2}\right)       \label{unitar:ND:elastic:g}\\
   A_L(s) &= \left(\frac{B(\mu)}{D} + \log\frac{s}{\mu^2}\right) D s^2 = \pi g(-s)Ds^2   \label{unitar:ND:elastic:AL}
\end{align}
\end{subequations}
are built from the $B$ and the $D$ factors defined by  
\begin{subequations}
\begin{align}
   A^{(0)}(s) &= Ks \label{def:A:LO}\\
   A^{(1)}(s) &= \left(B(\mu) + D\log\frac{s}{\mu^2} + E\log\frac{-s}{\mu^2}\right)s^2\ . \label{def:A:NLO}
\end{align}
\end{subequations}
These are computed in perturbation theory and have been reported earlier in~\cite{Delgado:2015kxa}.
The amplitudes are $\mu$-independent because $B(\mu)$ runs in such a way as to absorb the dependence coming from the logarithms.

\subsection{Invariant amplitude and differential cross section}

The non-vanishing matrix elements can be reconstructed from the (unitarized) partial waves by
\begin{subequations}
\begin{align}
\tilde T_{Ip}     &= 64\pi^{3/2}\cdot Y_{0,0}(\Omega)\cdot\tilde P_{I0}     &
\tilde T_I^{+-}   &= 64\pi^{3/2}\cdot Y_{2,2}(\Omega)\cdot\tilde P_{I2}     \\
\tilde R_{0p}     &= 64\pi^{3/2}\cdot Y_{0,0}(\Omega)\cdot\tilde R_0        &
\tilde T_I^{-+}   &= 64\pi^{3/2}\cdot Y_{2,-2}(\Omega)\cdot\tilde P_{I2} ,
\end{align}
\end{subequations}
where $I\in\{0,2\}$. $\tilde T_{I0}$ and $\tilde R_{00}$ are related with the positive parity state $(\Ket{++}+\Ket{--})/\sqrt{2}$ by means of the definition
\begin{subequations}
\begin{align}
  \tilde T_{Ip}&\equiv\frac{1}{\sqrt{2}}(\tilde T_I^{++}+\tilde T_I^{--}) = \sqrt{2}\tilde T_I^{++} \\
  \tilde R_{0p}&\equiv\frac{1}{\sqrt{2}}(\tilde R_0^{++}+\tilde R_0^{--}) = \sqrt{2}\tilde R_0^{++} .
\end{align}
\end{subequations}
Since we have 4 possible $\gamma\gamma$ initial states, the differential cross section for $\gamma\gamma\to\omega\omega$ will be
\begin{equation}\begin{split}\label{crosssecggww}
\frac{d\,\sigma_{\gamma\gamma\to\omega\omega}}{d\,\Omega} &= %
    \frac{1}{64\pi^2 s_{\gamma\gamma}}\cdot\frac{1}{4}\cdot\sum_j\lvert M_j\rvert^2 \\ %
    &= \frac{16\pi}{s_{\gamma\gamma}}\sum_{I\in\{0,2\}}\left[%
             \left\lvert\tilde P_{I0}\cdot Y_{0,0}(\Omega)\right\rvert^2
            +\left\lvert\tilde P_{I2}\cdot Y_{2,2}(\omega)\right\rvert^2 
            +\left\lvert\tilde P_{I2}\cdot Y_{2,-2}(\omega)\right\rvert^2 
    \right] \\
    &= \frac{16\pi}{s_{\gamma\gamma}}\left[ %
        \left(\lvert\tilde P_{00}\rvert^2 + \lvert\tilde P_{20}\rvert^2\right)\cdot\lvert Y_{0,0}(\Omega)\rvert^2 %
      +2\left(\lvert\tilde P_{02}\rvert^2 + \lvert\tilde P_{22}\rvert^2\right)\cdot\lvert Y_{2,2}(\Omega)\rvert^2 %
     \right]
\end{split}\end{equation}
And, for $\gamma\gamma\to hh$,
\begin{equation}
\frac{d\,\sigma_{\gamma\gamma\to hh}}{d\,\Omega} = %
    \frac{16\pi}{s_{\gamma\gamma}}\left\lvert\tilde R_0\cdot Y_{0,0}(\Omega) \right\rvert^2
\end{equation}

In implementing these two equations, which are a backbone of the computation, we have employed the Inverse Amplitude Method extension in equation~(\ref{Omnes}) for the $J=0$ channels, as is it is the one which has been more extensively studied in low-energy chiral perturbation theory and its uncertainties are well understood. For the $J=2$ resonances, the Inverse Amplitude Method cannot be used as a parametrization as it would require knowing the NNLO amplitude in the HEFT. As this is not the case, we have compromised and used the $N/D$ method as laid out in Eq.~(\ref{unitar:ND:elastic}). 

By using the change of basis from the $\omega\omega$ isospin one, $\Ket{I,M_I}$, to the charge one, $\{\Ket{\omega^+\omega^-},\,\Ket{\omega^-\omega^+},\,\Ket{zz}\}$,
\begin{subequations}
\begin{align}
   \Ket{\omega^+\omega^-} &= -\frac{1}{\sqrt{6}}\left(\Ket{20}+\sqrt{2}\Ket{00}\right)-\frac{1}{\sqrt{2}}\Ket{10} \\
   \Ket{\omega^-\omega^+} &= -\frac{1}{\sqrt{6}}\left(\Ket{20}+\sqrt{2}\Ket{00}\right)+\frac{1}{\sqrt{2}}\Ket{10} \\
   \Ket{zz}               &= \frac{1}{\sqrt{3}}\left(\sqrt{2}\Ket{20}-\Ket{00}\right),
\end{align}
\end{subequations}
and taking into account that $\gamma\gamma$ states do not couple with $J=1$ gamma--gamma states, the unpolarized $\gamma\gamma\to\{\omega^+\omega^-,\,zz\}$ differential cross section can be written as
\begin{subequations}
\begin{align}
   \frac{d\,\sigma_{\gamma\gamma\to\omega^+\omega^-}}{d\,\Omega} &= \frac{d\,\sigma_{\gamma\gamma\to\omega^-\omega^+}}{d\,\Omega}  \nonumber \\  %
        &=\frac{16\pi}{s}\cdot\frac{1}{6}\cdot\left[%
            \left\lvert\tilde P_{20}+\sqrt{2}\tilde P_{00}\right\rvert^2\cdot\lvert Y_{0,0}(\Omega)\rvert^2 %
          +2\left\lvert\tilde P_{22}+\sqrt{2}\tilde P_{02}\right\rvert^2\cdot\lvert Y_{2,2}(\Omega)\rvert^2 %
        \right] \\
   \frac{d\,\sigma_{\gamma\gamma\to zz}}{d\,\Omega} &= %
        \frac{16\pi}{s}\cdot\frac{1}{3}\cdot\left[%
            \left\lvert\sqrt{2}\tilde P_{20}-\tilde P_{00}\right\rvert^2\cdot\lvert Y_{0,0}(\Omega)\rvert^2 %
          +2\left\lvert\sqrt{2}\tilde P_{22}-\tilde P_{02}\right\rvert^2\cdot\lvert Y_{2,2}(\Omega)\rvert^2 %
    \right]
\end{align}
\end{subequations}

If we take the SM limit as laid out at the end of subsec.~\ref{subsec:partwaves}, we find $\frac{d\,\sigma_{\gamma\gamma\to zz}}{d\,\Omega}\to 0$ and $\frac{d\,\sigma_{\gamma\gamma\to\omega^+\omega^-}}{d\,\Omega}\to \frac{\pi \alpha^2}{s}\frac{\arrowvert Y_2^2\arrowvert^2}{3}$, respectively. 

A seeming puzzle with this expression is that the tree--level perturbative expression for $\gamma\gamma\to\pi\pi$  (discussed at length in chiral perturbation theory in~\cite{Bijnens:1987dc}), a pure scalar electrodynamics result, is given by 
\ba \label{BijnensCornet}
\frac{d\,\sigma_{\gamma\gamma\to\omega^+\omega^-}}{d\, \cos\theta} = \pi \frac{\alpha^2}{s} 
\ea
which is independent of the polar angle, and does not contain any factor $\arrowvert  Y_2^2\arrowvert^2$. This difference is an artifact of our partial wave expansion: if we wanted to recover the Born-like result of Eq.~(\ref{BijnensCornet}) we would need to resum the partial wave series. For example, the first few $P_{0J}$ with even $J=2\dots 12$ are
$\alpha/(6\sqrt{2})$ (given in Eq.~(\ref{A:scatter:gammagamma:partialWaves:F:alpha:1})), 
$\alpha/(6\sqrt{30})$, $\alpha/(6\sqrt{140})$, $\alpha/(6\sqrt{420})$, $\alpha/(6\sqrt{990})$, $\alpha/(6\sqrt{2002})$, and the first few $P_{2J}$ are
$\alpha/12$ (given in Eq.~(\ref{A:scatter:gammagamma:partialWaves:F:alpha:2})), 
$\alpha/(12\sqrt{15})$, $\alpha/(12\sqrt{70})$, $\alpha/(12\sqrt{210})$, $\alpha/(12\sqrt{495})$, $\alpha/(12\sqrt{1001})$. Each of these quantities multiplies the corresponding spherical harmonic in reconstructing the perturbative amplitude. The series is well behaved for any fixed angle $\theta$, but in truncating it, we introduce a spurious angle dependence.

We have not pursued the issue further since our aim is not to present precise off-resonance cross-sections for production of the EWSBS particles; this can be best computed by standard means (Feynman amplitudes not expanded in $J$). Both methods can also work together and part of us have recently assessed it, in a separate collaboration~\cite{Delgado:2017cls}, to implement in LHC Monte Carlo simulations.

Our goal here is to produce the resonance cross-sections; and near a BSM resonance, the dominance of its corresponding partial wave over all the other, perturbative ones, is warranted in the presence of experimental angular acceptance cuts that avoid any forward Coulomb divergence. Thus, in the figures that follow, one should pay attention to the differential cross-sections near the peak, and not take too seriously the background cross-sections that are affected by factors of order 1. The effect is lesser in directions perpendicular to the beam axis (low rapidity).

\subsection{Inverse process $\omega\omega\to\gamma\gamma$}
As an aside, and for completeness, we also give expressions for the process $\omega\omega\to\gamma\gamma$ (and for $hh\to\gamma\gamma$) that may be useful in the study of resonances decaying by the two photon channel. Assuming time reversal invariance $\Braket{i|T|j}=\Braket{j|T|i}$, and considering that we have $(2+1)^2 = 9$ possible initial states $\Ket{I,I_z}$, we obtain
\begin{align}
\frac{d\,\sigma_{\omega\omega\to\gamma\gamma}}{d\,\Omega} &= %
    \frac{4}{9}\frac{d\,\sigma_{\gamma\gamma\to\omega\omega}}{d\,\Omega} &
\frac{d\,\sigma_{hh\to\gamma\gamma}}{d\,\Omega} &= %
    \frac{4}{9}\frac{d\,\sigma_{\gamma\gamma\to hh}}{d\,\Omega}
\end{align}
Finally, since there is only 1 possible initial state, $\omega^+\omega^-\to\gamma\gamma$ and $zz\to\gamma\gamma$ can be written as
\begin{subequations}
\begin{align}
   \frac{d\,\sigma_{\omega^+\omega^-\to\gamma\gamma}}{d\,\Omega} &= %
       4\frac{d\,\sigma_{\gamma\gamma\to\omega^+\omega^-}}{d\,\Omega} &
   \frac{d\,\sigma_{zz\to\gamma\gamma}}{d\,\Omega} &= %
       4\frac{d\,\sigma_{\gamma\gamma\to zz}}{d\,\Omega}
\end{align}
\end{subequations}


\section{Production in $e^-e^+$ collisions} \label{sec:leptons}

\begin{figure}
\centering
\includegraphics[width=0.5\textwidth]{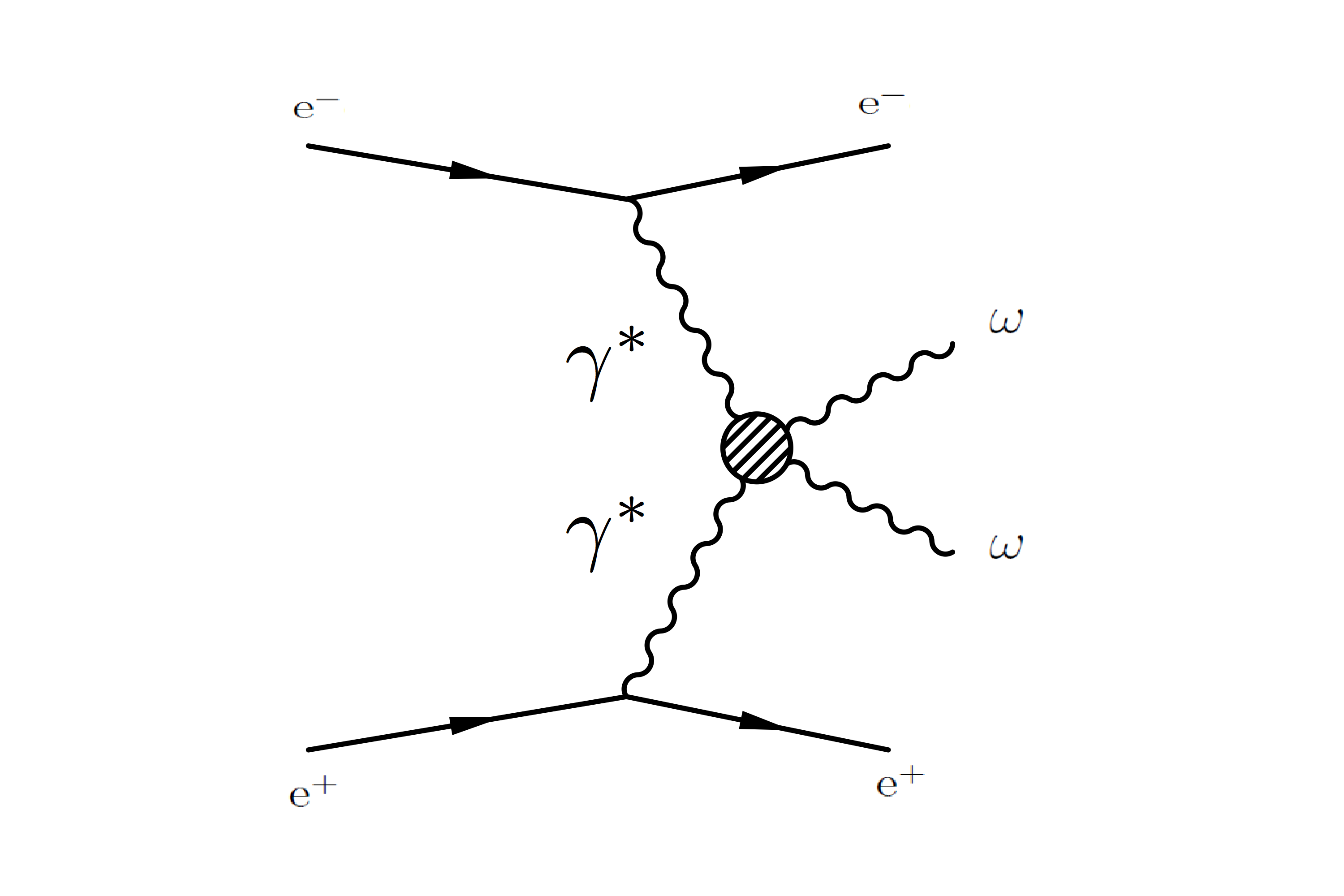}
\caption{Feynman diagram for EWSBS-pair photon-induced production at the ILC.
\label{fig:Feynmaneeprod}}
\end{figure}

The aim of this section is to study the differential cross section $\frac{d\sigma}{ds_{\gamma\gamma}dp_{T}^{2}}$ to photoproduce pairs of longitudinal $W_{L}$ electroweak bosons in $e^{-}e^{+}\rightarrow e^{-}e^{+}+\gamma\gamma\rightarrow e^{-}e^{+}+W_{_{L}}W_{L}$ at an energy of 1 TeV, the top of the energy range of the International Linear Collider and above. $\frac{d\sigma}{ds_{\gamma\gamma}dp_{T}^{2}}$ is obtained through the convolution of photon flux functions derived from the Equivalent Photon Approximation and the $\gamma\gamma\to \omega\omega$
cross section described in section~\ref{sigma:photo}. Fig.~\ref{fig:Feynmaneeprod} shows the characteristic Feynman diagram to be evaluated.

We will work in the $ee$ center of mass frame so that $p_{e^-}=(E;0,0,E)$, $p_{e^+}=(E;0,0,-E)$ and Mandelstam's invariant squared energy will be $s_{ee}=4E^2$. The only exception to this massless-electron kinematics will be in the regulation of $x$-integrals such as Eq.~(\ref{limitsxy}) below; thus, we actually work in the leading $\frac{m_e}{\sqrt{s}}$ approximation which leads to finite answers.

In the Equivalent Photon Approximation~\cite{Budnev:1974de}, the charged leptons can radiate collinear photons~\cite{Lyth:1973mm} (since their boosted Coulomb field is practically transverse to the lepton direction of motion) so that we may take the photon momenta as $p_{\gamma1}=(\omega_{1};0,0,\omega_{1})$ and $ p_{\gamma2}=(\omega_{2};0,0,-\omega_{2}) $; the corresponding invariant is $s_{\gamma\gamma}=s_{W_LW_L}=4\omega_1\omega_2$. If these two photons do enter into the EWSBS through a resonance, this is not produced in its rest frame as their momenta are not opposite in the laboratory. Instead, each photon carries a different fraction of its parent lepton momentum,
\begin{equation}
\begin{cases}
\omega_{1}=xE\\
\omega_{2}=yE \ .
\end{cases}\label{fracsmomentum}
\end{equation}
Substituting these two into $s_{\gamma\gamma}$ and eliminating $E$ for $s_{ee}$ we find the constraint
\begin{equation}
y=\frac{s_{\gamma\gamma}}{xs_{ee}}\label{xconstrainsy}\ .
\end{equation}

The variables $x$ and $y$ are bound above (by the maximum energy available from the electron),
\begin{equation}
y_{\rm max}=x_{\rm max}=1-\frac{m_{e}}{E}=1-\frac{2m_{e}}{\sqrt{s_{ee}}}\label{xmaxbound}\ .
\end{equation} 
Then Eq.~(\ref{xconstrainsy}) also gives a lower bound 
\be\label{limitsxy}
x_{\rm min} = y_{\rm min} = \frac{s_{\gamma\gamma}}{s_{ee}-2m_e\sqrt{s_{ee}}}
\ee
so that integration over the photon momentum fractions never hits the end points and is regular.

Neglecting all masses and photon virtualities we can interpret the Mandelstam $\gamma\gamma$ variables in the center of mass of the $\gamma\gamma\to \omega\omega$ subsystem as usual
\be
s_{\gamma\gamma} =  4p_{cm}^2\ \ \ \ \  t_{\gamma\gamma\to\omega\omega} = -2p_{cm}^2(1-cos\theta)
\ee
and can trade  $t_{\gamma\gamma}$ (the variable in terms of which our Feynman amplitudes are expressed) for the more directly measurable $p_t^2$ as 
\begin{equation}
p_{T}^{2}=-t\left(1+\frac{t}{s_{\gamma\gamma}}\right)\label{pttot}\ .
\end{equation}
Then, it immediately follows that
\ba
\frac{dp_{T}^{2}}{dt}=-1-\frac{2t}{s_{\gamma\gamma}}\\ \nonumber
\frac{dt}{d\Omega}=-\frac{s_{\gamma\gamma}}{4\pi} 
\ea
so the $\Omega$ angular dependence of the cross-section can be traded for one in $p_t^2$. Then, Eq.~(\ref{crosssecggww}) becomes 
\be
\frac{d\sigma_{\gamma\gamma\rightarrow\omega\omega}}{dp_{T}^{2}}=\frac{d\sigma_{\gamma\gamma\rightarrow\omega\omega}}{d\Omega}\ldotp\frac{d\Omega}{dt}\cdotp\frac{dt}{dp_{T}^{2}}=\frac{4\pi}{s_{\gamma\gamma}}\cdotp\frac{1}{1+\frac{2t}{s_{\gamma\gamma}}}\frac{d\sigma_{\gamma\gamma\rightarrow\omega\omega}}{d\Omega}
\ee

The photon virtualities are also bound~\cite{Budnev:1974de}
\ba
Q_{min}^{2}=\frac{m_{e}^2\omega^{2}}{E\left(E-\omega\right)}=\frac{m_{e}^2x^{2}}{1-x}\nonumber \\ \label{Qsqbounds}
Q_{max}^{2}=4E\left(E-\omega\right)=4E^{2}\left(1-x\right)=s_{ee}(1-x) 
\ea
and these bounds limit the interval of validity of the photon number density per unit energy and virtuality,
Eq.~(D.4) from~\cite{Budnev:1974de},
\begin{equation}
dn_{i}=\frac{\alpha}{\pi}\frac{d\omega_{i}}{\omega_{i}}\frac{d(Q_i^{2})}{\left\Vert Q_{i}^{2}\right\Vert }\left[\left(1-\left\Vert \frac{Q_{i,min}^{2}}{Q_{i}^{2}}\right\Vert \right)\left(1-\frac{\omega_{i}}{E}\right)D_{i}+\frac{\omega_{i}^{2}}{2E^{2}}C_{i}\right]
\end{equation}
where $C$ and $D$ are two constants that parametrize the internal structure of the charged particle (and, as usual, $Q^2=-q^2$). 

\begin{figure}
\centering
\includegraphics[width=0.48\textwidth]{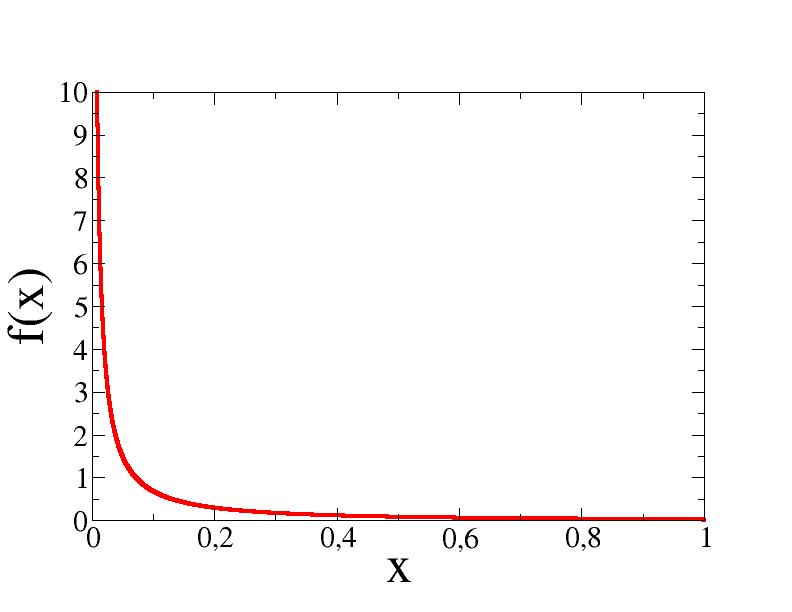}
\includegraphics[width=0.48\textwidth]{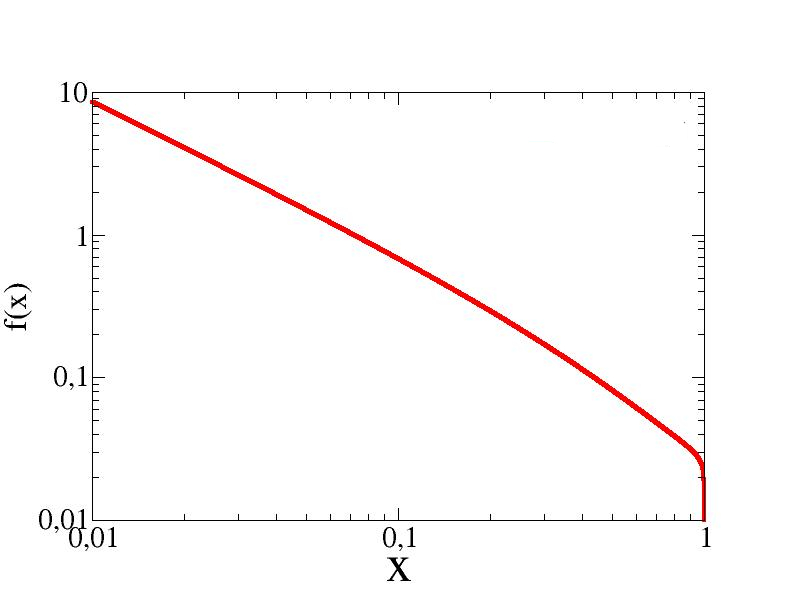}
\caption{\label{fig:flux}
Photon number density per unit $x$ (fractional energy taken from the electron), in linear (left) and logarithmic (right) scales. We fixed $s_{ee}$ to 1 TeV$^2$.}
\end{figure}

In the case of pointlike elementary particles such as $e^-e^+$, $C=D=1$, the photon flux can be integrated over virtuality to interpret it in a manner analogous to a parton distribution function,
\begin{equation}
\frac{dn}{dx}=f(x)=\frac{\alpha}{\pi x}\int_{Q_{min}^{2}}^{Q_{max}^{2}}\left[\frac{Q^{2}-Q_{min}^{2}}{Q^{4}}(1-x)+\frac{x^{2}}{2Q^{2}}\right]dQ^{2}\ .
\end{equation}
The integral over $Q^2$ can be performed analytically, yielding
\begin{equation}
f(x)=\frac{\alpha}{\pi x}\left\{ -1+\frac{x+(\frac{m_{e}^{2}}{s_{ee}}-1)x^{2}}{1-x}+\left[2-2x+x^{2}\right]\ln\left(\sqrt{s_{ee}}\frac{1-x}{m_{e}x}\right)\right\} \label{gammaine}
\end{equation}
which is represented in figure~\ref{fig:flux}.

With the photon fluxes at hand, we may now mount the cross section for the 
$\omega\omega$ production process, by means of
\begin{multline}
  \sigma\left(e^-e^+\to e^-e^+\gamma\gamma\to e^-e^+ \omega\omega\right) %
  = \\ %
  \int d\omega_{1}\int d\omega_{2}\frac{f_{\gamma/e^-}\left(\omega_{1}\right)}{\omega_{1}}\frac{f_{\gamma/e^+}\left(\omega_{2}\right)}{\omega_{2}}\sigma\left(\gamma\gamma\rightarrow\omega\omega\right)
\end{multline}
or, in differential form,
\begin{multline}
  \frac{d\sigma\left(e^-e^+\to e^-e^+\gamma\gamma\to e^-e^+ \omega\omega\right)}{dsdp_{T}^{2}}\left(s_{\gamma\gamma},\theta\right) %
  = \\ %
  \frac{1}{s_{\gamma\gamma}}\intop_{x_{min}}^{x_{max}}dx_{1}\frac{f\left(x_{1}\right)}{x_{1}}f\left(\frac{s_{\gamma\gamma}}{s_{ee}x}\right)\frac{d\sigma_{\gamma\gamma\rightarrow\omega\omega}\left(s_{\gamma\gamma},\theta\right)}{dp_{T}^{2}}\label{seceefinal}\ .
\end{multline}

\subsection{Some numerical examples}

We exemplify Eq.~(\ref{seceefinal}) with a set of parameters characteristic of the EWSBS in the presence of new physics. For simplicity we will decouple the $hh$ channel setting $b\simeq a^2$. We keep the LO parameter $a=0.81$ fixed and separating from its SM value (that would be $a=1$). This particular value is chosen because it is just under the $2\sigma$ recently proposed exclusion bound~\cite{Buchalla:2015qju}. Those authors report $a=0.98\pm 0.09\,(1\sigma)$ from current LHC data.

We generate elastic $\omega\omega$ resonances by means of the $a_4$ and $a_5$ NLO parameters, fixing all others to zero at NLO and higher; in this way, the entire coupling to the $\gamma\gamma$ sector is provided by the electron squared charge in $\alpha_{em}$. We have chosen the sets $a_4=10^{-3}$, $a_5=0$ and $a_4=10^{-3}$, $a_5=10^{-3}$, that have increasing BSM strength at NLO and generate resonances at decreasing $s_{\gamma\gamma}$. All these parameters are understood to be taken at the renormalization scale $\mu=3\,{\rm TeV}$ (their running to other scales can be found in our earlier work~\cite{Delgado:2013loa,Delgado:2015kxa}). They are basically unconstrained except for the current absence of BSM resonances. The sets we use do provide resonance in the energy region just above $1\,{\rm TeV}$.

We have chosen $p_t=50,100,200$ GeV that would pass typical experimental cuts\footnote{For example, CMS demands $p_t(e\mu)$ above 30 GeV to suppress background from $\tau\tau$ production when searching for $WW$, and $p_t(2(l^-l^+))$ above 40 GeV to suppress quarkonium when seeking $ZZ$.} and future $e^-e^+$ machines will similarly impose $p_t$ cuts at trigger time.

\begin{figure}
\includegraphics[width=0.48\textwidth]{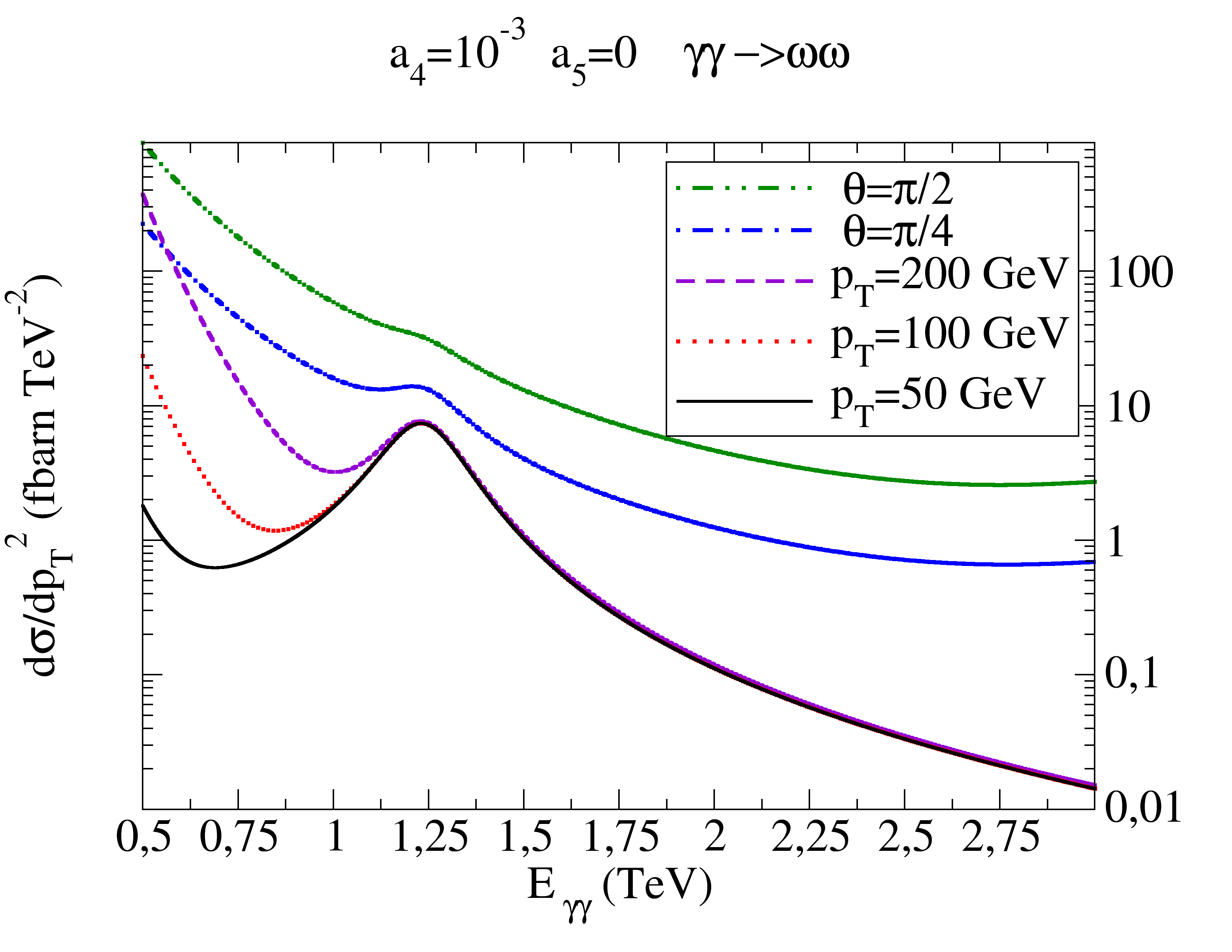}
\includegraphics[width=0.48\textwidth]{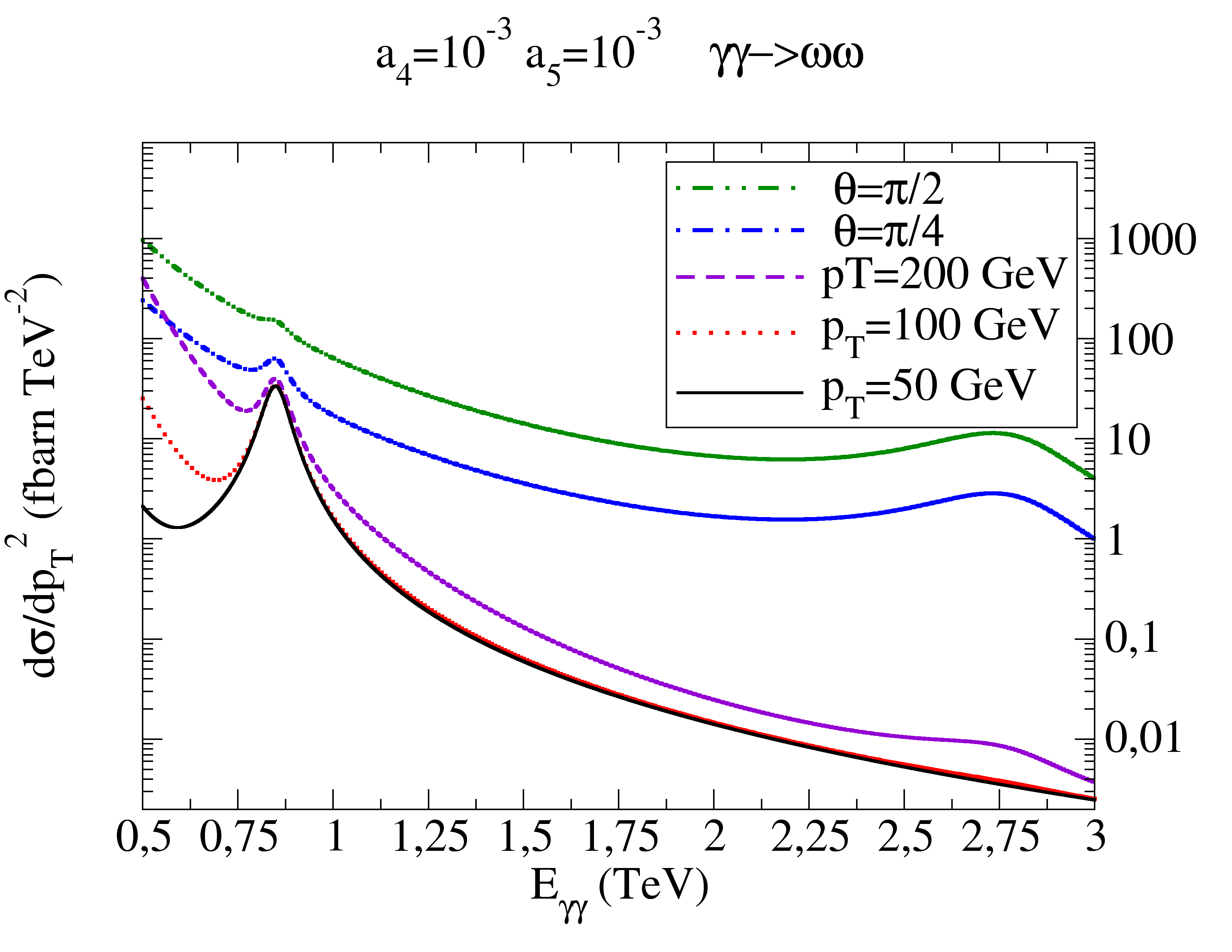}
\caption{\label{fig:Xsecgammagamma}  
Differential cross section for $\gamma\gamma\to W_LW_L/Z_LZ_L$ at the indicated energy.
Left: $a_4=10^{-3}$, $a_5=0$. Right: $a_5=a_4=10^{-3}$ as indicated. Both sets induce resonances around 1 TeV; the one at higher mass in the right plot is narrower ($f_2$-like) and the one common to both plots is an $f_0$-like, broader structure.  
We show both fixed-angle (in the CM) and fixed-$p_t$ scattering.
}
\end{figure}

Fig.~\ref{fig:Xsecgammagamma} represents the differential cross section per unit squared $p_t$ as function of the $\gamma\gamma$ (viz. $\omega\omega$) center of mass energy ($\sqrt{s_{\gamma\gamma}}$) (and also as function of the scattering angle $\theta$ in the cm frame of $\gamma\gamma$). The left and right plots have been produced with each of the two parameter combinations and show a clear resonance around a TeV.

Both plots show similar features. The fixed-angle cross section is larger and falls slower with the energy than the fixed $p_t$ one. Resonances are however clearer at fixed $p_t$, and their line shape is the better resolved the lighter they are (with the parameters chosen, a strong scalar resonance appears around 1~TeV). 

\begin{figure}
\includegraphics[width=0.48\textwidth]{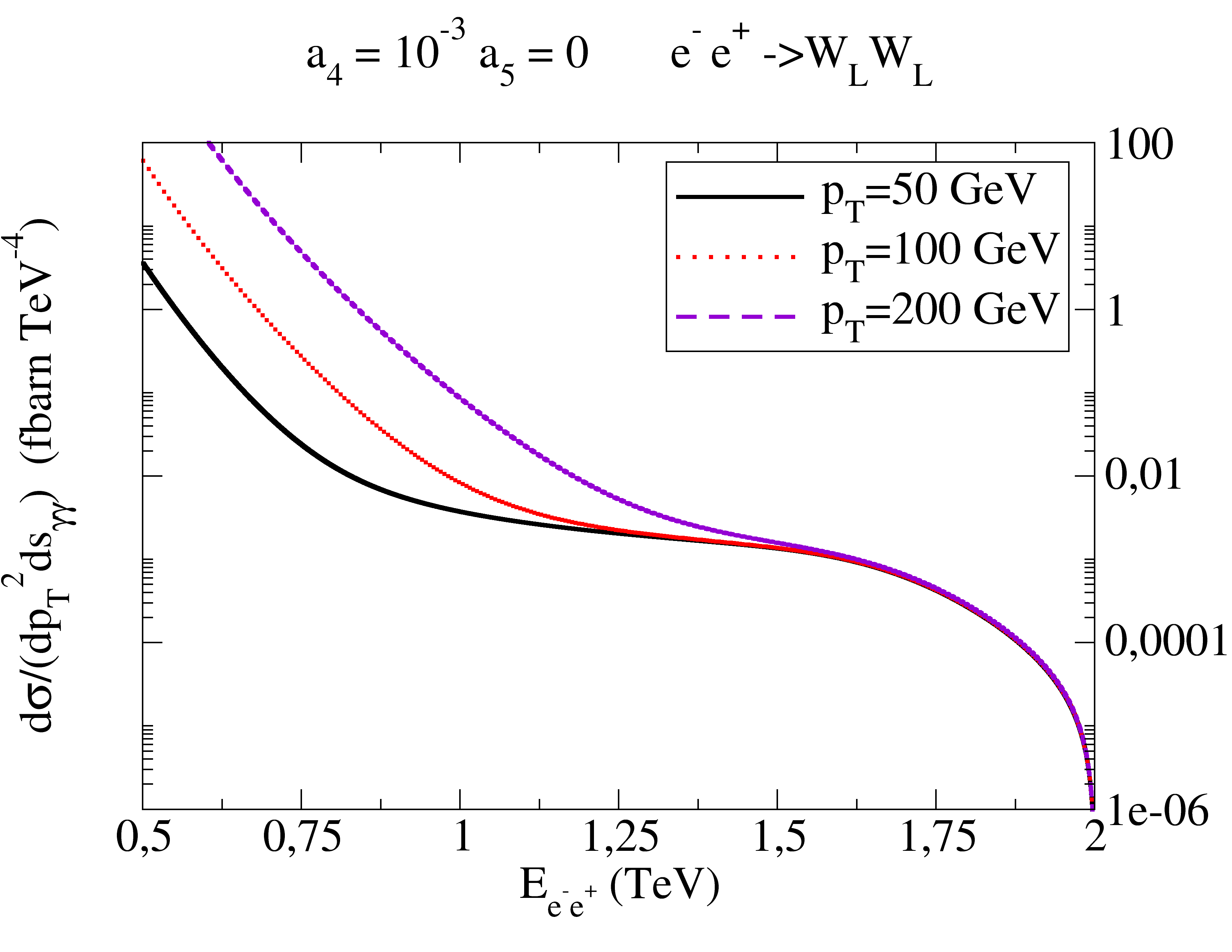}
\includegraphics[width=0.48\textwidth]{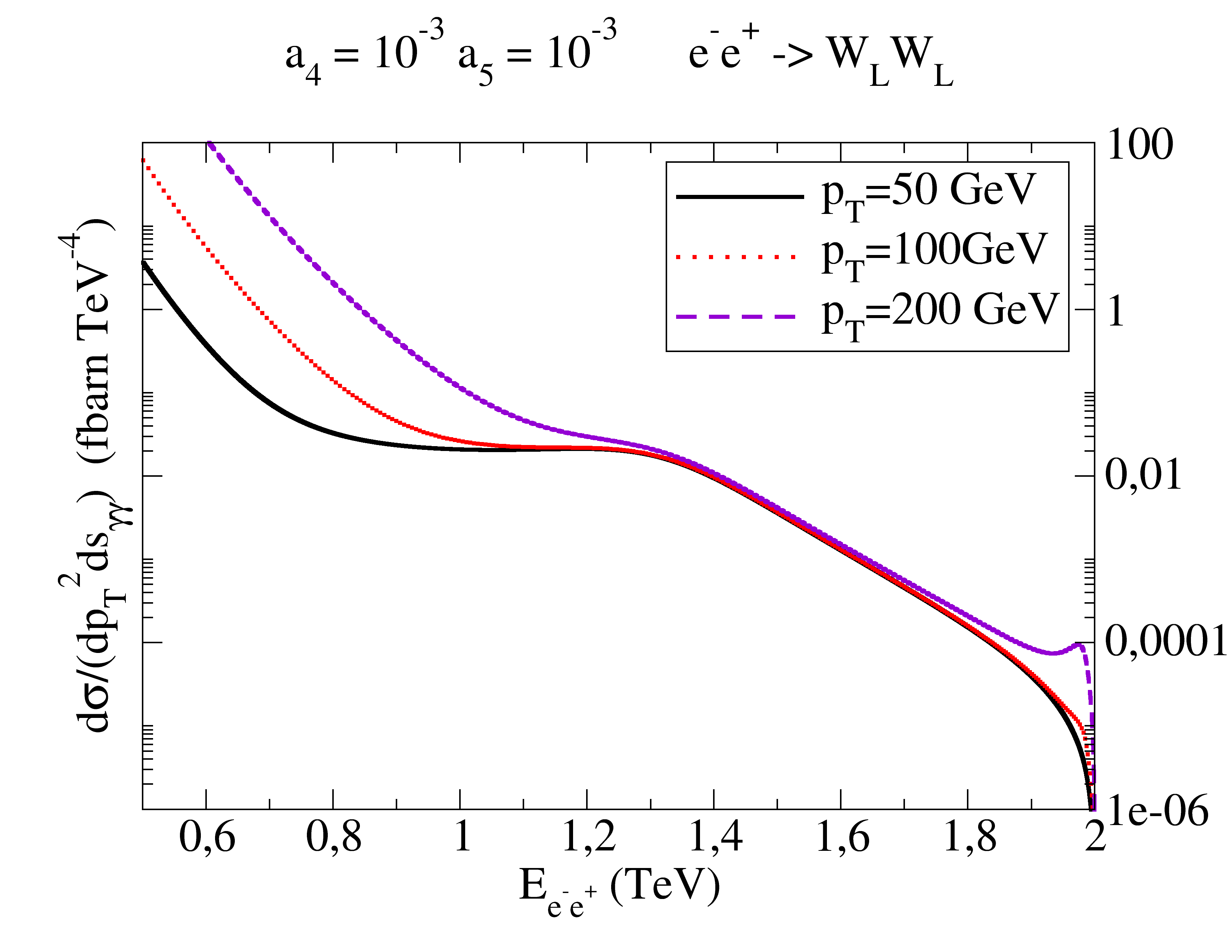}
\caption{\label{fig:Xsecee}  
Cross section for $e^-e^+\to e^-e^+ \gamma\gamma\to e^-e^+ \omega \omega$,
differential respect to $p_t^2$ and the cm energy of the photon ($\omega$) pair. 
Left and right plots with the same parameters as Fig.~\ref{fig:Xsecgammagamma}. The $e^- e^+$ energy is fixed at 2 TeV. 
}
\end{figure}

Fig.~\ref{fig:Xsecee} shows the convolution of the cross section $\gamma\gamma\to \omega\omega$ in Fig.~\ref{fig:Xsecgammagamma} with the photon flux factors, to yield
the $e^-e^+$ production cross section that can be readily obtained in experiment.

This we present in doubly differential form, respect to $E_{e^-e^+}$ and respect to $p_t^2$ (of each produced $\omega$).
If the resonance of the EWSBS is above 1 TeV, as shown in the left plot of the figure, the resonance shape is not so distinct (especially in the presence of experimental errors), but the line shape exposes a clear knee with a change of power-law slope and is shifted to higher values after the resonance. On the other hand, a resonance below 1 TeV is more clearly visible and can be better reconstructed if $p_t$ is modest. For larger $p_t\sim 200$ GeV, the behavior of the line shape is similar to that of a higher energy resonance. 

The reader may be intrigued by the growth of the cross-section with $p_t$ for small $s$.
One should remember that the underlying Lagrangian is Effective Field Theory-based and thus, coupled derivatively. Therefore, an increase of the transverse momentum provides larger amplitudes at the $\gamma\gamma\to W_LW_L$ level. For larger $s$ such that unitarity is saturated and for larger $p_t$ this effect diminishes and the usual kinematic effects lower the cross-section.

\section{Production in $pp$ collisions}
In this section we revisit $\gamma\gamma \rightarrow W^+_LW^-_L$ at hadron colliders, focusing on the LHC ($pp$ initial state), so the complete reaction is $pp\rightarrow ppW^+W^-$ (through $\gamma\gamma$). 

We estimate the cross section $\frac{d\sigma(pp\rightarrow ppW^+_LW^-_L)}{dp_t^2 ds_{\gamma\gamma}}$  from that for the inelastic $\frac{d\sigma(\gamma\gamma \rightarrow \omega \omega)}{dp_t^2}$  in analogy to the case of lepton colliders; the only difference is that now the photon flux is not directly calculable with arbitrary precision, but partly fit to data. Additionally, one or both of the outgoing protons might be disrupted, giving several situations: elastic ($pp$ in the final state), semielastic (only one of the two protons remains), totally inelastic (both protons break up). Additionally, the inelasticity can be few-body, with momentum transfer in the resonance zone, or deeply inelastic, with large $Q^2$ and accepting a parton-level description.

Theoretical studies of photoinduced production at the LHC date more than a decade~\cite{Nystrand:2004vn}, at least for hadron resonances; the cross sections found for those are of course huge in comparison to the electroweak sector. Though the experimental identification of this process is rather difficult in the noisy environment of a hadron collider, taking into account that the photons are not detected and must be inferred, at least 13 events have already been reported by CMS~\cite{Khachatryan:2016mud}. ATLAS also sees candidate events~\cite{Aaboud:2016dkv} and while some are expected for the proposed Future Circular Collider~\cite{dEnterria:2017jmj}, its energy is too low to be of interest for the resonance region.

Moreover, two projects, CT-PPS (CMS-TOTEM Precision Proton Spectrometer~\cite{totem}) and AFP (the ATLAS Forward Proton detector~\cite{Giacobbe:2016miz}) aim at detecting elastically scattered protons near the beampipe. These experiments take data at 200 meters (CT-PPS) and 210 meters (AFP) from their respective interaction points. Both employ the LHC magnets to separate the scattered protons by their momentum slightly different from that of the beam, and detect them downstream.

The kinematics for the process closely parallels the discussion in section~\ref{sec:leptons} through Eq.~(\ref{xconstrainsy}), just substituting $e\to p$ as necessary.

Given the photon flux in the proton beam as, once more, $f(x)$, and in the collinear photon approximation, we may write~\cite{d'Enterria:2013yra}
\ba \label{factorizacioncolineal}
\frac{d\sigma_{pp\rightarrow ppW^+W^-}}{ds_{\gamma\gamma}}
&=& \int dx dyE_p^2 \frac{f(x)}{E_px}\frac{f(y)}{E_py}\sigma_{\gamma \gamma \rightarrow  W^+_LW^-_L}\delta(s_{\gamma\gamma}-4xyE_p^2) \\
&=&
\int dx \frac{1}{s_{\gamma\gamma}}\frac{f(x)}{x}f(\frac{s_{\gamma\gamma}}{4E_p^2x})\sigma_{\gamma \gamma \rightarrow  W^+_LW^-_L}
\ea
where in the second step we have used the relation analogous to Eq.~(\ref{xconstrainsy}).

If the proton is left intact (elastic photon emission) then the flux factor $f(x)$ is calculable from the electromagnetic  form factor of the proton. In the deep inelastic regime, we can speak of the photon as a parton of the proton; and in the intermediate region, the proton is left in an excited state (one of several resonances), $f(x)$ then being a nontrivial structure function. 

As the photons are collinear with the proton, the angular dependence of the $WW$ emission comes from the reaction $\gamma\gamma\to WW$, as in $e^-e^+$ collisions. Again, $p_t^2$ is, unlike $\Omega$, invariant under longitudinal boosts (and is easily measured) so we take it as second variable and write 
\be
\frac{d\sigma_{\gamma \gamma \rightarrow  \omega^+\omega^-}}{dp_t^2}=\frac{1}{1+2t/s_{\gamma\gamma}}\frac{4\pi}{s_{\gamma\gamma}}\frac{d\sigma}{d\Omega}_{CM},
\ee
where the first factor stems from the variable change $t\rightarrow p_t^2$ and the second from $\Omega\rightarrow t$.

We may then write Eq.~(\ref{factorizacioncolineal}) in double differential form,
\be \label{ecuacion}
\frac{d\sigma_{pp\rightarrow ppW^+W^-}}{dp_t^2 ds_{\gamma\gamma}}=
 \frac{1}{s}\int_{x_{min}} dx\frac{f(x)}{x}f(\frac{s_{\gamma\gamma}}{4E_p^2x})\frac{d\sigma_{\gamma \gamma \rightarrow  W^+W^-}}{dp_t^2}
\ee

\subsection{Photon flux in the proton}
We need to convolute photon-level cross-sections with the collinear photon flux in the proton, with $f_{\gamma|p}(E_\gamma)\equiv f(x)$ ($x=E_\gamma/E_p$) computed under two kinematic regimes which can be distinguished by experimental triggers.

In the first, we take the absorption cross-section for real photons as not too different from that for virtual photons (small virtuality). On top of this approximation, there is the mild assumption that the cross-section must fall quickly after a certain energy. Then one can find, for elastically scattered protons, an expression in terms of the Sachs electromagnetic form factors $G_E$ and $G_M$~\cite{Budnev:1974de,Kniehl:1990iv}  
\be f(x)=\frac{\alpha}{\pi x}\int_{Q^2_{min}}^\infty dQ^2(1-x)  \frac{Q^2-Q^2_{min}}{Q^4}D+ \frac{x^2}{2Q^2}C \label{flujo}
\ee
\be C=G_M^2    \qquad D=\frac{4M_p^2G_E^2+Q^2G_M^2}{4M_p^2+Q^2}
\ee
with lower integration limit $Q^2_{min}=(M_px)^2/(1-x)$. 

Alternatively, a second kinematic regime is the deep inelastic kinematics; $f(x)$ is directly taken as the parton distribution function. We now show parametrizations of both elastic and deeply inelastic photon fluxes.

First to mention is the very crude parametrization employed by Drees and Zeppenfeld~\cite{Drees:1988pp} and also recently adopted in~\cite{Esmaili:2016enf}.
In this high-energy application, the authors neglect the lower limit $Q^2_{min}$, and the magnetic form factor $G_M(Q^2)$, and parametrize the electric one $G_E(Q^2)$ by a simple dipole form
\begin{equation} 
G_E=1/[1+Q^2/(770\,{\rm GeV})^2]^2\ .
\end{equation}
We will plot the resulting photon flux in Fig.~\ref{fig:flujo}, and further include (a) the simple improvement of~\cite{Nystrand:2004vn} that considers the minimum $Q^2$ and (b) the parametrization of Kniehl~\cite{Kniehl:1990iv} that now includes both $G_E$ and $G_M$ at the proton-photon vertex.

In addition to those classic works, we will also work with more modern parametrizations that reflect progress in hadron physics in the last two decades and help better characterize systematic errors.  We will try the low-energy parametrization of Lorenz and Mei\ss ner~\cite{Lorenz:2014vha} based on a conformal coordinate change from $Q^2$ to $z$, 
\be z(Q^2)=\frac{\sqrt{t_{cut}+Q^2}-\sqrt{t_{cut}}}{\sqrt{t_{cut}+Q^2}+\sqrt{t_{cut}}}
\ee
with $t_{cut}=4m_\pi^2$ the charged pion-pair threshold. This allows a Taylor power series expansion in $z$ of the form factors
\be 
G_{E,M}= \sum_{k=0}^{10}a_k z(Q^2)^k \label{factor}
\ee
in terms of  free $a_k$ constants that were adjusted to low-$Q^2$ data (up to about $1\,{\rm GeV}^2$). The expansion converges rather well.

This low energy form should provide a very accurate fit only below that scale, but since the form factor is integrated in Eq.~(\ref{flujo}) to obtain the flux, contributing all the way to $Q^2\simeq6$ GeV, we need to supplement this parametrization with a high-energy contribution. Thus, for $Q^2>1\,{\rm GeV}$ we adopt the simplest Brodsky-Lepage~\cite{Lepage:1979za} form factor that follows the power-law counting of QCD~\cite{Brodsky:1973kr}, that yields for large $Q^2$
\be 
G_M(Q^2)= \frac{C^2 32 \pi^2 \alpha_s^2(Q^2)}{9 Q^4} [\ln(Q^2/\Lambda^2)]^{-12} \label{factorQ}
\ee
$$\alpha_s= \frac{4\pi}{9} \ln(Q^2/\Lambda^2)$$
where $\Lambda=200\,{\rm MeV}$ and $C^2$ is a constant that we use to match continuously with the low-$Q^2$ parametrization; this we do at $Q^2_{\rm match}=850\,{\rm GeV}^2$. As for the electric form factor, in this simple parametrization it is obtained from $G_E=G_M/\mu_p$.

\begin{figure}[h]
	\centering
	\includegraphics[width=0.48\textwidth]{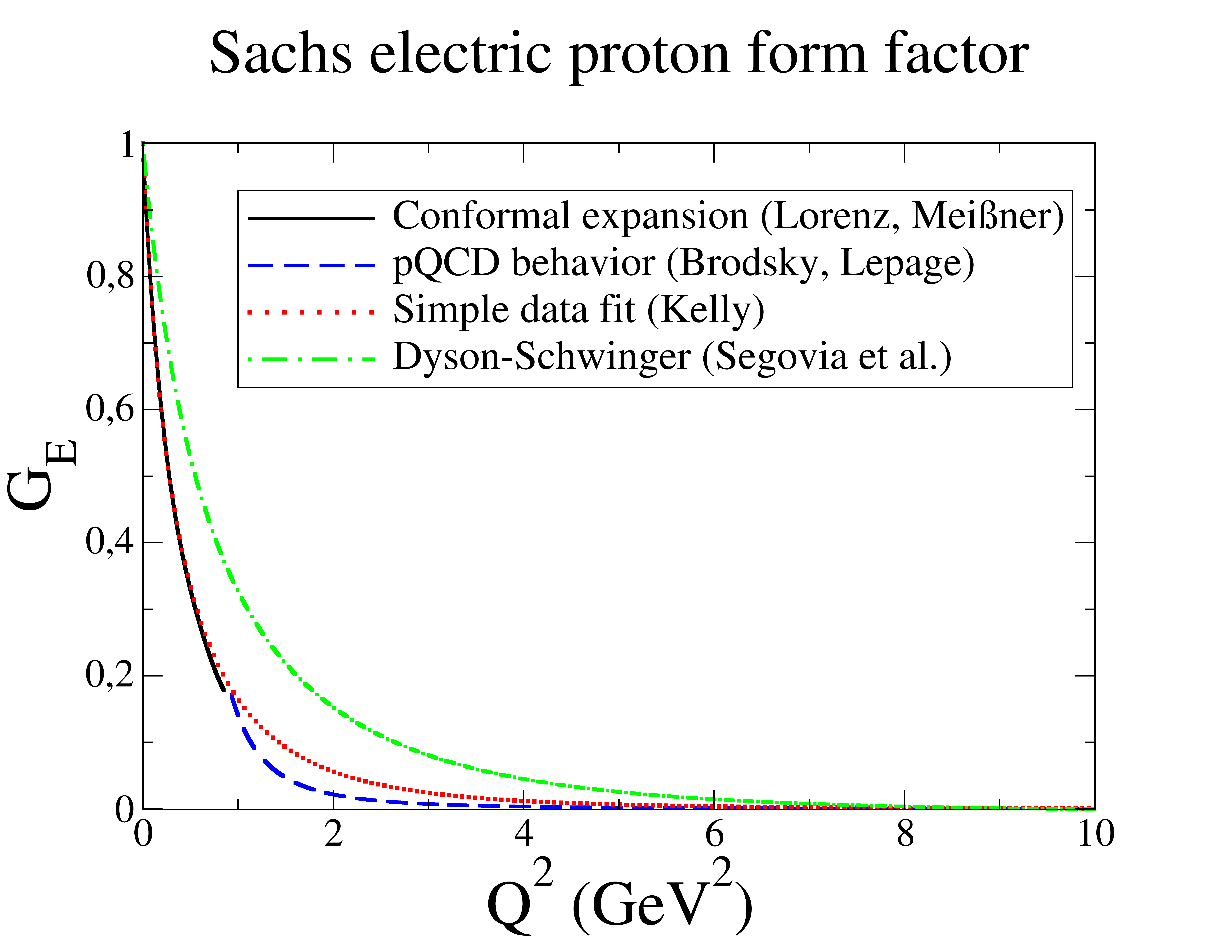}
	\includegraphics[width=0.48\textwidth]{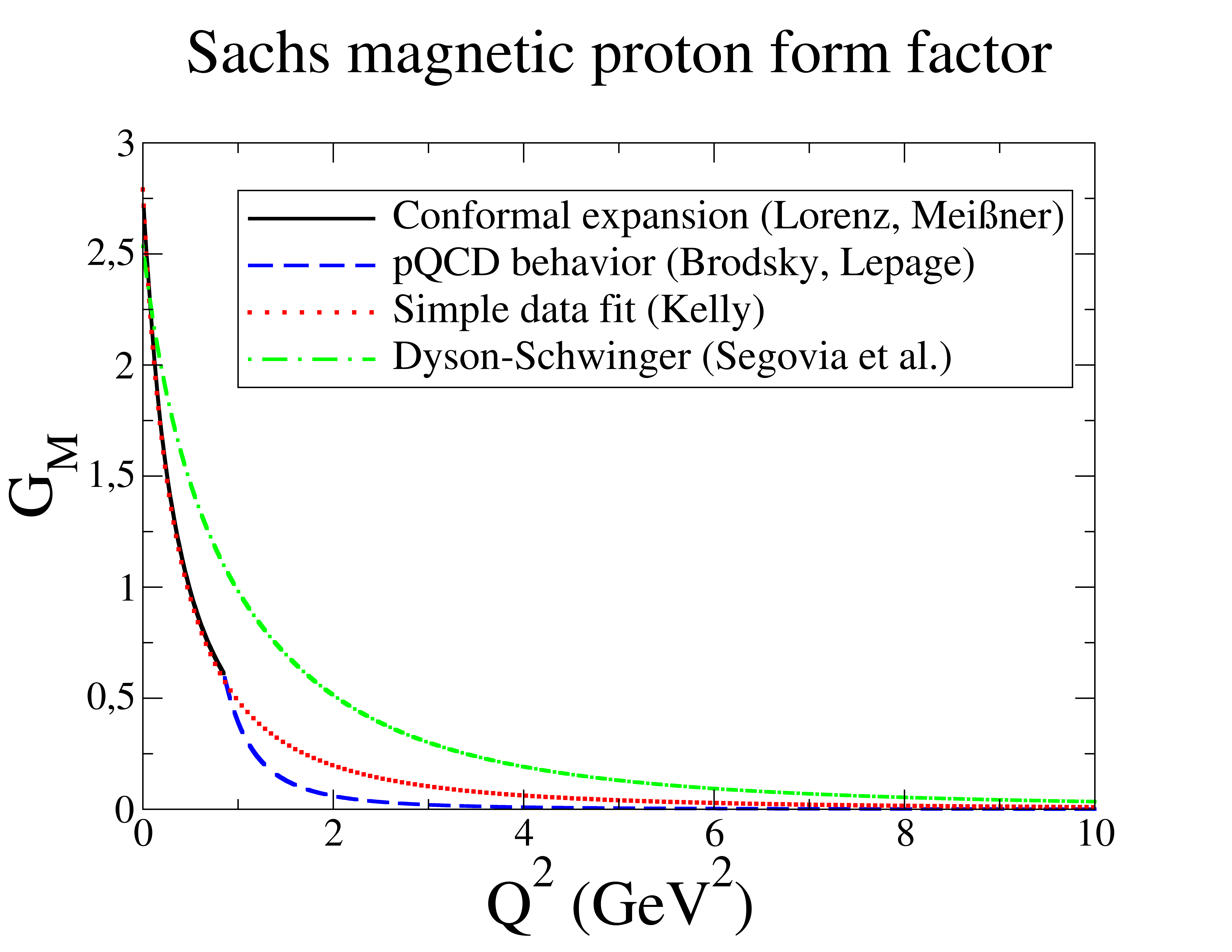}
	\caption{Some form factor parametrizations and theoretical computations as functions of $Q^2$ taken from the ample literature. Solid and dashed line: matching Eq.~(\ref{factor}) at low momentum transfer with the Brodsky-Lepage asymptotic form. Additionally, simple fit to the data by Kelly (dotted line) and Dyson-Schwinger computation by Segovia and others (dotted-dashed line).
}
\label{figfactor}
\end{figure}

\begin{figure}[h]
	\centering	
	\includegraphics[width=0.6\textwidth]{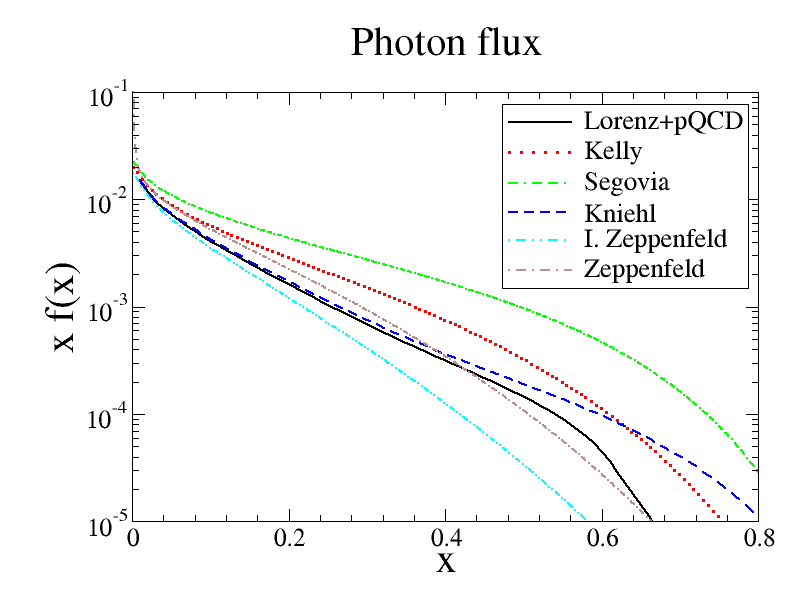}
	\caption {Photon flux (multiplied by $x$) against $x$ to be used for elastic scattering (the proton exits the collision intact). Continuous black line: 
parametrization matching dispersion relations at low momentum and counting rules at high $Q^2$. Dotted line (red online): the flux as computed by Drees and Zeppenfeld~\cite{Drees:1988pp}. Dashed-dotted line (green online): improvement over this that includes the minimum $Q^2$ cutoff~\cite{Nystrand:2004vn}. Dashed line (blue online): flux factor of Kniehl \cite{Kniehl:1990iv} including both $G_E$ and $G_M$.}
	\label{fig:flujo}
\end{figure}

Fig.~\ref{figfactor} represents the form factor obtained by matching these low-energy data fit and asymptotic parametrization. The use of this form factor does not significantly change the results obtained with those of~\cite{Kniehl:1990iv,Drees:1988pp}, and \cite{Nystrand:2004vn}.

Finally, we also include in the figure two more contemporary parametrizations of these form factors. One is the data-oriented fit of Kelly~\cite{Kelly:2004hm} and the other, a theoretical computation by Segovia {\it et al}. based on the Dyson-Schwinger equations~\cite{Segovia:2014aza}.

\begin{figure}
	\centering
	\begin{minipage}[t]{.48\textwidth}
		\includegraphics[width=1\textwidth]{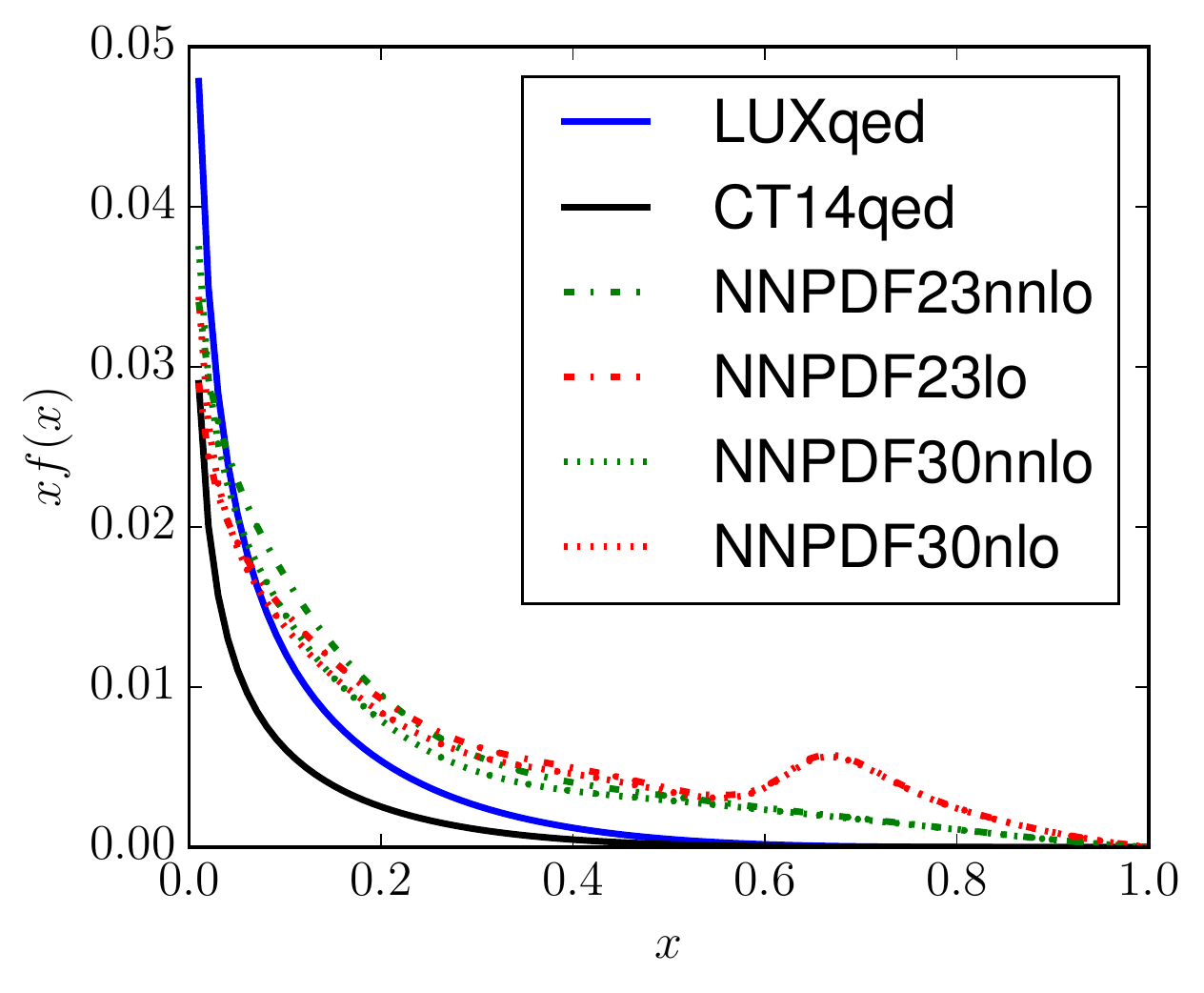}
	\end{minipage}
	\begin{minipage}[t]{.49\textwidth}
		\includegraphics[width=1\textwidth]{FIGS.DIR/PDF_compare_Q100.pdf}
	\end{minipage} \\
	\centering
	\begin{minipage}[t]{.48\textwidth}
		\includegraphics[width=1\textwidth]{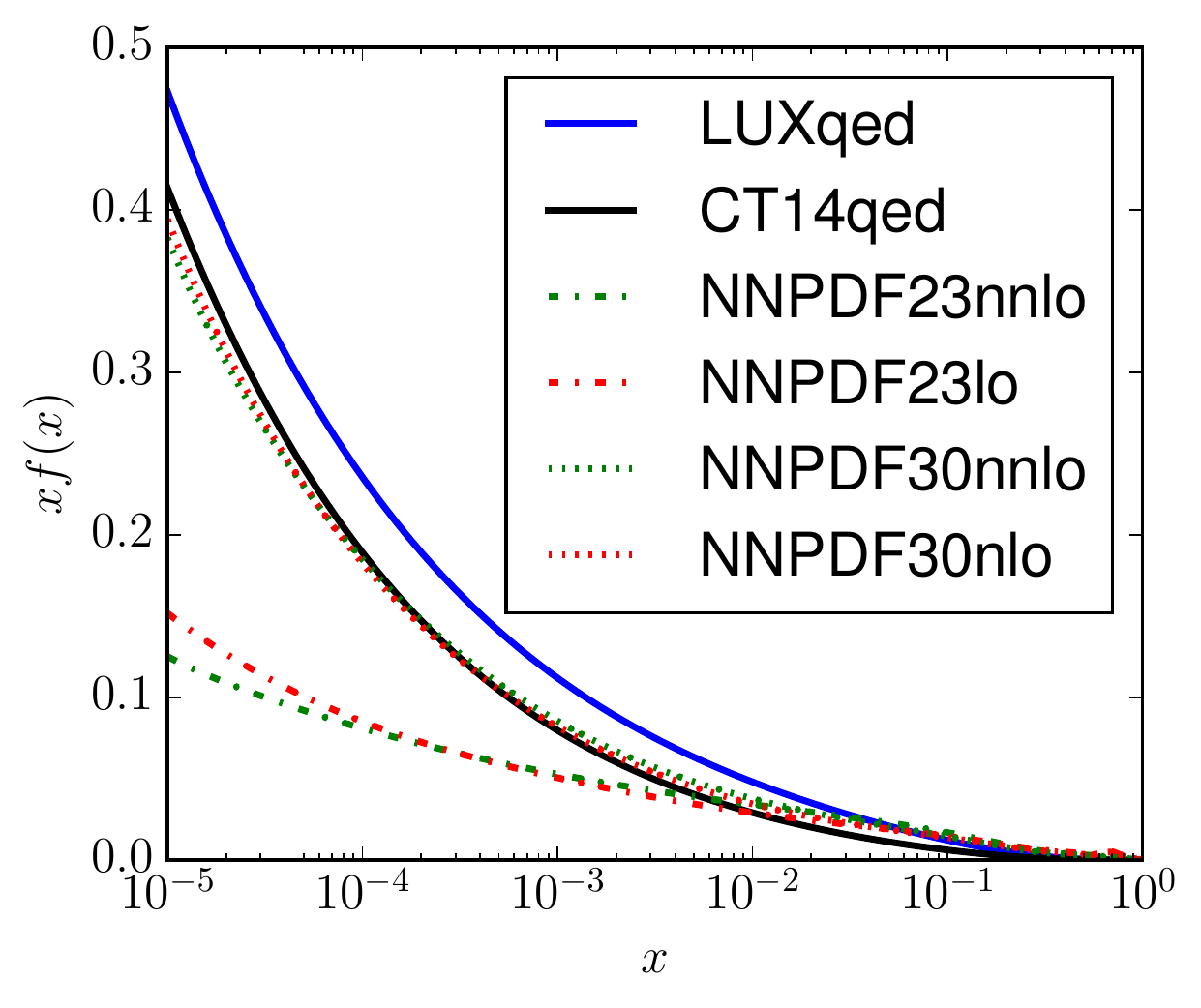}
	\end{minipage}
	\begin{minipage}[t]{.51\textwidth}
		\includegraphics[width=1\textwidth]{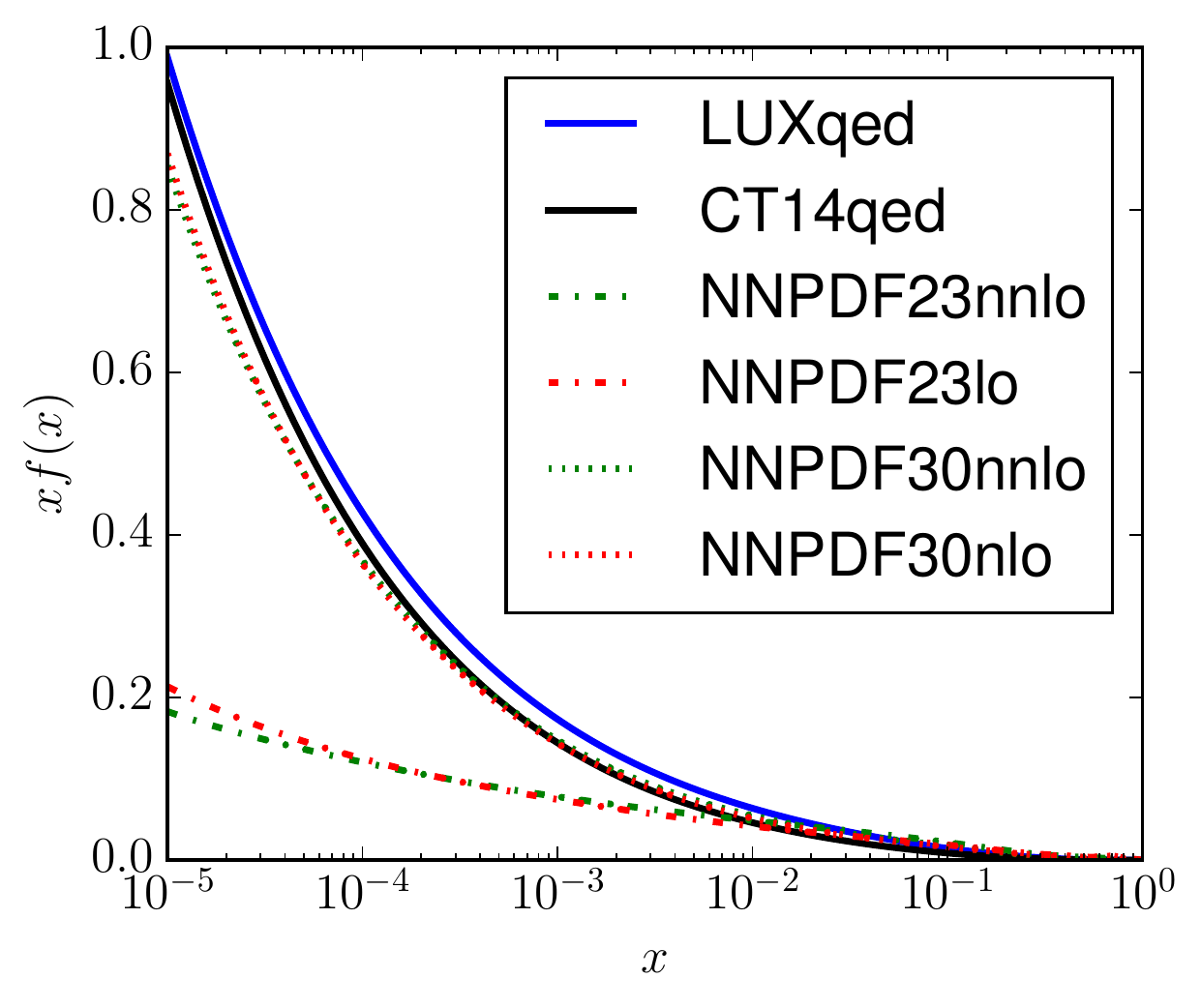}
	\end{minipage}
	\caption{\label{fig:pdfslinear}
{\it Top panel:} Product of the photon distribution function by $x$. {Dashed lines, NNPDF2.3; dotted lines, NNPDF3.0 (colors online on both cases for LO, NLO and NNLO). Solid black line: CT14qed set. Solid gray line (blue online): effective ``pdf'' for the resonance (not necessarily DIS) region, LUXqed.} The left plot takes a characteristic energy scale $\mu=100\,{\rm GeV}$, the right plot $\mu=1\,{\rm TeV}$. Note that $\mu\sim\sqrt{s_{\gamma\gamma}}$ can be much larger than the actual energy scale of the photoproduction process, $Q$. See explanation on page~\pageref{Discuss_energy_scale_mu}. 
{\it Bottom panel:} logarithmic scale to appreciate the low-$x$ behavior.
}
\end{figure}

We now turn to the deeply inelastic cross sections. We have at our disposal several different photon distribution functions in the proton, published respectively by the collaborations 
CT14QED (or, for shortness, CTQ14 in what follows)~\cite{Schmidt:2015zda}, NNPDF3.0QED~\cite{Ball:2014uwa}, NNPDF2.3QED~\cite{Ball:2013hta} and MRST2004QED (or just MRST)~\cite{Martin:2004dh}. Additionally, the LUX photon~\cite{2016PhRvL.117x2002M} ``pdf'' has also been included in the comparisons. Note that LUX is not only a proper pdf when the photon virtuality is large, but also an effective way of encoding the Weizs\"acker-William photon flux of the proton at large energy (even at moderate and low $Q^2$), in particular 
accounting for the resonance region $Q^2\sim\mO(1\,{\rm GeV}^2)$ (not necessarily DIS), as will be ellaborated upon in subsec.~\ref{LUXsection}.

Two of these sets, MRST and CTQ14, are obtained with a similar analysis and their results are consequently also similar, with the difference between the two sets falling with $Q^2$. Therefore we will show our results for the CTQ14 set only, with those obtained from the MRST pdf sets being numerically close. The CTQ14 collaboration has fit isolated photon production in DIS in the interval $10\,{\rm GeV}^2<Q^2<350\,{\rm GeV}^2$, and we expect that the pdf parametrizations will be usable in this momentum squared range.

The earlier NNPDF photon distributions were rather different from those of CTQ14 (and MRST) as can be appreciated from Fig.~\ref{fig:pdfslinear}, especially so at low $x$. The difference might have been attributable to NNPDF excluding the direct DIS information on the photon as discussed in~\cite{Gao:2017yyd} which perhaps makes its uncertainties unnecessarily large. Newly for NNPDF3.0, this difference with CTQ14 almost vanishes on the low-$x$ region, as can be seen in the bottom panel of Fig.~\ref{fig:pdfslinear} (where the newest set, the dotted line, is very close to the LuxQED and CTQ14 solid lines) and also in appendices~\ref{app:sets:NNPDFs} and~\ref{app:sets:CT14}. In any case we employ all these sets so we can explore the systematic uncertainty in the cross section estimates. 

\newpage 

\subsection{Some numerical examples}\label{elasticnum}

\begin{figure}[t]
\includegraphics[width=0.48\textwidth]{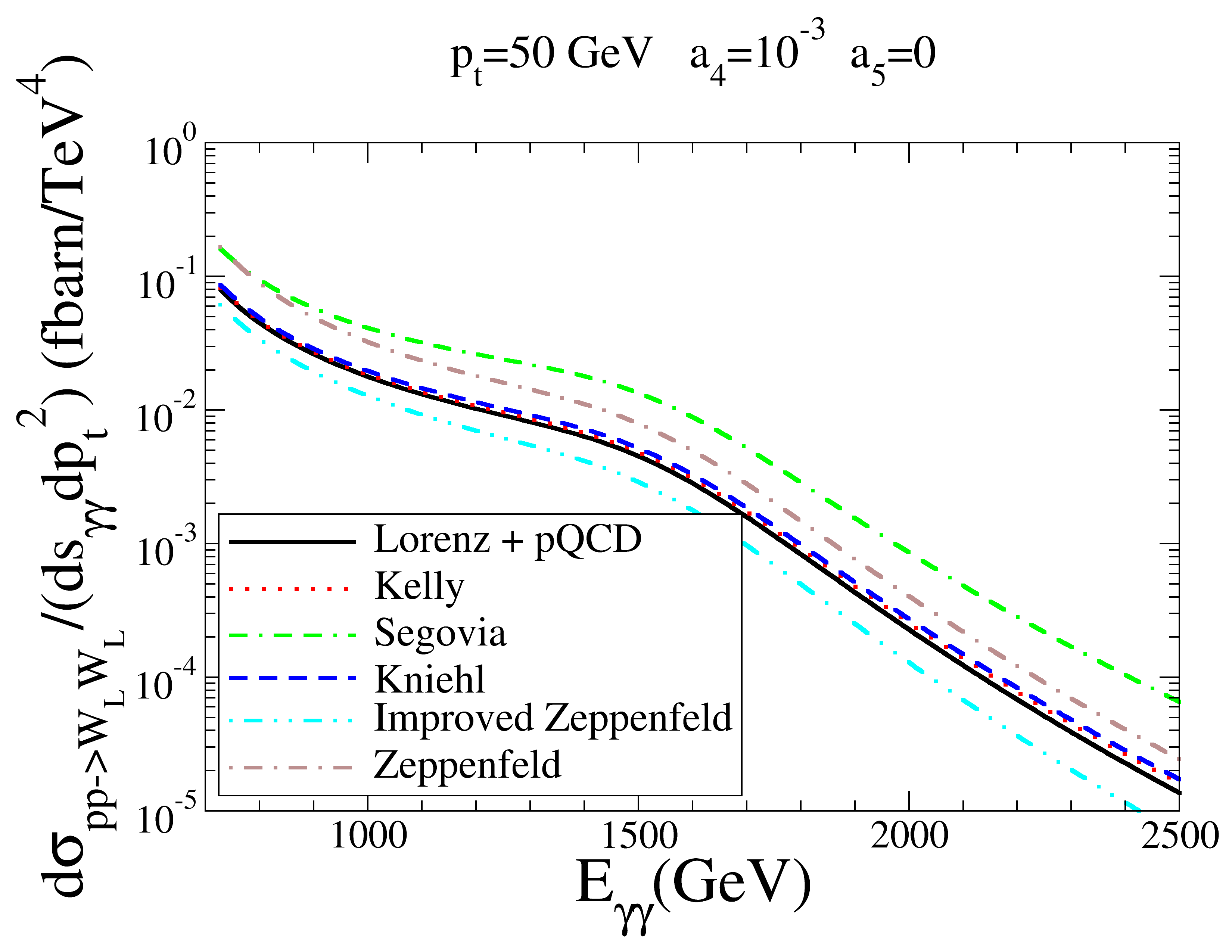}
\includegraphics[width=0.48\textwidth]{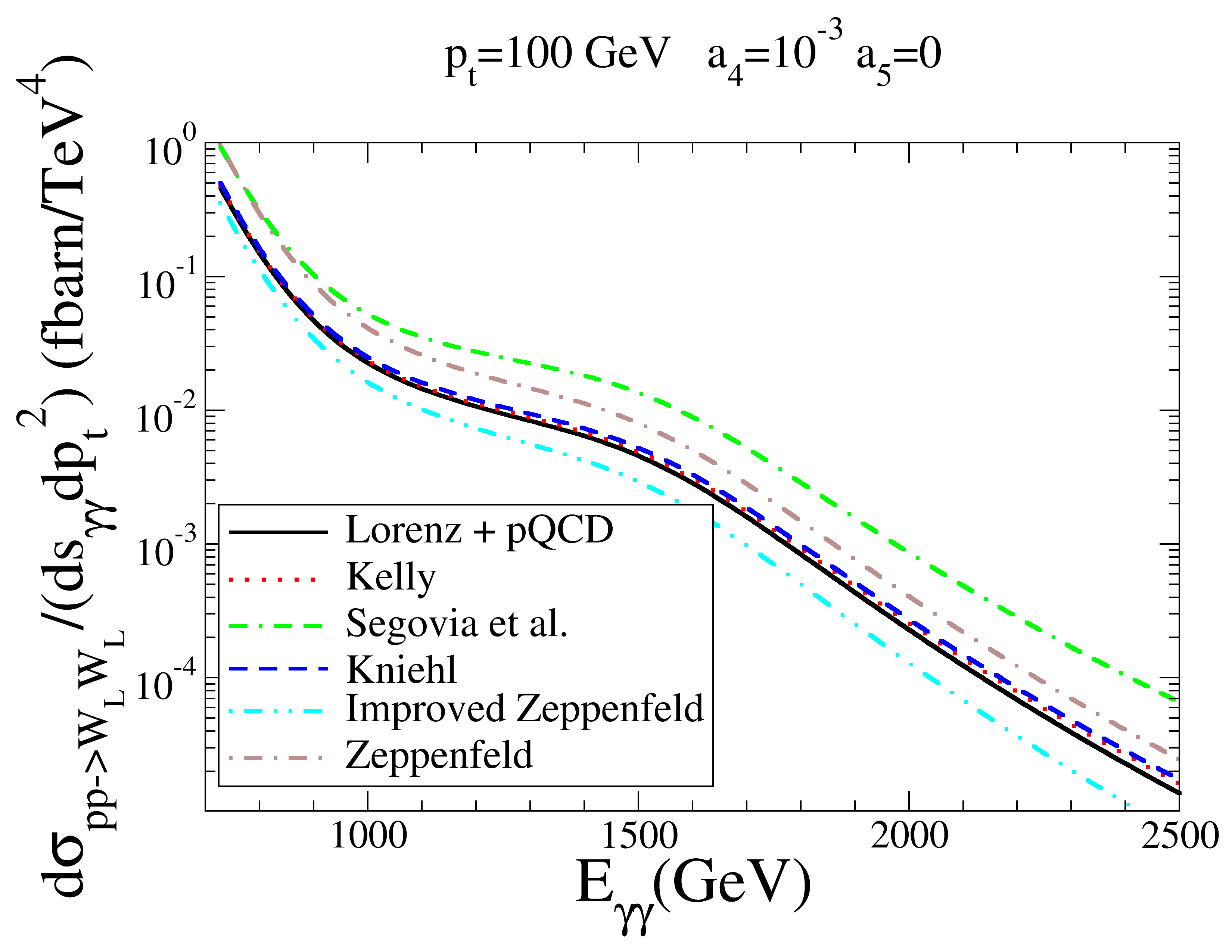}
\centering
\includegraphics[width=0.48\textwidth]{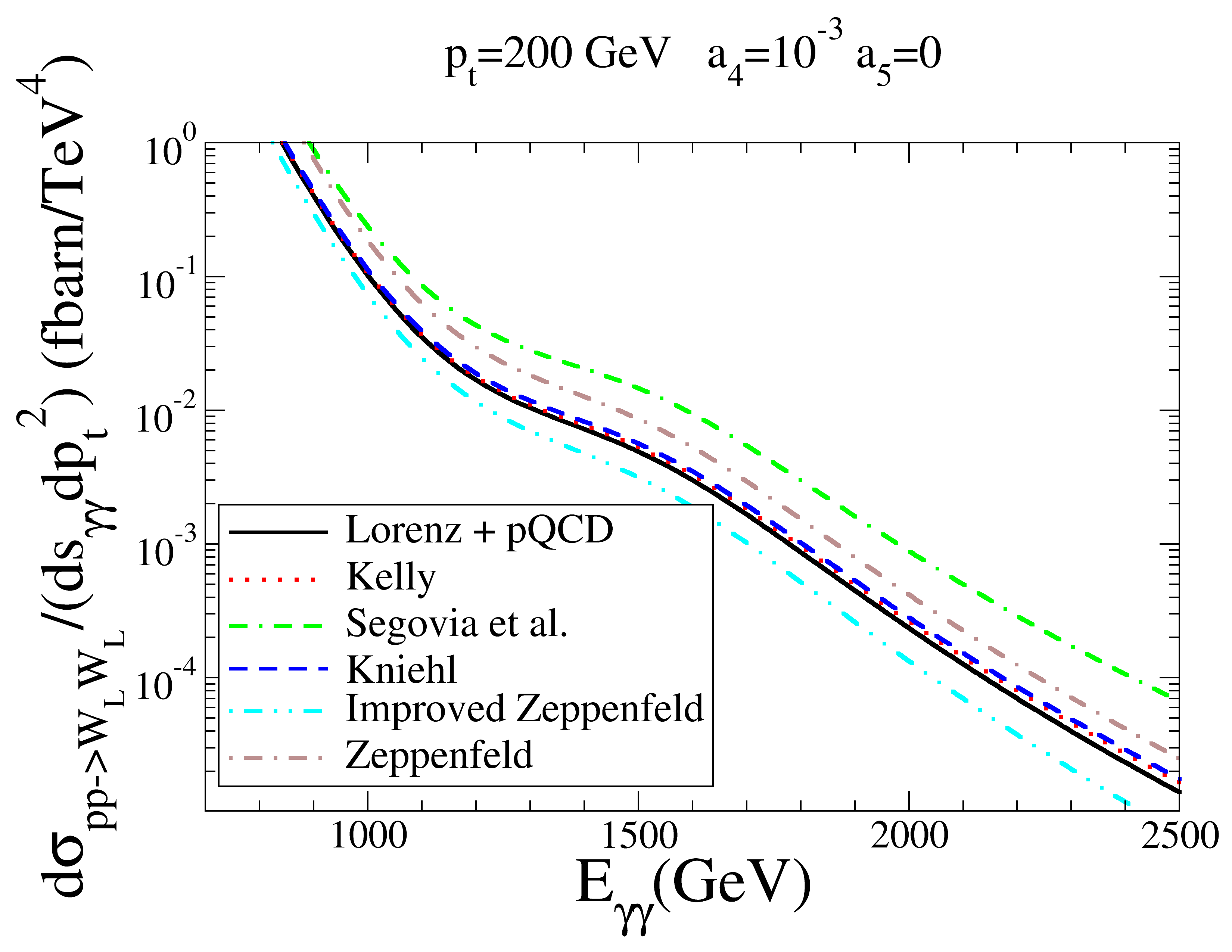}
\caption{\label{fig:secppElastic} Cross section for production of $W_LW_L$ pairs via photon-photon fusion in pp collisions, with both protons elastically scattered. The indicated NLO HEFT parameter $a_4=10^{-3}$ injects a resonance of the EWSBS around $1\,{\rm TeV}$. We show three different transverse momenta to show sensitivity to possible experimental cuts.}
\end{figure}
We now have all ingredients needed to estimate proton-proton cross sections that produce $W_LW_L$ or $Z_LZ_L$ by means of intermediate $\gamma\gamma$ states.

\begin{figure}[h]
\centerline{
\includegraphics[width=0.6\textwidth]{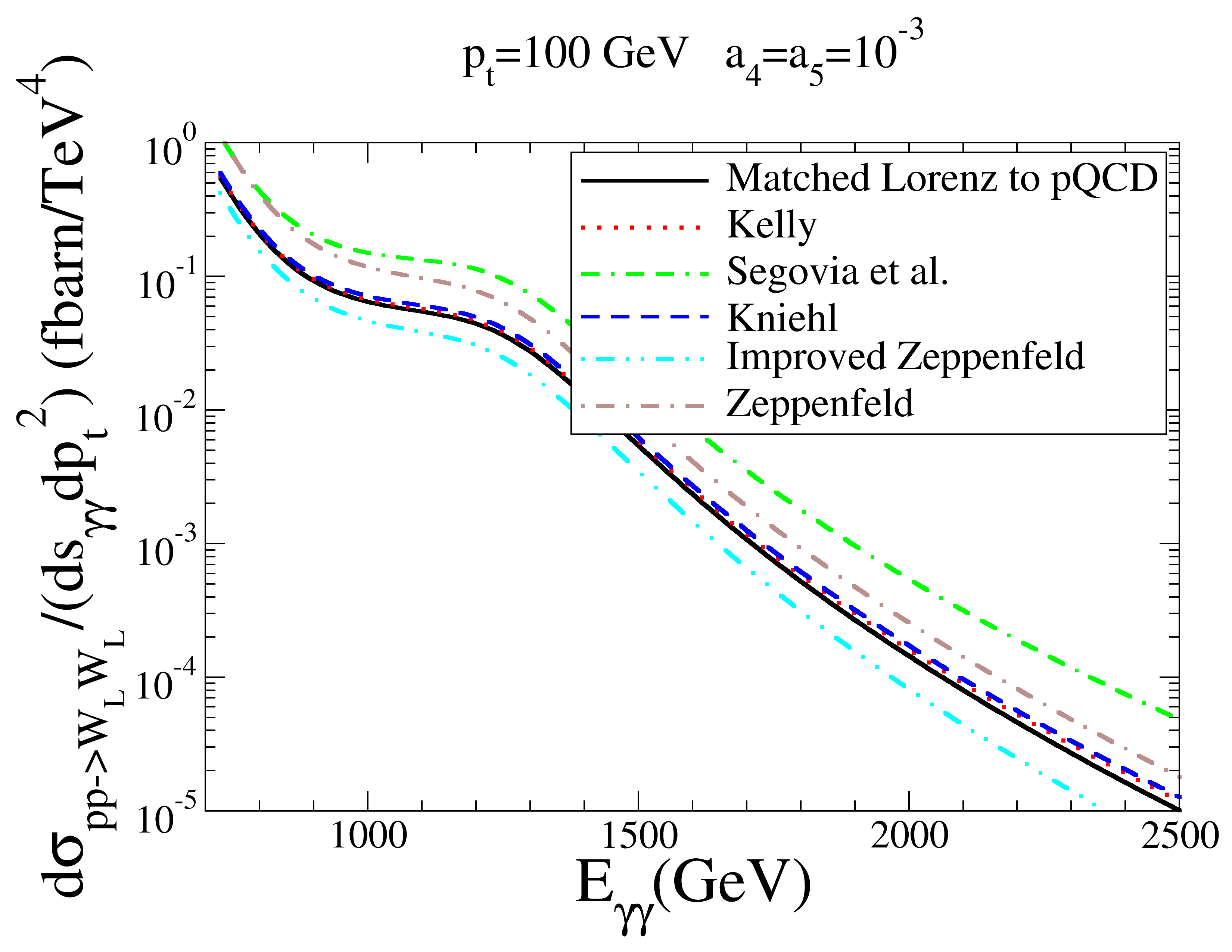}}
\caption{Computation with the elastic photon flux, similar to Fig.~\ref{fig:secppElastic}, but adding the $a_5$ NLO parameters.\label{fig:secppElastic2}}
\end{figure}

In Figs.~\ref{fig:secppElastic} and~\ref{fig:secppElastic2} we put to use the elastic photon fluxes computed above and shown in Fig.~\ref{fig:flujo} to compute the cross sections for TeV-EWSBS resonance production with intermediate photon states, leaving the protons unharmed. 

From the figures, it appears that the cross section is small and since it increases slightly with $p_t$, not much harm is done by imposing experimental cuts thereon that exclude low-lying quarkonia or $\tau\tau$ production.

It also appears (see Fig.~\ref{fig:secppElastic2}) that if a resonance would exist below $1\,{\rm TeV}$ (which we can achieve by increasing $a_4$ or adding a contribution from $a_5$ as done in the figure), the cross section would increase significantly.

\begin{figure}[h]
\centerline{
\includegraphics[width=0.8\textwidth]{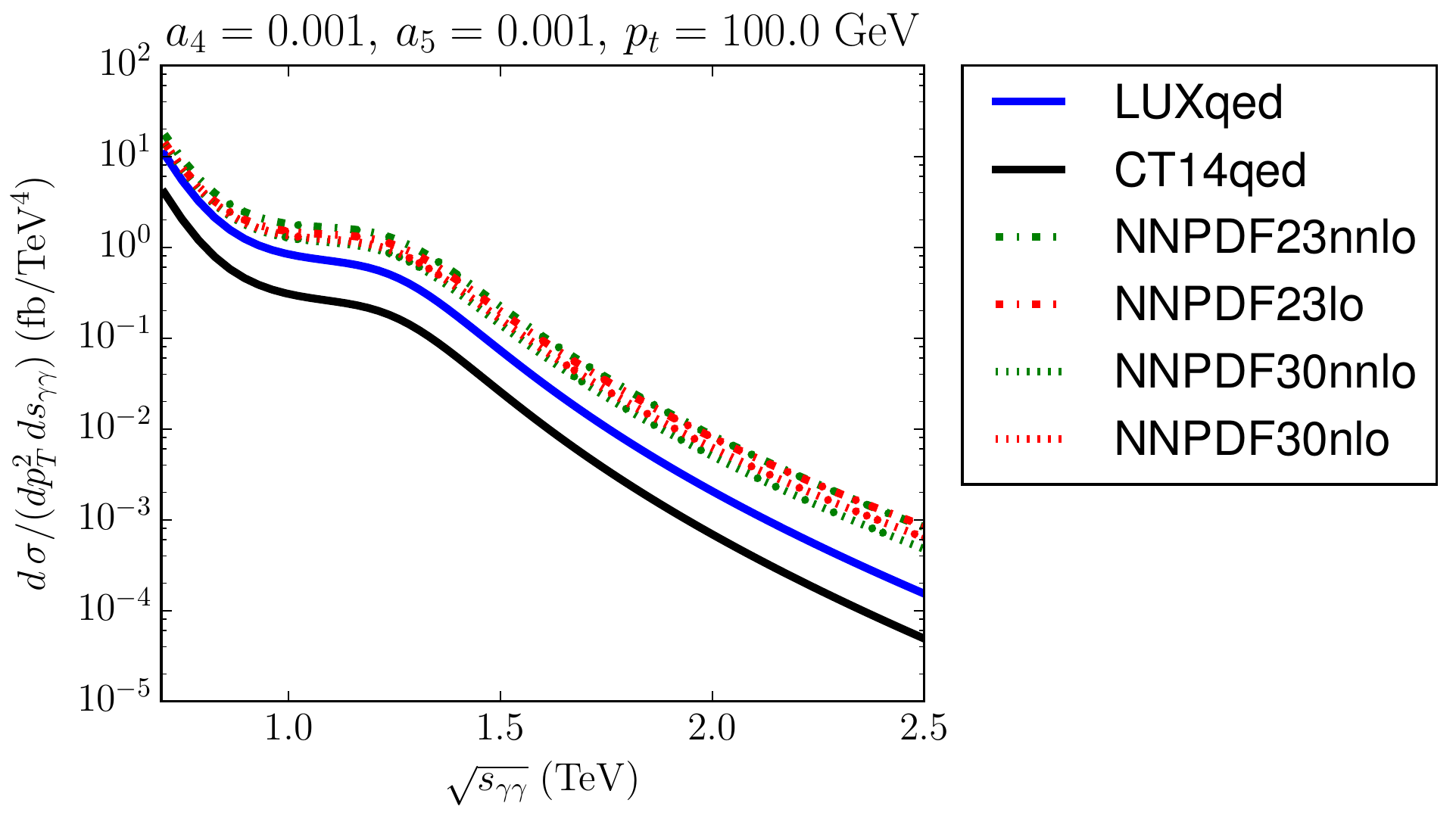}}
\caption{Same as Fig.~\ref{fig:secppElastic2} but both protons dissociate in the deeply inelastic regime. The (effective) PDF energy scale is $\mu^2=s_{\gamma\gamma}$. \label{fig:DISproduction2}}
\end{figure}

\begin{figure}[h]
\centerline{\includegraphics[width=0.8\textwidth]{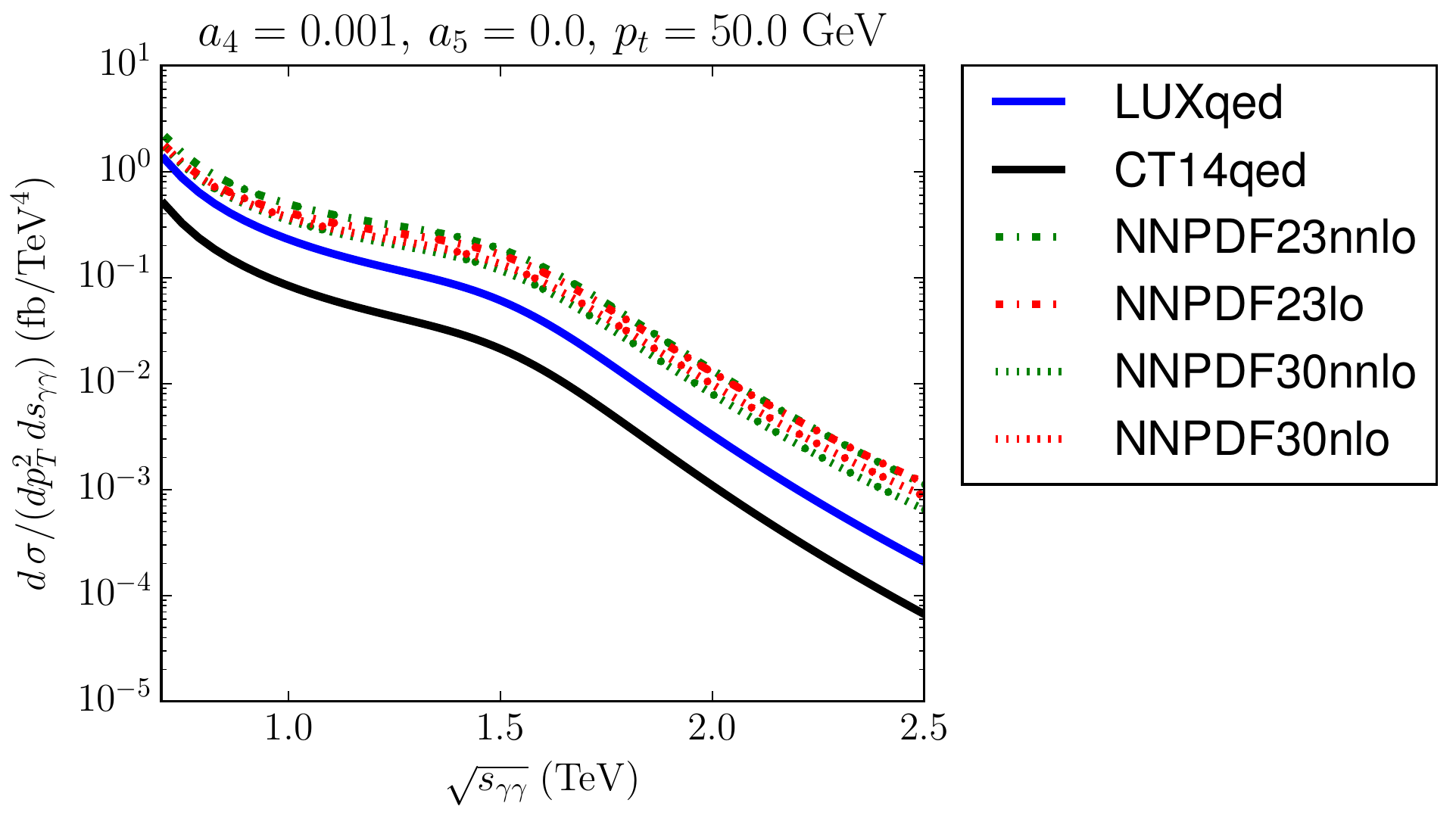}}
\centerline{\includegraphics[width=0.49\textwidth]{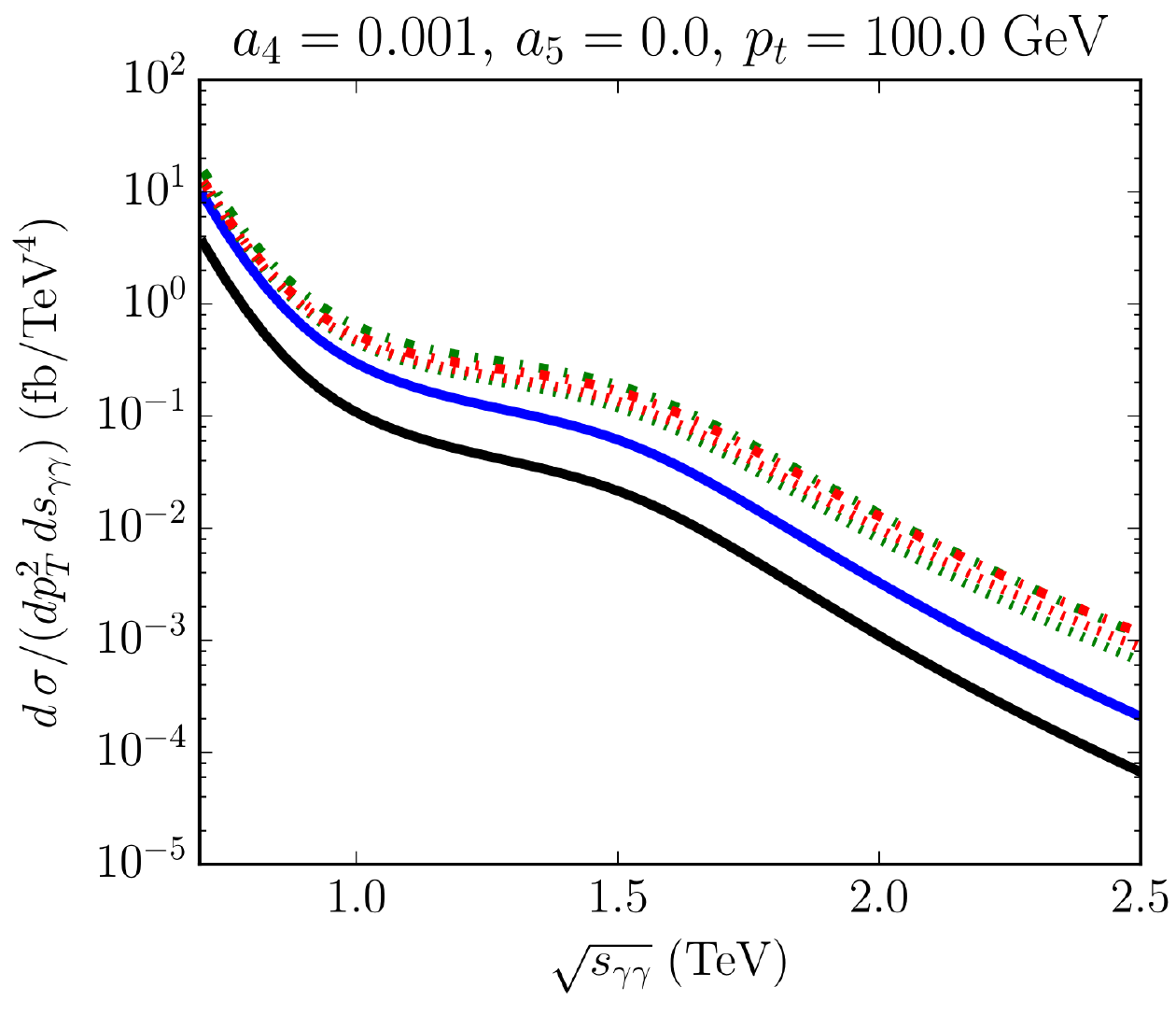}\ \ 
\includegraphics[width=0.49\textwidth]{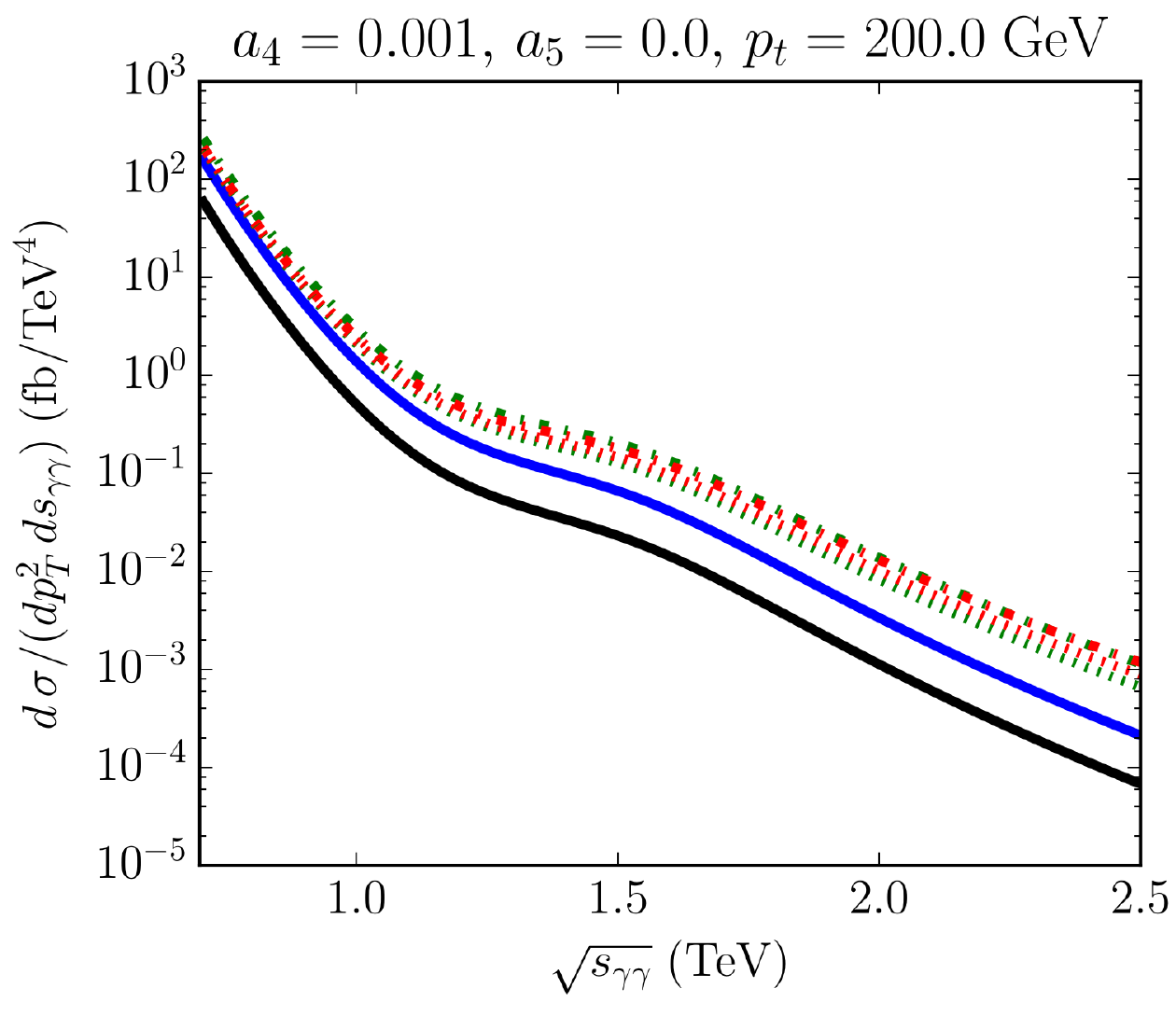}}
\caption{Same as Fig.~\ref{fig:secppElastic} but both protons dissociate in the deeply inelastic regime. Note, by comparing with fig.~\ref{fig:secppElastic}, that this can be up to one order of magnitude more likely than elastic scattering. The (effective) PDF energy scale is $\mu^2=s_{\gamma\gamma}$.
\label{fig:DISproduction}}
\end{figure}

As we do not find very strong signals, we need to be comprehensive and increase the kinematic range  with the inelastic regime but not DIS (that is, lift any restrictions on the fate of the final state protons which we will perform in the next subsection). A very easy computation can be carried out with the DIS pdfs in Fig.~\ref{fig:pdfslinear}, where both protons dissociate (there is no difficulty in computing, for example, the instances in which one proton is dispersed elastically and the other dissociated, by combining the different fluxes, all at hand). In Figs.~\ref{fig:DISproduction2} and~\ref{fig:DISproduction} we show just this computation.

The cross section obtained from the NNPDF set is quite larger than that from the CT14 one as NNPDF is the largest of the two for higher $x$ (a $1-2\,{\rm TeV}$ resonance in a $13\,{\rm TeV}$ collider requires $x\sim 0.1-0.2$).

The cross section for these DIS events can easily be 5 times larger than the elastic one, but they are very difficult to reconstruct as they can leave charged tracks in the central tracker that would not pass the cuts to reduce background.

Therefore, an interesting strategy would be to search for inelastic, but not deeply inelastic, events where one or both protons are dissociated in the $1-3\,{\rm GeV}$ resonance region.

\subsection{Inelastic regime (not necessarily DIS)}\label{LUXsection}

The cross sections reported so far in pp collisions, elastic and deeply inelastic are rather small, and there is small hope of measuring the later in pp because it would probably leave activity in the central barrel, so that it would not be easily identifiable over background. See~\cite{Harland-Lang:2016apc} for extensive discussion on how to incorporate various rapidity-gap cuts that assist event identification into theoretical calculations. Actually, we expect most of the cross section not to be in those extreme regimes, but rather correspond to an intermediate, inelastic but not deeply inelastic proton recoil (in the baryon resonance energy region).

A full theory description of that region ($1-2.5\,{\rm GeV}$) is beyond our ability, as many resonances of various spins populate it and likely contribute. Therefore, we resort once more to a data-driven description, adopting a photon flux $f$ that incorporates information from Jefferson laboratory and other mid-energy facilities.  A convenient parametrization of the photon content of the proton useful for $pp\to \gamma\gamma +X $ is provided by the LUX photon~\cite{2016PhRvL.117x2002M} ``pdf'' that is precisely the photon flux that we need~\footnote{We adopt the set {\tt LUXqed\_plus\_PDF4LHC15\_nnlo\_100} herein.}, describing low-$Q^2$ data from A1, CLAS and Hermes GD11-P. 
In addition to low-energy baryon resonances, the authors of~\cite{2016PhRvL.117x2002M} also incorporate the elastic form factors and DIS functions that we have examined above into their photon flux. The high-$Q^2$ flux\footnote{See the breakup of parameter space $(x,Q^2)$ in Fig.~1 of~\cite{2016PhRvL.117x2002M}.} is a proper pdf for the photon evolved at NNLO and fit to standard data. Casting form factors and inelasticities in the language of parton distribution functions makes all the pieces fit into the standard Monte Carlo collider machinery.

Note that, in this framework, the energy scale $\mu^2$ at which the pdf is set (and that enters into the well-known LHAPDF library~\cite{Gomes:2013qza}) differs from the virtuality of the actual $\gamma$ emission process, $Q^2$. This can be checked in Eq.~(6) of Ref.~\cite{2016PhRvL.117x2002M}, where $\mu^2$ appears as a cutoff of the integration over $Q^2$. Indeed,  Fig.~2 of Ref.~\cite{2016PhRvL.117x2002M} exposes that, for $\mu=100\,{\rm GeV}$ and $x>0.05$, more than half of the \emph{effective pdf} comes from physics at an energy scale $Q^2<(1\,{\rm GeV})^2$. Hence, the requirement $\mu>10\,{\rm GeV}$ of the pdf~\footnote{For example, in the set {\tt LUXqed\_plus\_PDF4LHC15\_nnlo\_100}.} is meant to limit applicability of the photon flux to collider phenomenology at a center of mass energy of $s>(10\,{\rm GeV})^2$ (in the spirit of the Weizs\"acker-Williams approximataion), but not as a limitation on the virtuality of  the  emission process that can be soft as in $p\to\gamma^*p^*$. The parameter $\mu$ should be set at the scale of the large $\gamma\gamma$ scattering energy.\label{Discuss_energy_scale_mu}

We have also examined an alternative work~\cite{Martin:2014nqa} that also parametrizes effective PDFs (more properly, photon fluxes to be used with the Weizs\"acker-Williams approximation) via elastic (and $\Delta(1232)$--inelastic) form factors 
\begin{equation}
  f_{\gamma\ {\rm el}}^p (x,Q_0^2) = \frac{\alpha^{\rm QED}}{2\pi}\frac{[1+(1-x)^2]}{x}\int_0^{\lvert x\rvert < Q_0^2} dQ_t^2\frac{Q_t^2}{(Q_t^2+x^2m_p^2)^2}F_1^2(t),
\end{equation}
where $Q_t$ is the photon transverse momentum, $t=-(Q_t^2+x^2m_p^2)/(1-x)$, and $F_1$ the Dirac electromagnetic proton form factor (multiplying $\gamma_\mu$ at the photon-proton vertex). Note the curious absence of the Pauli (helicity non-conserving) form factor $F_2$~\footnote{These are related to the Sachs form factors below Eq.~(\ref{flujo}) via~\cite{Gluck:2002fi} $G_E = F_1 - \tau F_2$, $G_M = F_1 + F_2$, $G_E$ and $G_M$.}. The contribution of $F_2$ is included in~\cite{Gluck:2002fi}, but that work is limited to the elastic contribution whereas \cite{Martin:2014nqa} gives an analytical expression accounting for the lowest possible proton excitation, $\Delta(1232)$.

In any case, we employ these works for cross--checks and show the outcome produced with the newer and more complete 
LUX NNLO $\gamma$--flux~\cite{2016PhRvL.117x2002M}. 

\begin{figure}[h]
\centerline{\includegraphics[width=0.47\textwidth]{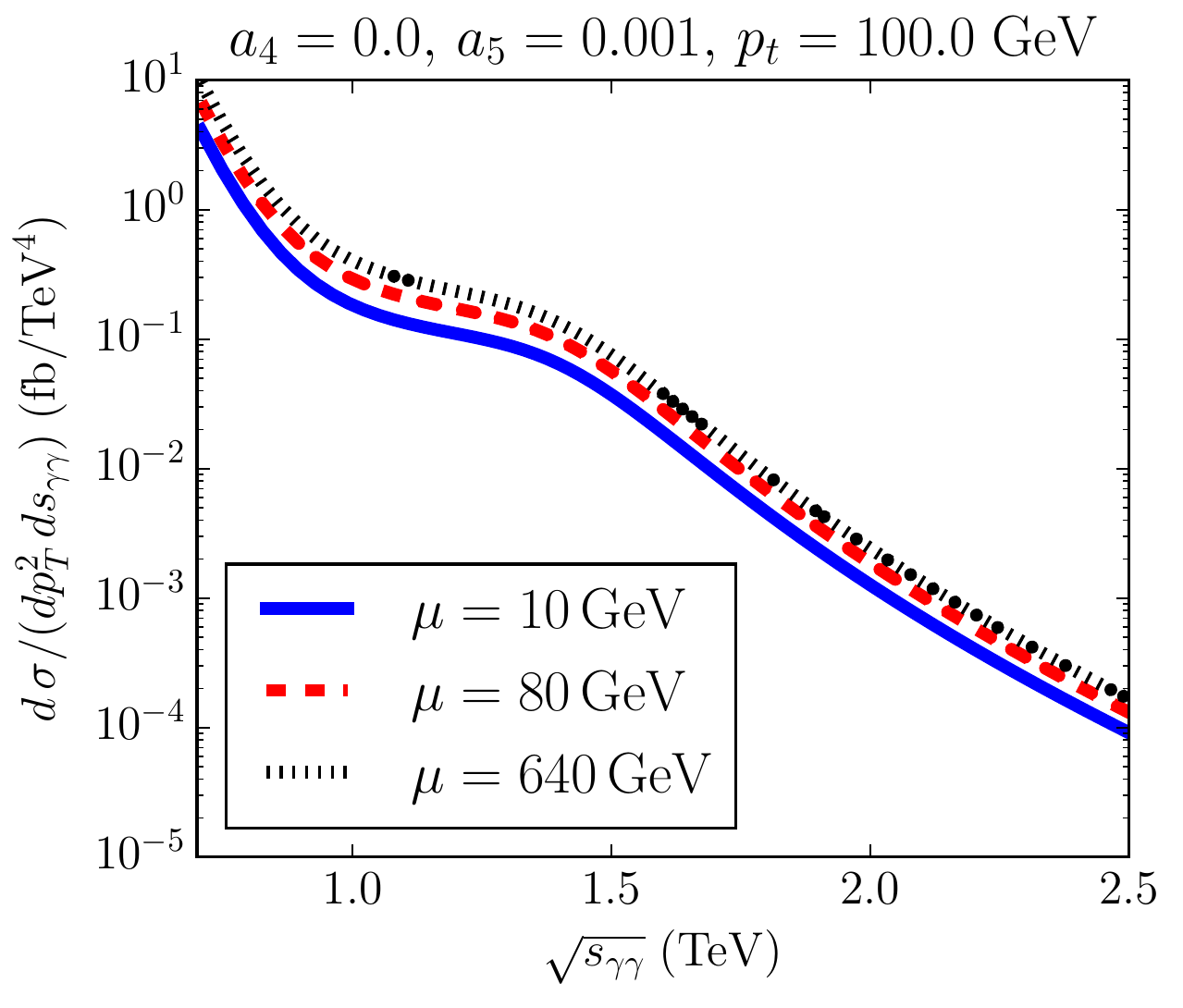}\ \ 
\includegraphics[width=0.47\textwidth]{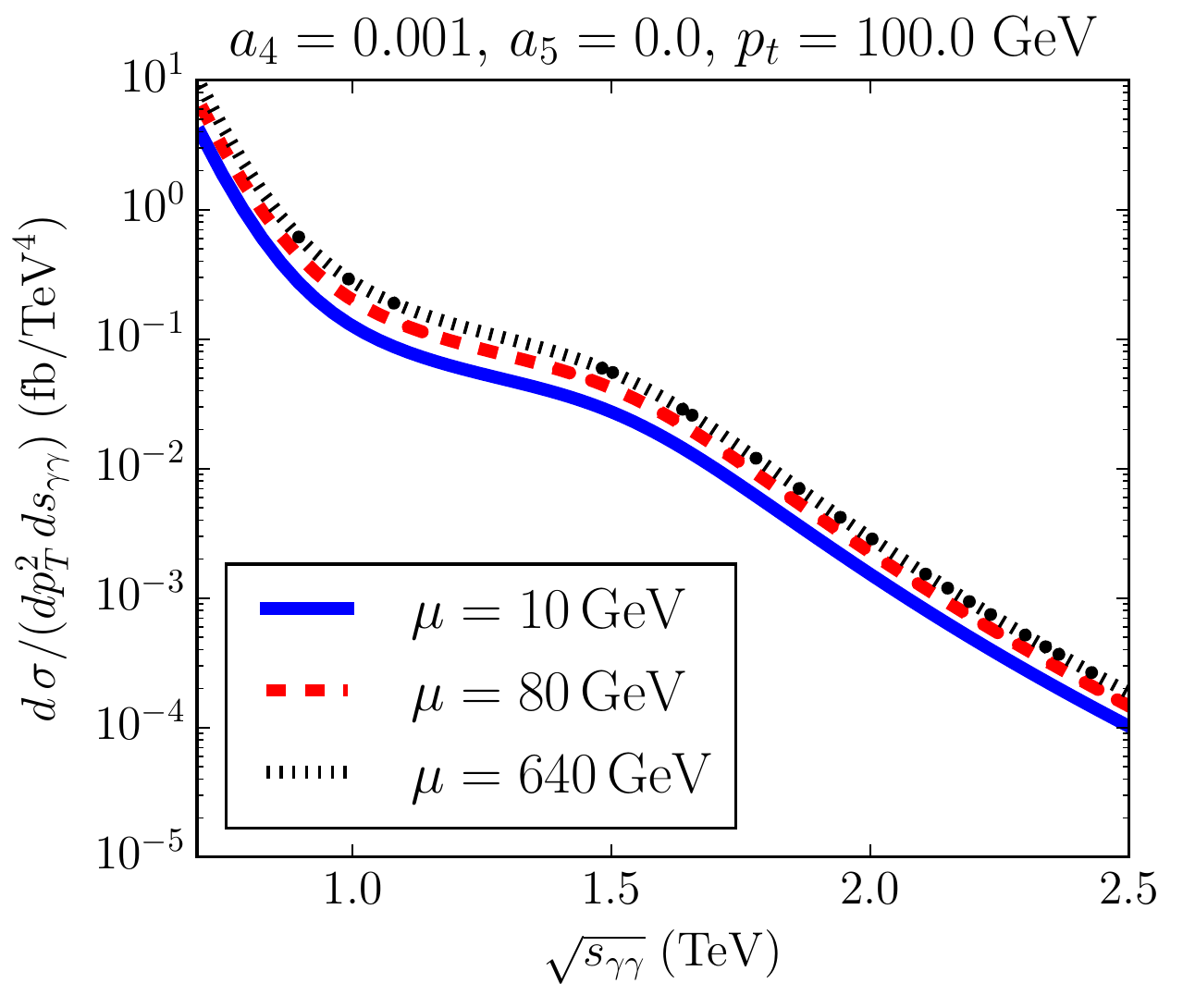}}
\centerline{\includegraphics[width=0.5\textwidth]{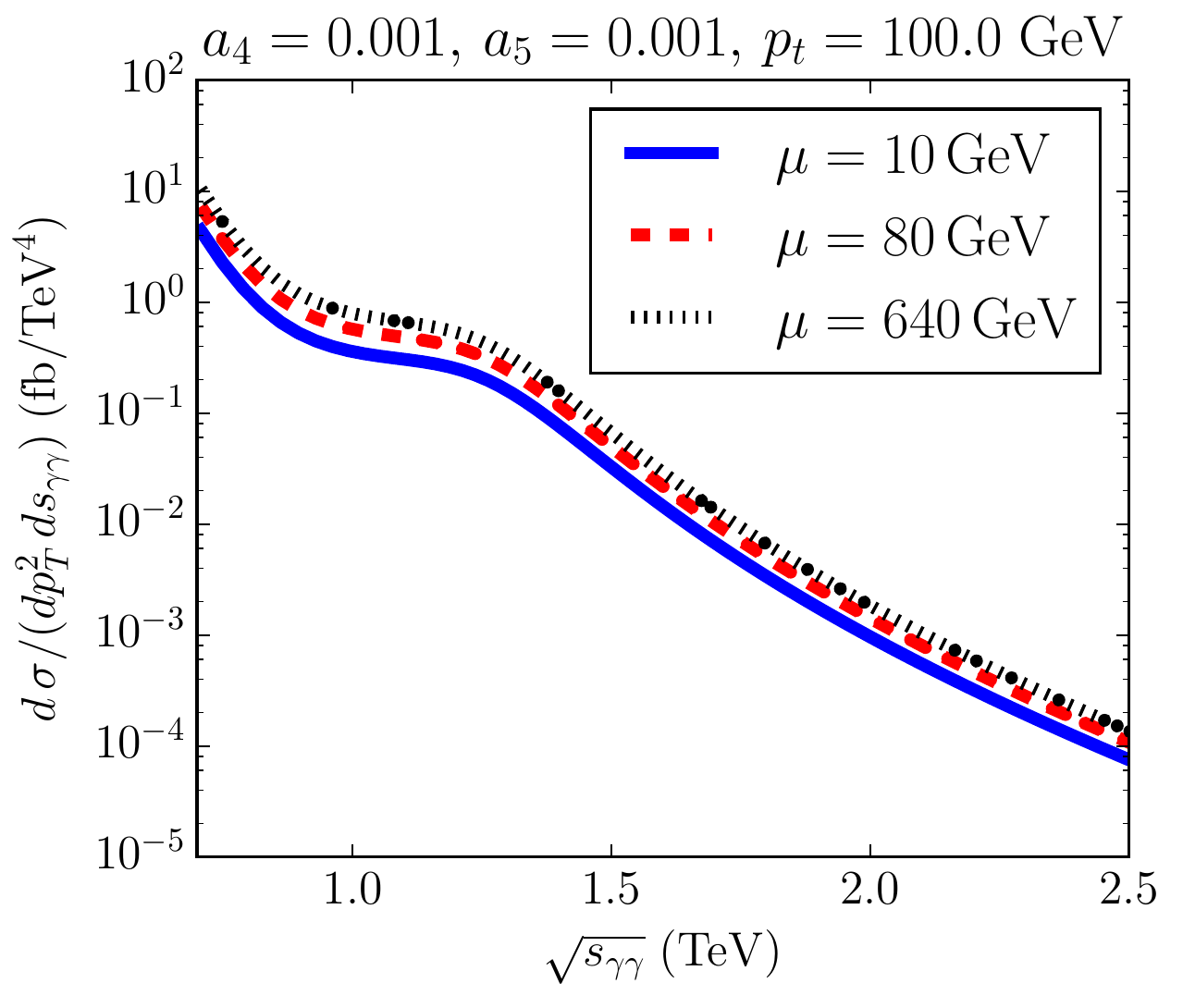}}
\caption{Cross section for $pp\to\gamma\gamma p^* p^*\to W_LW_L + X$ for three values of $\mu$ (the scale at which the LUX~\cite{2016PhRvL.117x2002M} photon flux factor is evaluated), differential respect to the produced $s_{W_L W_L}$ and the squared transverse momentum. The NLO parameters, visible in the plots, are chosen so that a resonance in the EWSBS amplitudes is present around 1.2--1.5 TeV. After convolution with the photon flux, only a broad shoulder is visible.
\label{inelastic}}
\end{figure}

Proceeding then as in Eq.~(\ref{ecuacion}), we obtain the cross section reported in Fig.~\ref{inelastic}. Because the LUX photon flux requires a $\mu$ scale (as it incorporates inelastic structure functions of the proton), we vary this in the graph over a reasonable range. We have taken as parameters $a^2=b=0.81^2$, $c_\gamma = 10^{-4}$, and $a_4=10^{-3}$, consistently with our previous sets. All the other NLO parameters from both the EWSBS and the photon sector are set to zero, namely $g=d=e=a_1=a_2=a_3=a_5= 0$.  This set yields a typical resonance around $1.5\,{\rm TeV}$.

The figure shows what happens to this resonance of the EWSBS after convolution with the inelastic photon fluxes: it becomes a broad shoulder, experimentally challenging after accounting for statistical data uncertainties.

\begin{figure}[h]
\centerline{\includegraphics[width=0.6\textwidth]{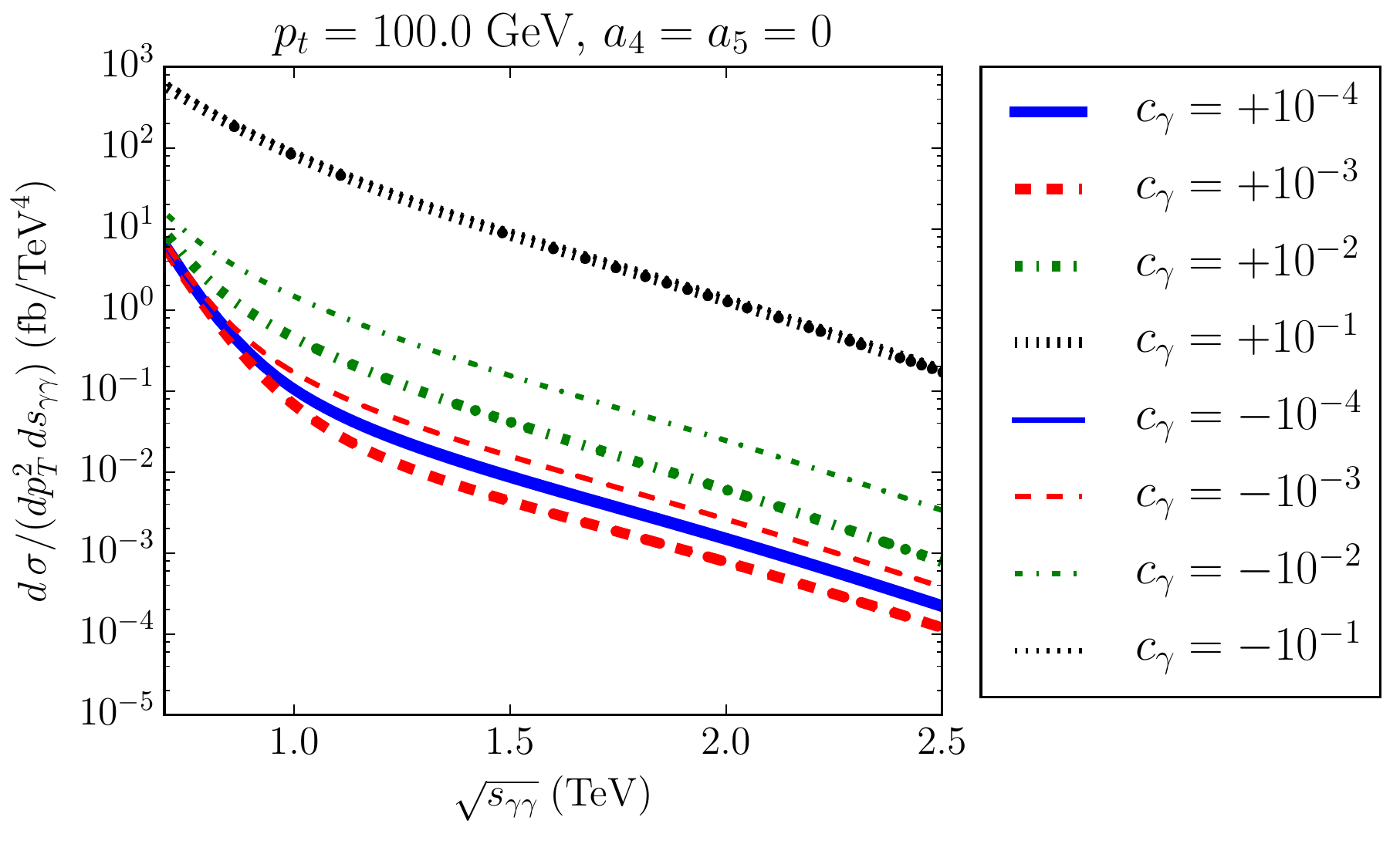}}
\centerline{\includegraphics[width=0.59\textwidth]{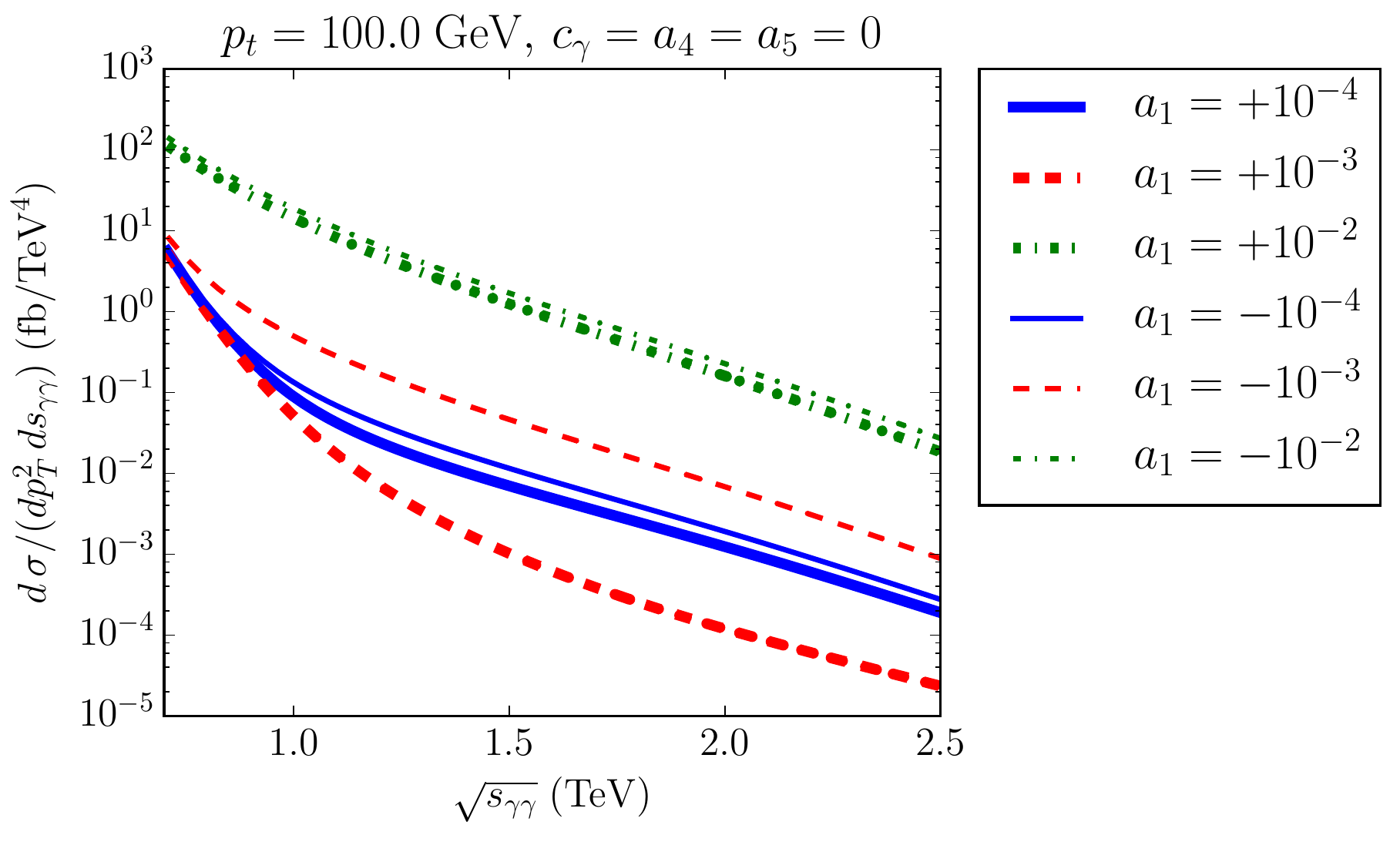}
\includegraphics[width=0.39\textwidth]{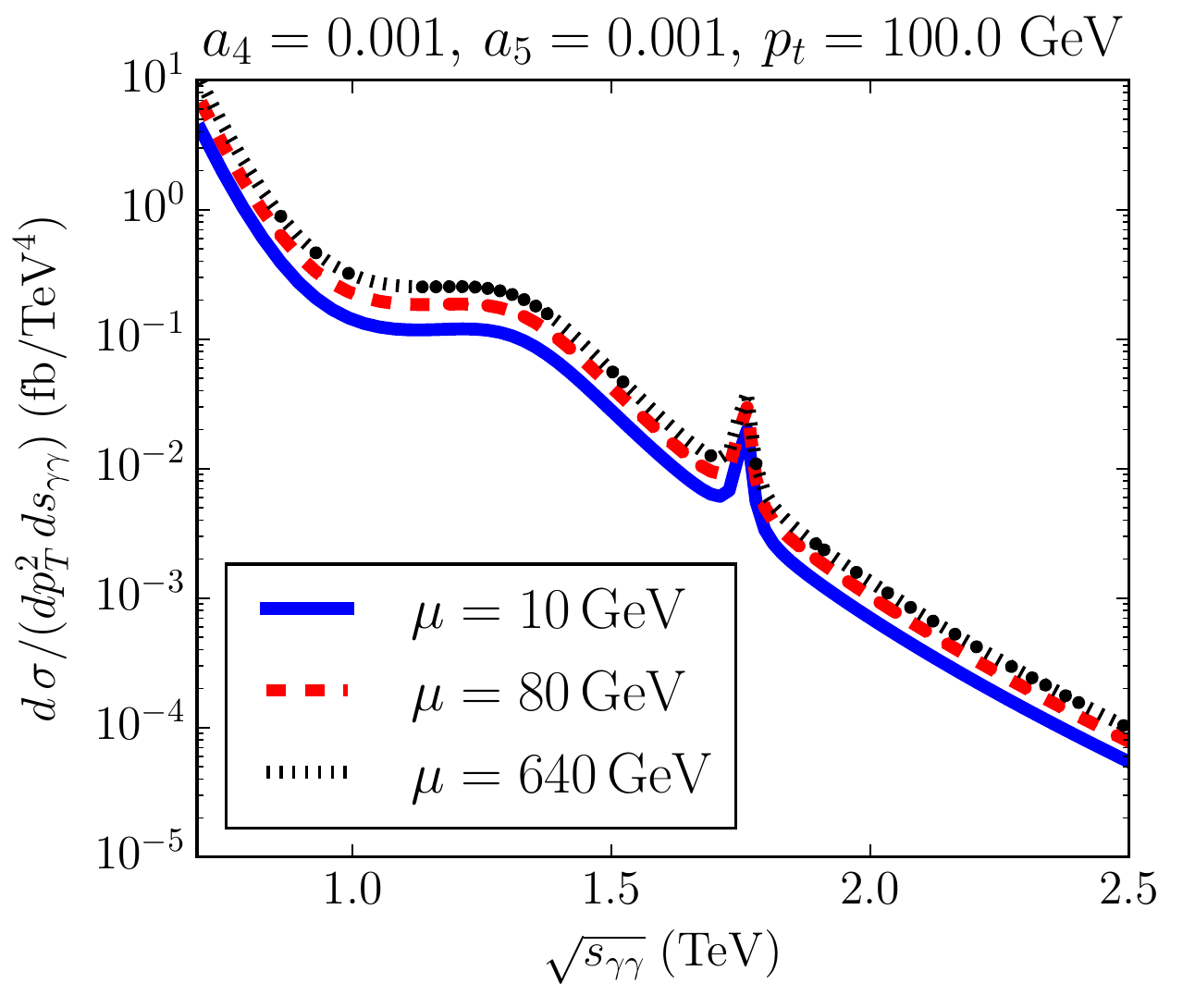}}
\caption{Cross section for $pp\to\gamma\gamma p^* p^*\to W_LW_L + X$ with different values of $c_\gamma$ (top) and  $a_1$ (bottom--left). The bottom--right graph corresponds with $a^2=b=0.95$, $a_4=a_5=10^{-3}$. In all the cases, the LUX scale is $\mu=640\,{\rm GeV}$.%
\label{inelastic_additional}}
\end{figure}

In Fig.~\ref{inelastic_additional}, we scan over  $c_{\gamma}$ (top) and $a_1$ (bottom left), also with $a^2=b=0.81^2$ and $a_4=a_5=0$. For completeness, we have also included a case with $a^2=b=0.95^2$ and $a_4=a_5=10^{-3}$ (bottom right graph of Fig.~\ref{inelastic_additional}), consistent with LHC constraints~\cite{deBlas:2018tjm}. This set of values introduces a clear resonance at $\sqrt{s}\sim 1.8\,{\rm TeV}$ that is narrower (and thus, dominated by NLO parameters).

\begin{figure}
\centerline{\includegraphics[width=0.9\textwidth]{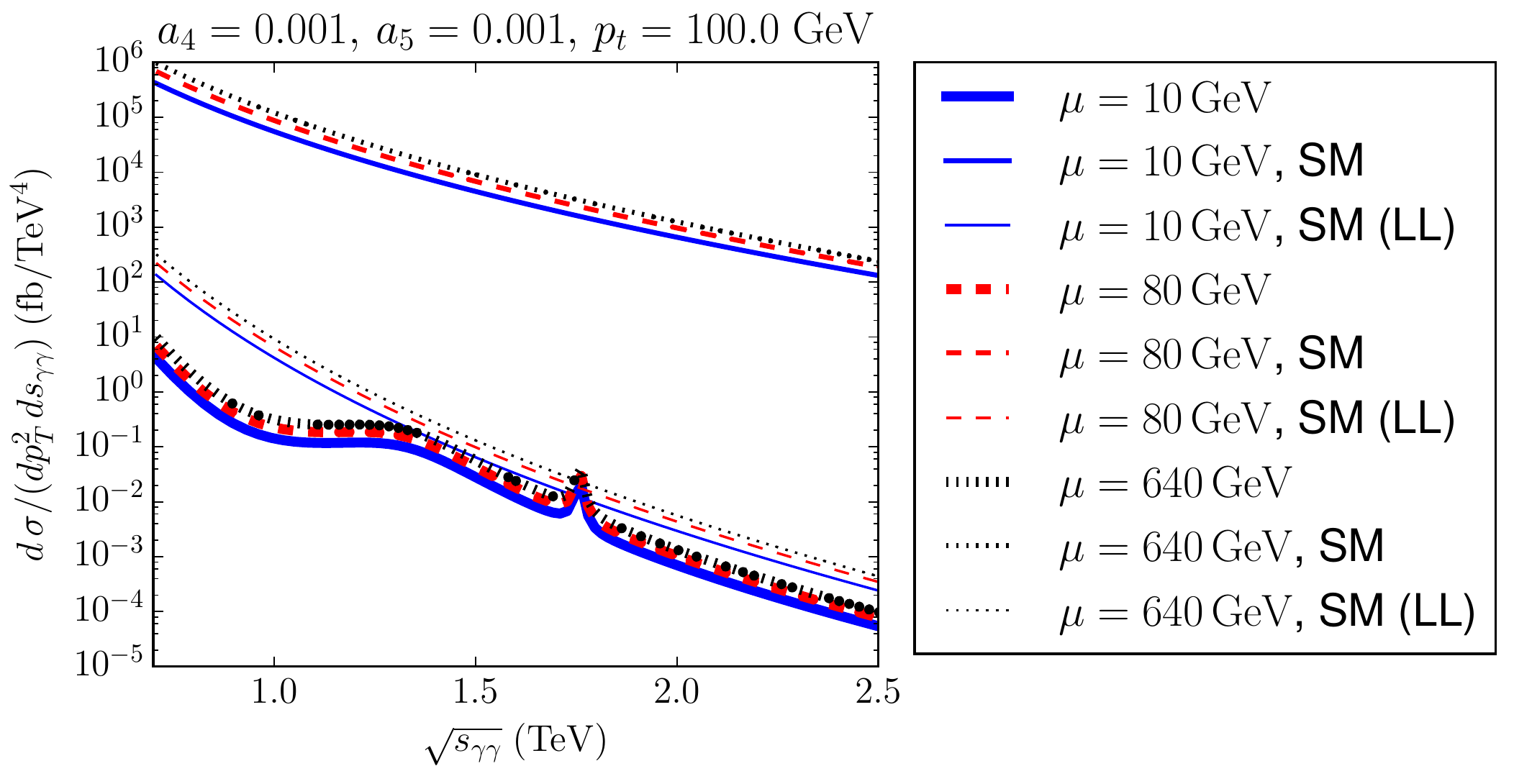}}
\centerline{\includegraphics[width=0.9\textwidth]{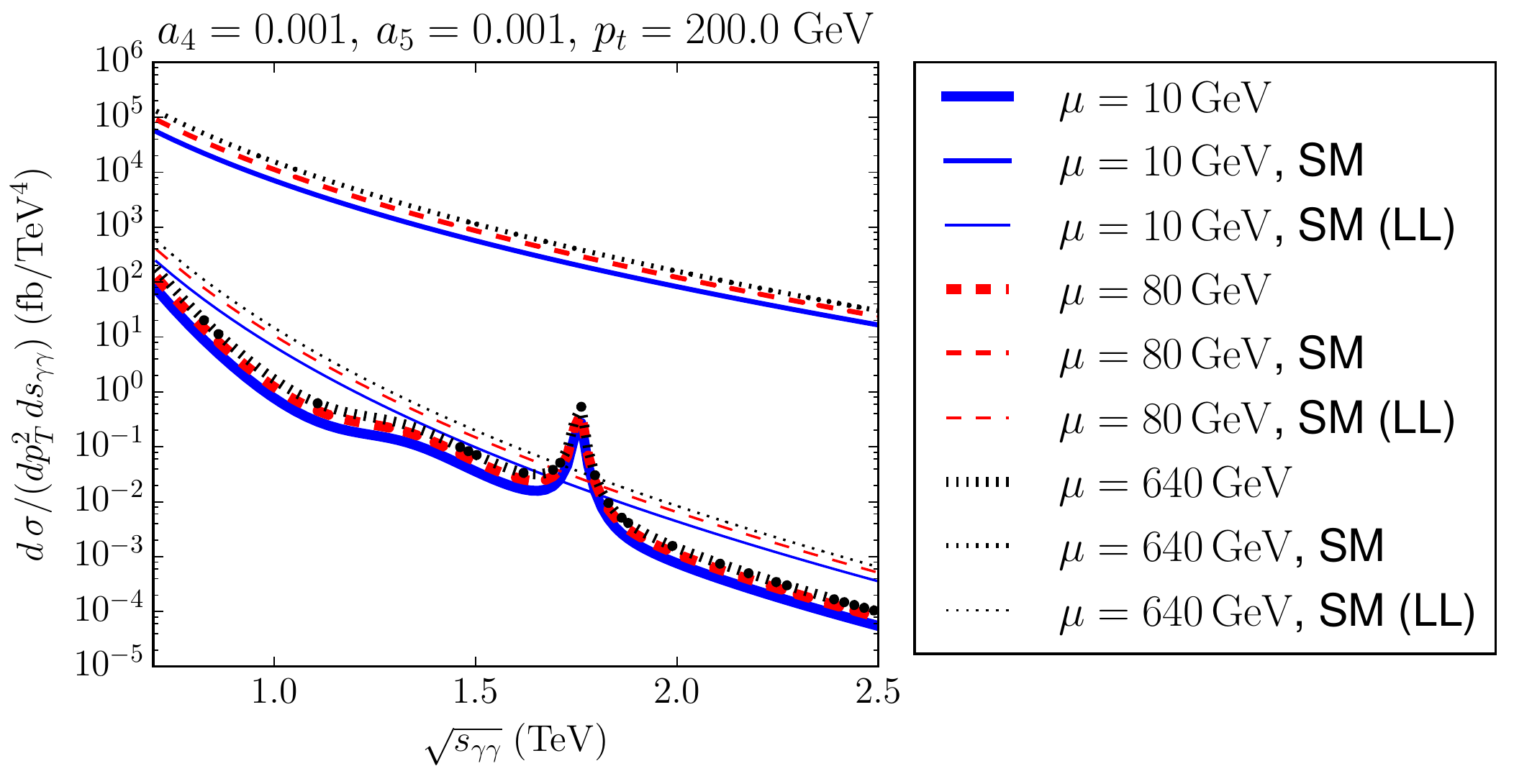}}
\centerline{\includegraphics[width=0.9\textwidth]{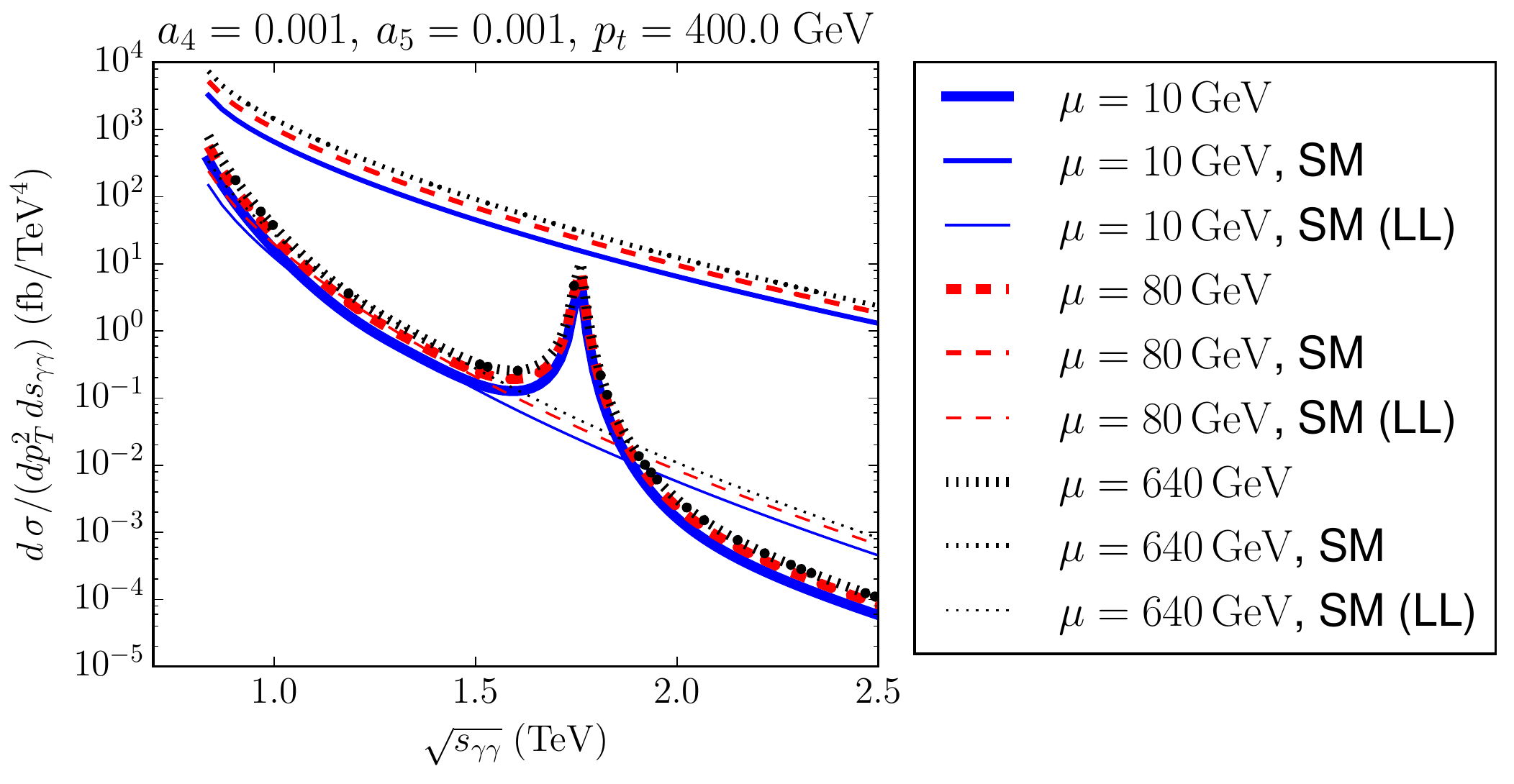}}
\caption{Cross section for $pp\to\gamma\gamma p^* p^*\to\omega_L\omega_L + X$ ($a_4=a_5=10^{-3}$) versus SM background~\cite{Yehudai:1991az,Denner:1995jv}. From top to bottom, $p_T=100,\,200,\,400\,{\rm GeV}$; in each graph we show the scale dependence. The SM background only includes the $WW$ signal, since $\gamma\gamma\to ZZ$ vanishes at LO. \texttt{SM(LL)} stands for the SM background $\gamma\gamma\to W_LW_L$.%
\label{inelastic_wSM}}
\end{figure}

Finally, in Fig.~\ref{inelastic_wSM} we compare the signal with $a^2=b=0.95^2$, $a_4=a_5=10^{-3}$, with the SM background $pp\to\gamma\gamma p^* p^*\to W^+_LW^-_L + X$. Note that $\gamma\gamma\to ZZ$ vanishes at LO in the SM. The SM computations have been taken from Refs.~\cite{Yehudai:1991az,Denner:1995jv}. Note the big backround coming from the transverse modes. However, such a background can be decreased by looking for events at high $p_T$.

The cross sections that we find are not larger than those in subsec.~\ref{elasticnum}, in spite of including further kinematic windows. It may be that the older parametrizations from MRST, CT14QED or NNPDF overestimated the photon flux.

\section{Discussion and outlook}

Photon-photon induced production of electroweak resonances, if they exist, is an alternative to their production from gluon-gluon interactions. While the cross sections are naturally small, as we have quantified, they are very clean if the outgoing elastically scattered protons can be tagged (see figures~\ref{fig:secppElastic} and~\ref{fig:secppElastic2}). Such searches are complementary to two-photon reconstruction in the final state of a central collision, a method that is already constraining the EWSBS below about 900~GeV~\cite{Khachatryan:2016yec}.

Our approach to assess the EWSBS from two-photon collisions is based on the EFT formalism supplemented with unitarity. We are not able to distinguish specific models~\cite{Ghosh:2017jbw} as long as they are cast at low energy in the symmetry mold of the Standard Model: new physics enters through the low-energy EFT couplings only. 

We have shown the Standard Model background production of $W^+W^-$ from expressions available in the literature~\cite{Luszczak:2014mta} and incorporated into standard Monte Carlo programs (see figure~\ref{inelastic_wSM}). The Leading Order production is easy to understand in our setup: basically, use 
Eq.~(\ref{BijnensCornet}) as opposed to our expressions. To take the limit continuously is less trivial as we are relying in a truncated partial wave expansion, which fails at forward/backward angles for which it is not designed, and at low energies where terms of order $m_h$, $M_W$ are not negligible (e.g. the SM Higgs potential if it is at work).
Still, taking the limit of $M_W\to 0$ of the background calculated in figure~\ref{inelastic_wSM} and of the analytical expressions of~\cite{Denner:1995jv} would eventually allow to match with our calculations with all parameters taken at the SM values, if the partial wave expansion is put aside and Feynman amplitudes are used for the comparison.

We have computed the elastic-elastic cross section (both protons intact), which is the cleanest experimental channel. The number of events to be found increases with $p_t$, for modest values thereof. If a new resonance was around $E_{\gamma\gamma}=1\,{\rm TeV}$, we have shown in Fig.~\ref{fig:secppElastic}, for example, that the cross section would be rather flat in energy and around $10^{-2}\,{\rm fbarn/TeV}^4$ or somewhat more. This means that an integrated luminosity of $300\,{\rm fbarn}^{-1}$ at the LHC run II would prove insufficient to gather enough events at this high invariant boson-boson mass, specially when only certain diboson decay channels are experimentally reconstructed, further reducing the cross section by their branching fraction. Further small reductions are due to absorption effects~\cite{Lebiedowicz:2015cea} in the photon debris.

Thus, looking for inelastic processes to increase the cross section seems mandatory. We have shown the deeply inelastic cross sections in which both protons dissociate (figure~\ref{fig:secppElastic}), but those events are difficult to isolate because the non-photon-initiated background is too large, leaving activity in the central silicon trackers. The resonance-mediated inelastic (but not deeply inelastic) events where the proton dissociates but mostly in the forward direction are, therefore, more promising. But precise predictions are here difficult because we find quite some systematic difference due to the chosen pdf set; one can opt for the newest LUXQED set.

The situation is a bit better for resonances below $1\,{\rm TeV}$, that may be detectable with this method as the cross sections are an order of magnitude larger. Additionally, for resonances of larger mass there may be hope in collisions involving heavy ions: for example, lead-lead induced $\gamma\gamma$ collision cross-sections are enhanced by a factor $(Z=82)^2$ if collisions are incoherent which is unfortunately diminished by a factor 2000 smaller luminosity than in proton-proton collisions with the current LHC machine, so perhaps p-Pb collisions are the optimal search channel. At small momentum transfer, the entire nucleus can interact collectively and then the $Pb-Pb$ reaction is enhanced by $Z^4$ which is more promising; but EFT interactions grow derivatively, so this strategy works only at somewhat large $s$ or the underlying scattering amplitude is in turn decreasing the cross sections.

We have similarly predicted example cross sections for a future electron-positron collider operating in the TeV region. If we take as reference the proposed luminosity of the CLIC collider, that could conceivably accumulate about $650\,{\rm fbarn}^{-1}$ per year, our resonance cross sections of order $10^{-3}\,{\rm fbarn}$ will only yield a couple of events per year. Thus we find that, while CLIC may be apt for exploring vector resonances that couple in an $s$-wave to $e^-e^+$, it will fall short in luminosity to be a practical tool for scalar or tensor resonances in $\gamma\gamma$ physics. 

The inclusion of all computations reported in Monte Carlo simulations of the LHC (or ILC) detectors by interested collaborations should be possible and is encouraged. 

\section*{Acknowledgements}
We thank useful conversations with S.J. Brodsky, J.J.~Sanz-Cillero, D.~Espriu, M.J.~Herrero, and the members of the UPARCOS unit at UCM. Work supported by the Spanish grants MINECO:FPA2014-53375-C2-1-P, MINECO:FPA2016-75654-C2-1-P (A.D. and  F.J.L.-E.) and  MINECO:BES-2012-056054,  FIS2013-41716-P and the ``Ram\'on Areces'' Foundation (R.L.D.).

\appendix
\section{Elecroweak Chiral Lagrangian}
In this appendix we very briefly discuss the chiral Lagrangian in the electroweak context, particularly its coupling to photons: its extension to HEFT has been presented in our other publications, most recently in~\cite{Dobado:2017lwg}. The scattering amplitudes $\gamma\gamma\to\{W_L^+W_L^-,\,Z_LZ_L\}$ used in this work have also been obtained in somewhat more detail in our previous work~\cite{Delgado:2014jda} and~\cite{Delgado:2015kxa,DelgadoLopez:2017ugq} (purely EW sector terms). They are additionally mentioned in the CERN Yellow Report of the Higgs Cross Section Working Group~\cite{deFlorian:2016spz}. The unitarization scheme that is used here is detailed in~\cite{Delgado:2016rtd}. In any case, for the sake of clarity, we quote some of those results here too.

The Electroweak Chiral Lagrangian up to dimension $\mO(p^4)$ is the backbone of the Lagrangian. It can be written as~\cite{Delgado:2014jda}
\begin{equation}
\mL_{\rm ECLh} = \mL_2 + \mL_4 +\mL_{\rm GF} +\mL_{\rm FP}\, ,
\end{equation}
where $\mL_2\sim\mO(p^2)$ and $\mL_4\sim\mO(p^4)$. The Landau gauge is adopted, so that the gauge-fixing and non-Abelian Fadeev-Popov terms ($\mL_{\rm GF}$ and $\mL_{\rm FP}$) can be neglected~\cite{Appelquist:1980vg}. $\mL_2$ and $\mL_4$ can be written as
\begin{subequations}
\begin{align}
  \mL_2 &= -\frac{1}{2g^2}{\rm Tr}(\hat{W}_{\mu\nu}\hat{W}^{\mu\nu})-\frac{1}{2 g^{'2}}{\rm Tr}(\hat{B}_{\mu\nu}\hat{B}^{\mu\nu})\nonumber\\
        &  +\frac{v^2}{4}\left[1+2a\frac{h}{v}+b\frac{h^2}{v^2}\right]{\rm Tr}(D^\mu U^\dagger D_\mu U)
           +\frac{1}{2}\partial^\mu h\,\partial_\mu h+\dots\,\\
  \mL_4 &=  a_1{\rm Tr}(U\hat{B}_{\mu\nu}U^\dagger\hat{W}^{\mu\nu})
          +ia_2{\rm Tr}(U\hat{B}_{\mu\nu}U^\dagger[V^\mu,V^\nu])
          -ia_3{\rm Tr}(\hat{W}_{\mu\nu}[V^\mu,V^\nu]) \nonumber\\
        & -\frac{c_\gamma}{2}\frac{h}{v}e^2 A_{\mu\nu} A^{\mu\nu}
          +a_4{\rm Tr} (V_\mu V_\nu) {\rm Tr} (V^\mu V^\nu) %
          +a_5{\rm Tr} (V_\mu V^\mu) {\rm Tr} (V_\nu V^\nu) \nonumber\\
        & +\frac{g}{v^4}(\partial_\mu h\partial^\mu h)^2  %
          +\frac{d}{v^2}(\partial_\mu h\partial^\mu h){\rm Tr}[(D_\nu U)^\dagger D^\nu U] \nonumber\\
        & +\frac{e}{v^2}(\partial_\mu h\partial^\nu h){\rm Tr}[(D^\mu U)^\dagger D_\nu U] %
          +\cdots
\end{align}
\end{subequations}
where we have introduced
\begin{equation}
  -c_W\frac{h}{v}{\rm Tr}(\hat{W}_{\mu\nu}\hat{W}^{\mu\nu})-c_B\frac{h}{v}{\rm Tr}(\hat{B}_{\mu\nu}\hat{B}^{\mu\nu}) %
  =-\frac{c_\gamma}{2}\frac{h}{v}e^2 A_{\mu\nu}A^{\mu\nu}+\dots
\end{equation}
and the covariant derivative of the $U$ field is defined as
\begin{equation}
  D_\mu U = \partial_\mu U + i\hat{W}_\mu U - iU \hat{B}_\mu
\end{equation}
with
\begin{subequations}
\begin{align}
  \hat{W}_\mu &= g W_{\mu,i} \frac{\tau^i}{2}, &
  \hat{B}_\mu &= g' B_\mu\frac{\tau^3}{2}\\
  \hat{W}_{\mu\nu} &= \partial_\mu\hat{W}_\nu - \partial_\nu\hat{W}_\mu + i[\hat{W}_\mu,\hat{W}_\nu], &
  \hat{B}_{\mu\nu} &= \partial_\mu\hat{B}_\nu - \partial_\nu\hat{B}_\mu .
\end{align}
\end{subequations}
This Lagrangian leads to the Feynman rules computed in Ref.~\cite{Delgado:2014jda}. 
The amplitude elements $A(\gamma\gamma\to W_L^+W_L^-,Z_LZ_L)$ have been computed both with the spherical and linear representations of the $U$ field discussed next in appendix~\ref{sec:coset}, and yielding the same result~\cite{Delgado:2014jda}. The unitarized partial waves can be found on section~\ref{subsec:partwaves} of the present work.

\subsection{Spherical (or square--root) parametrization of the coset \label{sec:coset}}

Here we remind the reader of two possible  choices of the coset parametrization for $SU(2)_L \times SU(2)_R/ SU(2)_{L+R}$. The coordinates on that coset, three Goldstone boson fields, are of course not unique but $S$ matrix elements (on-shell amplitudes) do not depend on their choice. Very often one finds an exponential parametrization
\begin{equation}
U(x) = \exp\left(i\frac{\tilde{\pi}}{v}\right)\, ,
\end{equation}
with $\tilde{ \pi}= \tau^a \pi^a(x)$ and $\tau^a$ $(a=1,2,3)$ being Pauli matrices. This choice is well suited for $SU(3)$ chiral perturbation theory with three flavors.

However, in the electroweak sector (as well as in two-flavor ChPT in QCD) the coset is just the space $SU(2)$, isomorphic to the $S^3$ three-dimensional sphere. This suggests the use of simpler ``spherical'' coordinates:
\begin{equation}
U(x) = \sqrt{1-\frac{\omega^2}{v^2}}+ i \frac{\tilde\omega}{v} \, ,
\end{equation}
where again $\tilde{ \omega}= \tau^a \omega^a(x)$ and $\omega^2=\sum_a (\omega^{a})^2=\tilde \omega ^2$. The resulting  Feynman rules and Feynman diagrams are less numerous than for the exponential parametrization and thus, calculations are a bit simpler: yet the final answers are identical to the exponential parametrization, as we showed in~\cite{Delgado:2014jda}. There, we recalled how to change between the two sets of coordinates by rewriting the exponential as
\begin{equation}
U(x) = \cos \frac{\pi}{v}+ i \frac{\tilde \pi}{\pi} \sin \frac{\pi}{v},
\end{equation}
where $\pi =\sqrt{\pi^2}$ with $\pi^2=\sum_a (\pi^{a})^2$ and then comparing to the spherical parametrization to recover
\begin{equation}
\omega^a= \pi^a \frac{v}{\pi}\sin  \frac{\pi}{v},
\end{equation}
which implies $\omega^2= v^2 \sin^2(\pi/v)$. An expansion (formally, in powers of $\pi^2/v^2$) yields the series
\begin{equation}
\omega^a= \pi^a \left[1-\frac{1}{6}\bigg(\frac{\pi}{v}\bigg)^2
+\frac{1}{120}\bigg(\frac{\pi}{v}\bigg)^4-\frac{1}{5040}\bigg(\frac{\pi}{v}\bigg)^6+\dots
\right].
\end{equation}
The ``eaten'' Goldstone bosons that provide the longitudinal components of the  $W^{\pm}$ and $Z$ gauge bosons are then
 $\omega^{\pm}=(\omega^1\mp i\omega^2)/\sqrt{2}$, $\omega^0=\omega^3 (=z)$.

The Feynman rules involving less than four WBGBs are exactly the same in both parametrizations since they differ in terms at least quadratic  in the WBGBs. However the vertices with four WBGBs are indeed different in both parametrizations if the WBGBs are off-shell (but they coincide for on-shell amplitudes). The next section quotes the Feynman rules, that we do not rederive here.

\section{Uncertainty bands for the NNPDFs} \label{app:sets:NNPDFs}
\begin{figure}[h]
\centerline{\includegraphics[width=0.47\textwidth]{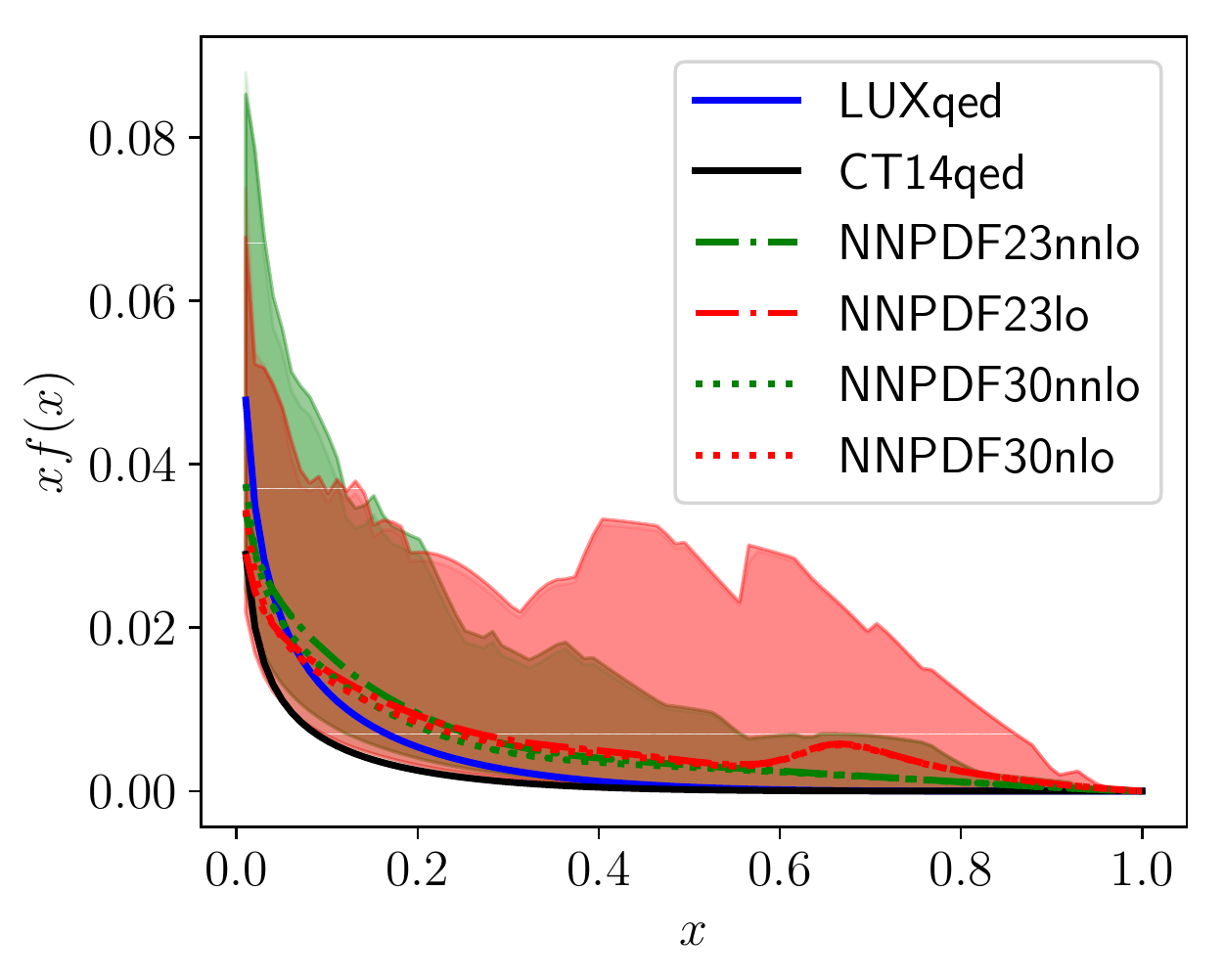}\ \ 
\includegraphics[width=0.47\textwidth]{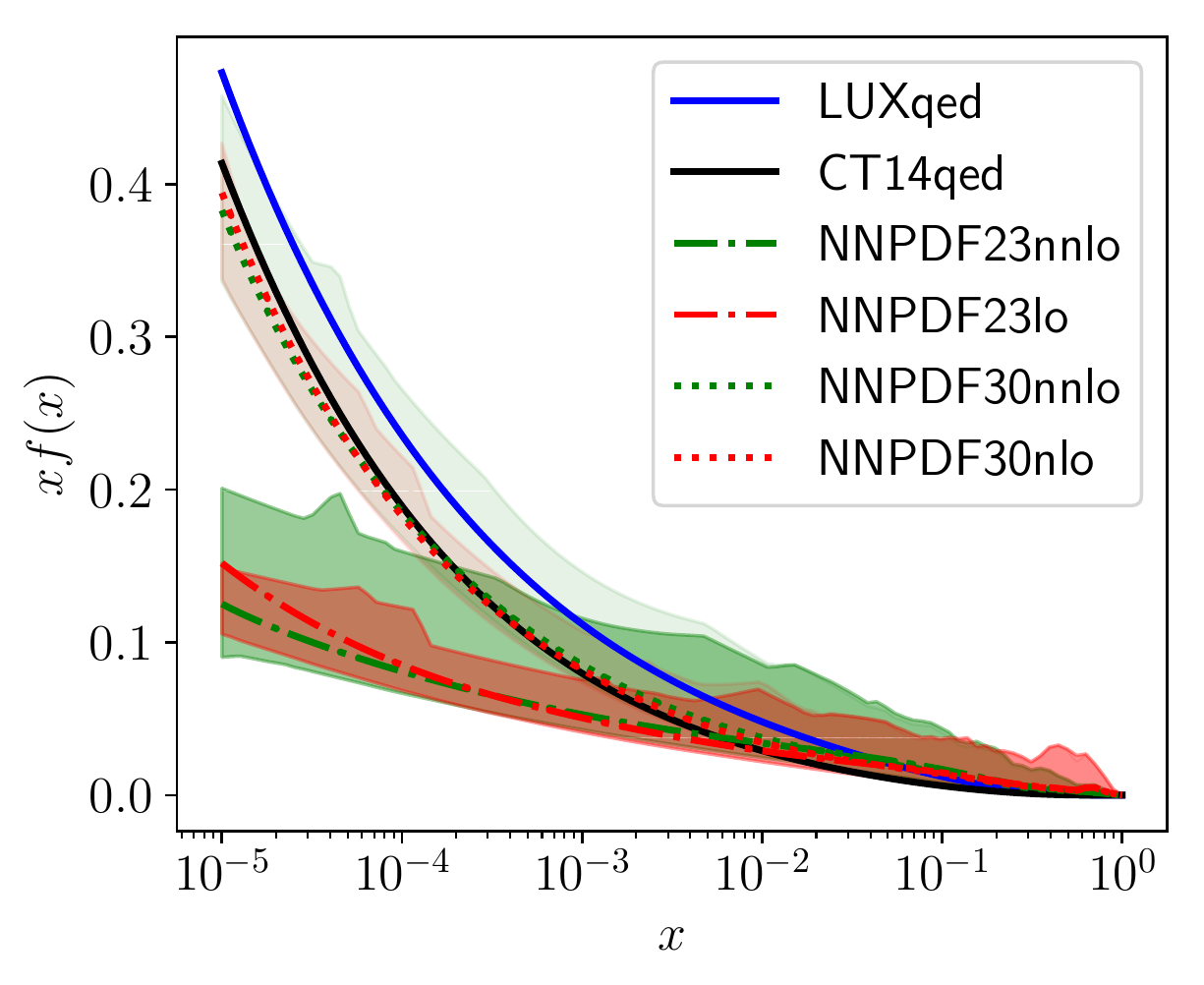}}
\caption{\label{fig:uncband1_NNPDFs} We compare the uncertainty band of the photon NNPDF sets shown, at a scale of 100 GeV, in linear (left) and logarithmic (right) scales. The latest 3.0 sets are compatible, within the uncertainty band, with the LUXQED and the CT14 determinations, though they have much leeway, especially near $x=0.6$ where they are not very well determined. }
\end{figure}

\begin{figure}[h]
\centerline{\includegraphics[width=0.47\textwidth]{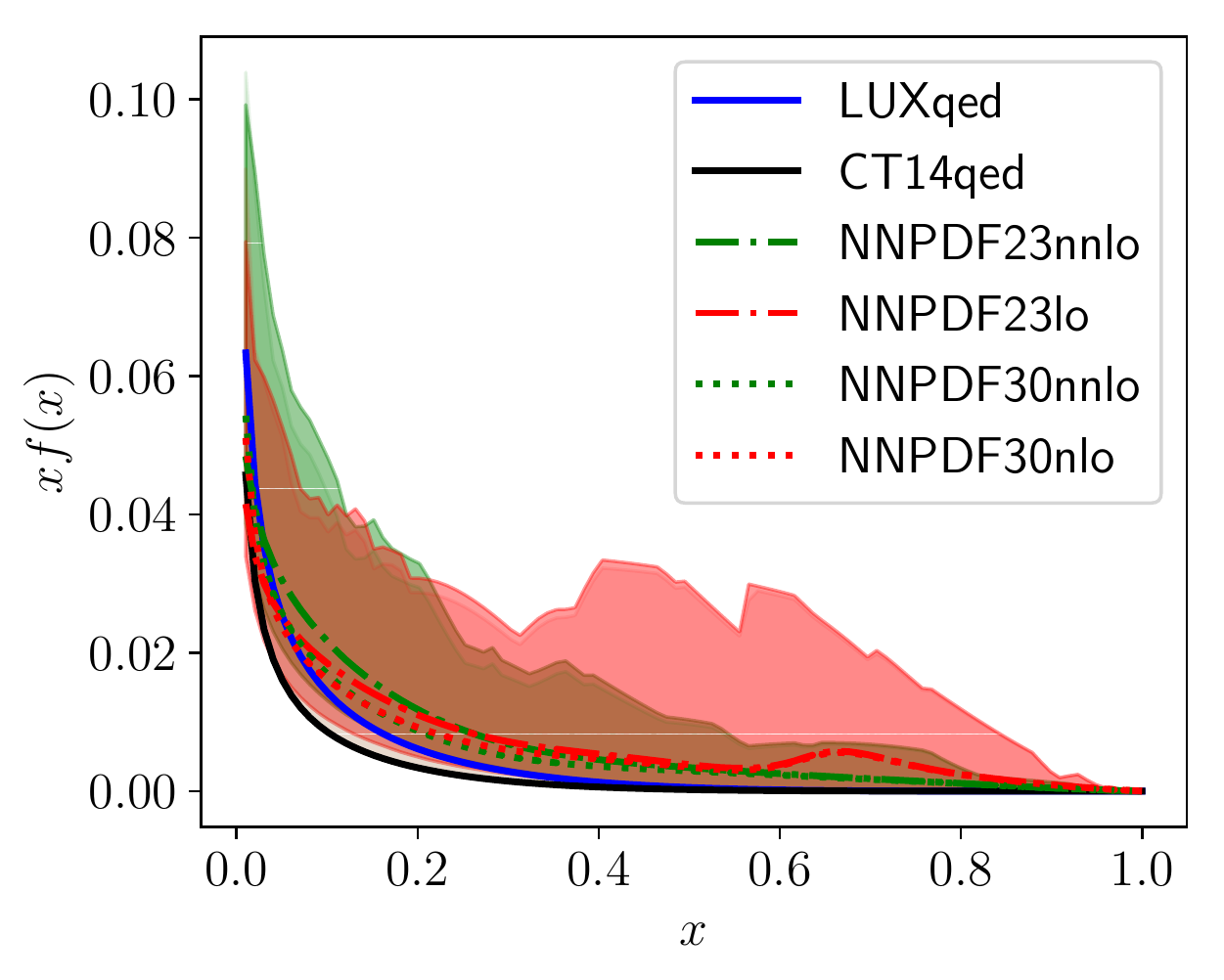}\ \ 
\includegraphics[width=0.47\textwidth]{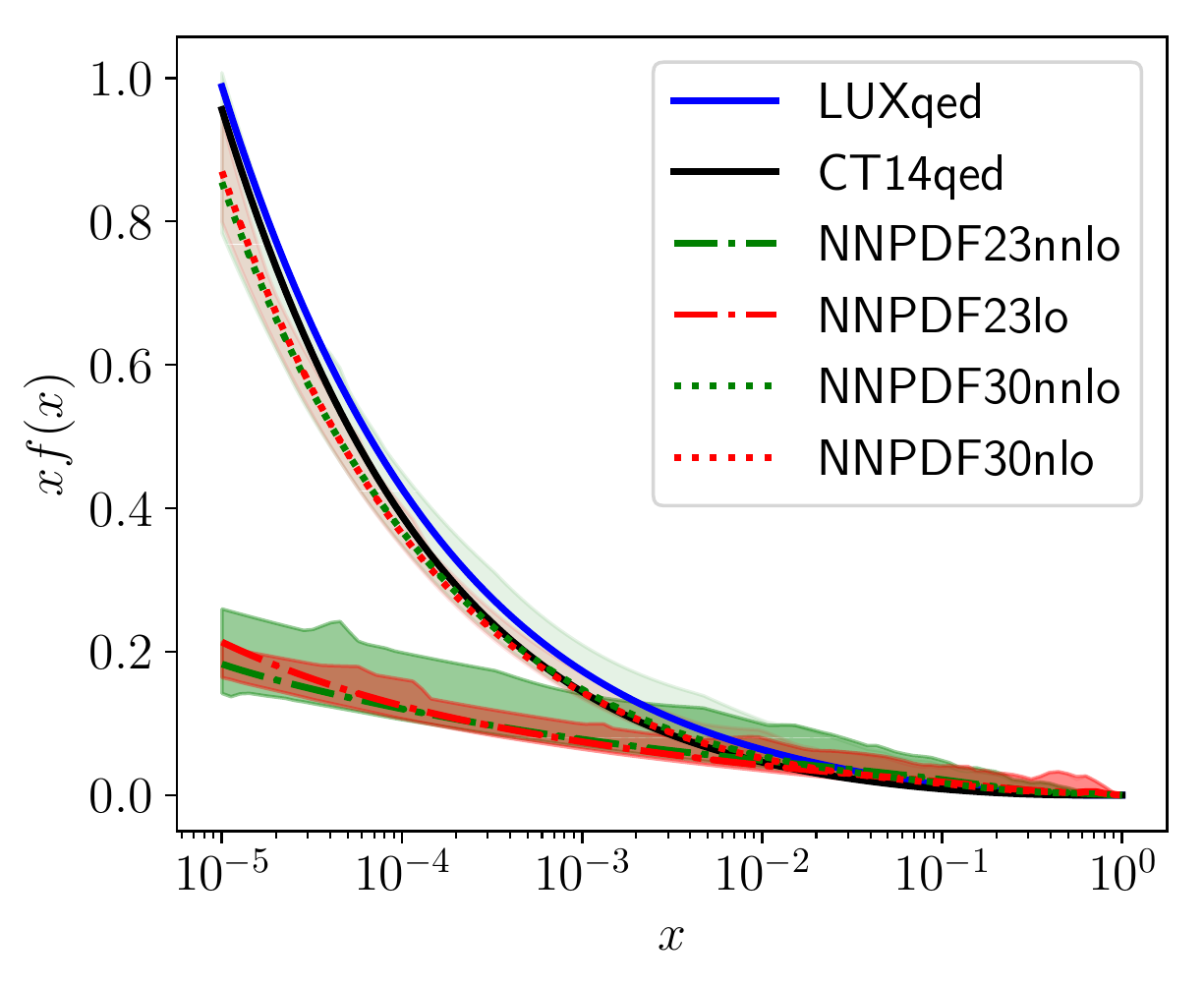}}
\caption{\label{fig:uncband2_NNPDFs} Same as in figure~\ref{fig:uncband1_NNPDFs} but for a scale of 1 TeV. The conclusion stands and the newest NNPDF sets are now in agreement with CT14 and LUXQED, within uncertainties.}
\end{figure}

\begin{figure}[h]
\centerline{\includegraphics[width=0.8\textwidth]{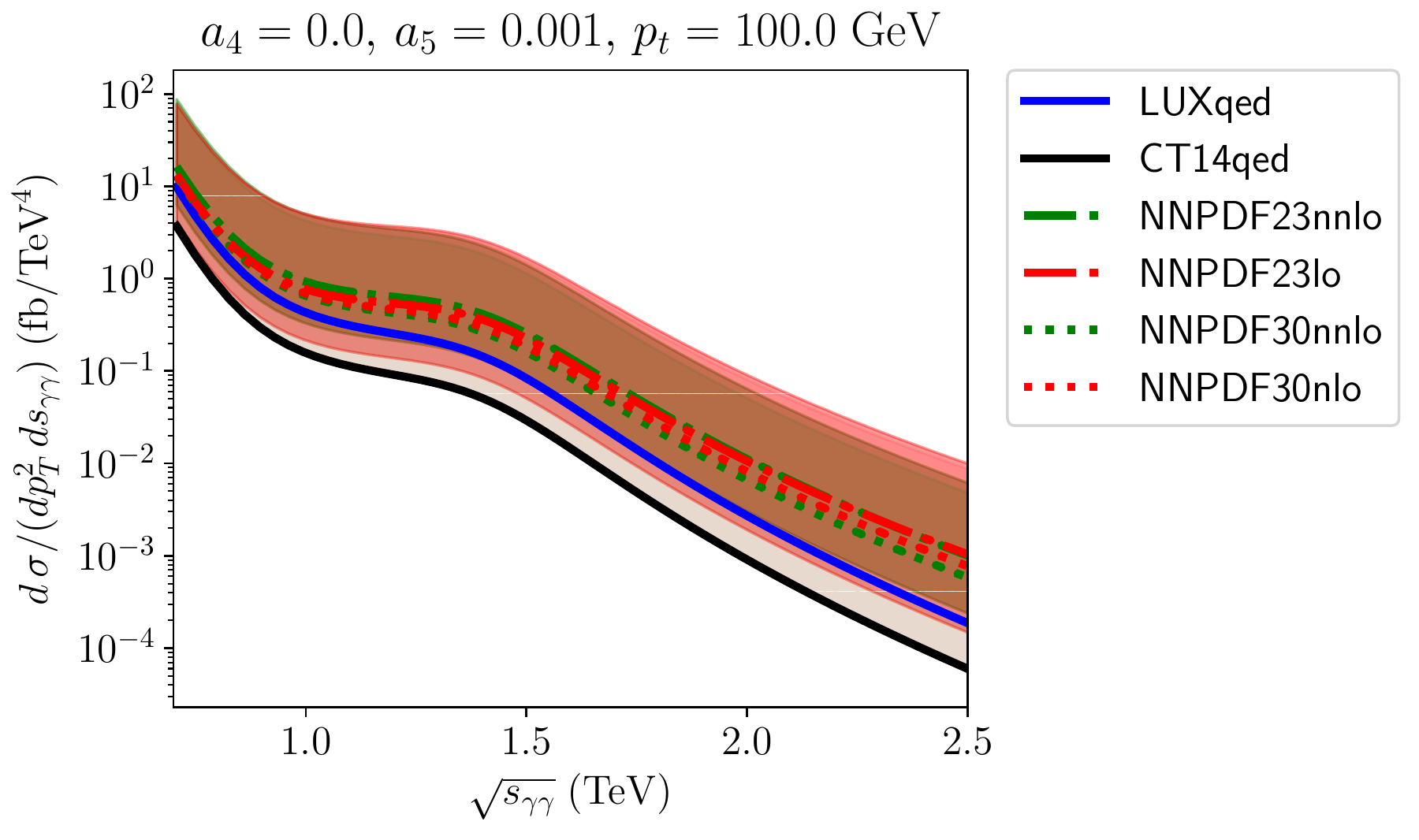}}
\centerline{\includegraphics[width=0.8\textwidth]{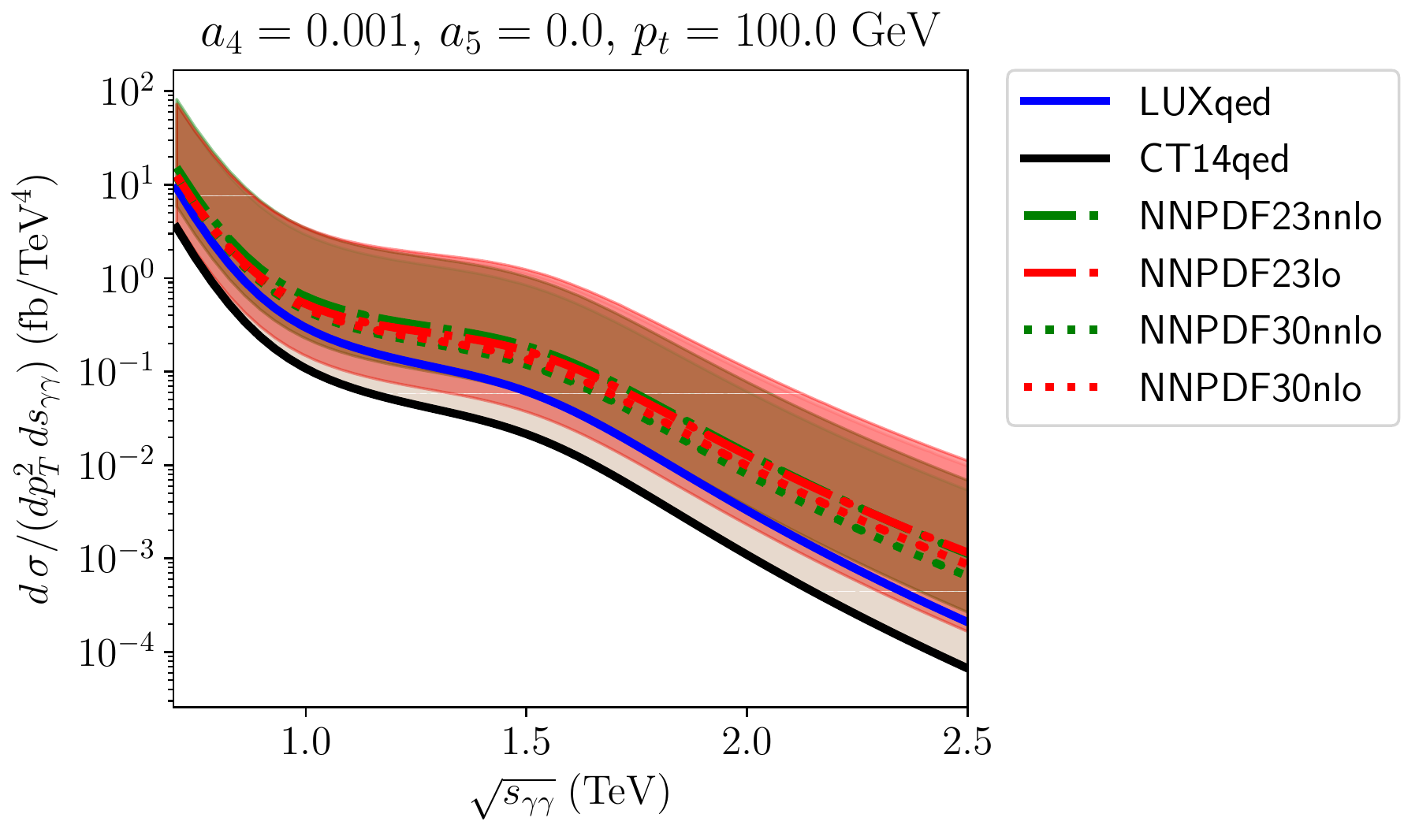}}
\centerline{\includegraphics[width=0.8\textwidth]{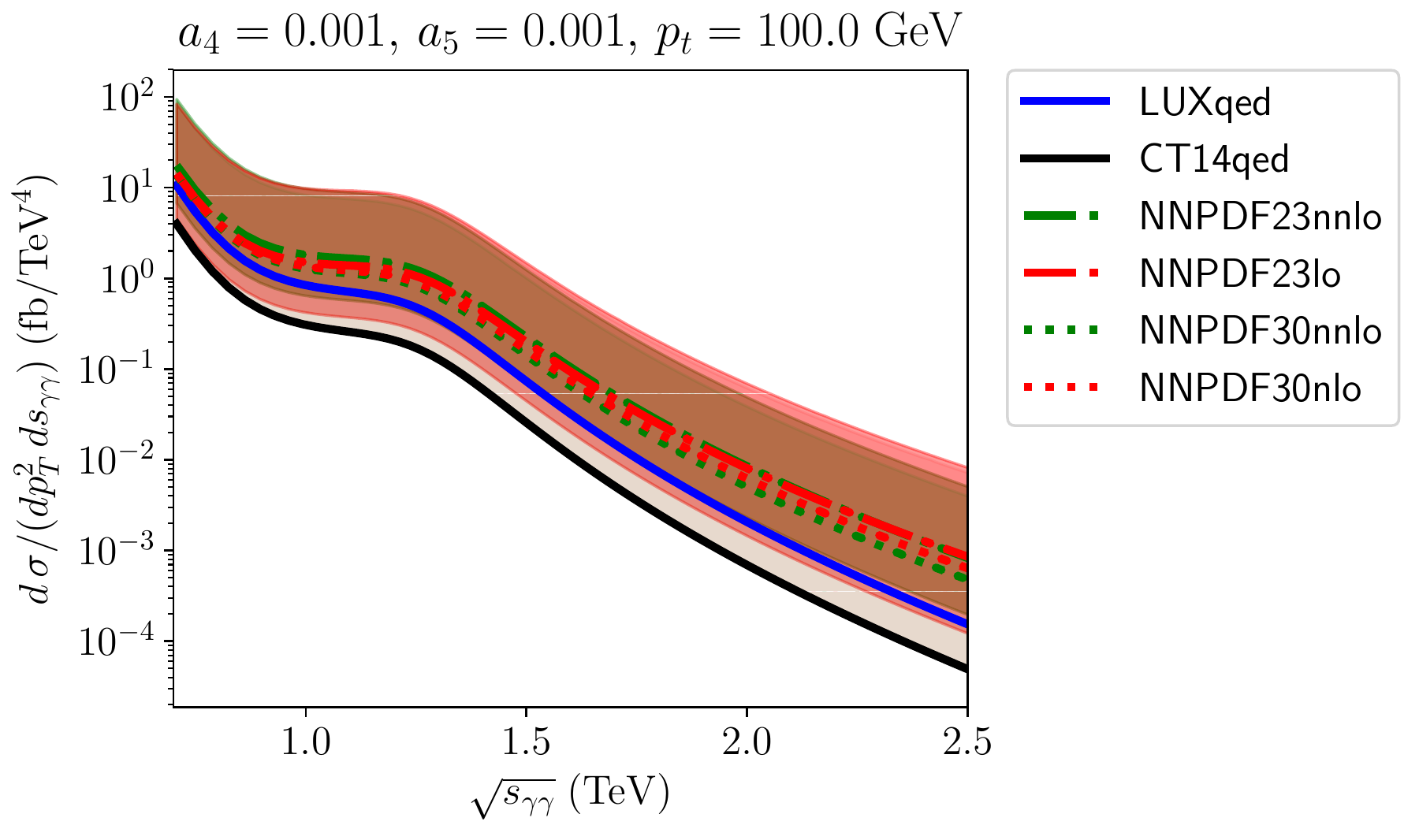}}
\caption{\label{fig:uncband_NNPDFs_DISproduction} Same as Fig.~\ref{inelastic}, with the NNPDF uncertainty bands. Both protons dissociate in the deeply inelastic regime. (Effective) PDF energy scale $\mu^2=s_{\gamma\gamma}$.}
\end{figure}

In this brief paragraph we plot the uncertainty bands for the NNPDF sets extracting a photon from the proton; we show that, within that uncertainty, the new sets are compatible with the CT14 and LUXQED pdf sets. Figures~\ref{fig:uncband1_NNPDFs} and~\ref{fig:uncband2_NNPDFs} display all the sets. This gives us some confidence in their use to predict photon-initiated cross-sections; the uncertainty bands for these are shown in figure~\ref{fig:uncband_NNPDFs_DISproduction}.

\clearpage

\section{Uncertainty bands for the CT14qed} \label{app:sets:CT14}
\begin{figure}[h]
\centerline{\includegraphics[width=0.47\textwidth]{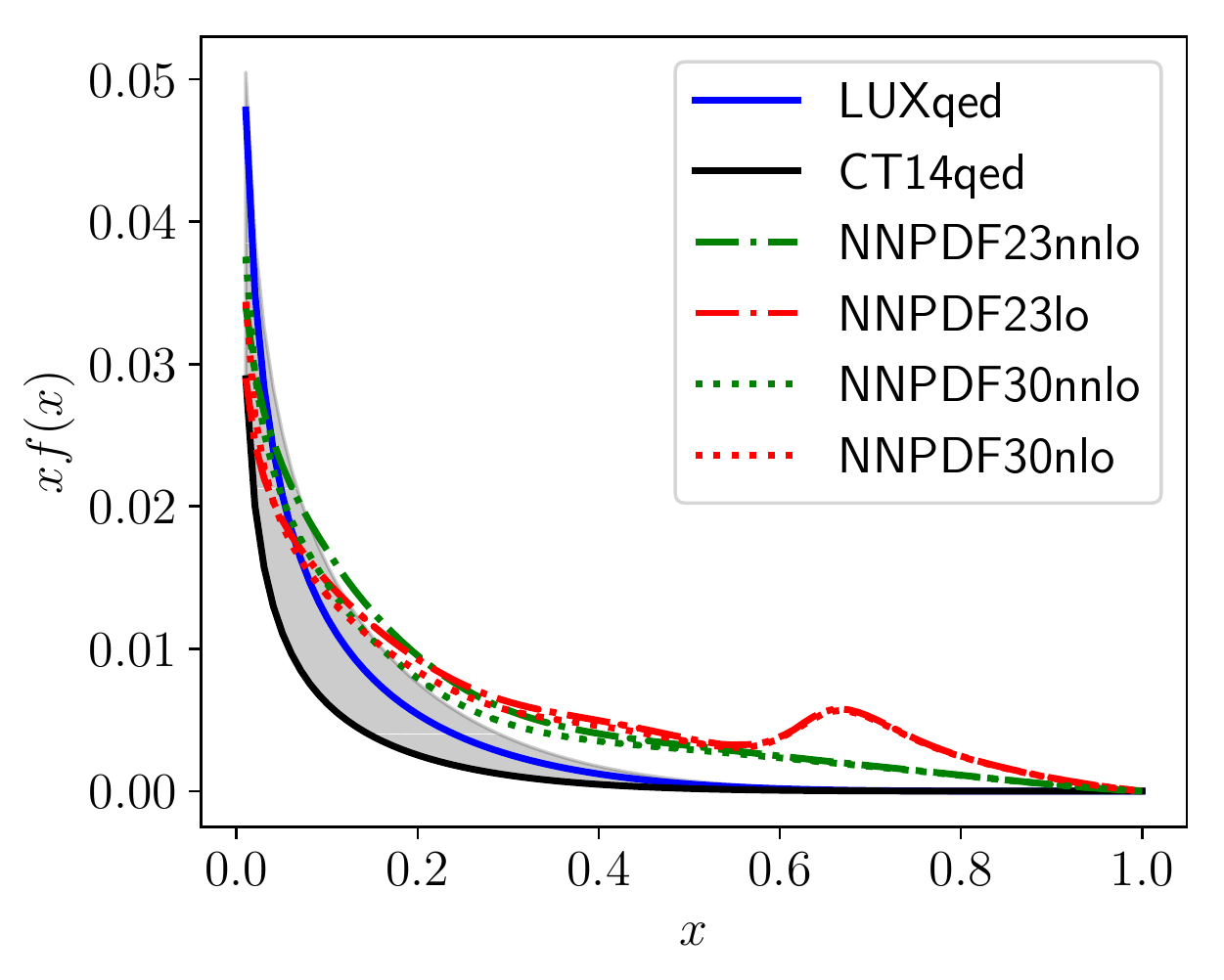}\ \
\includegraphics[width=0.47\textwidth]{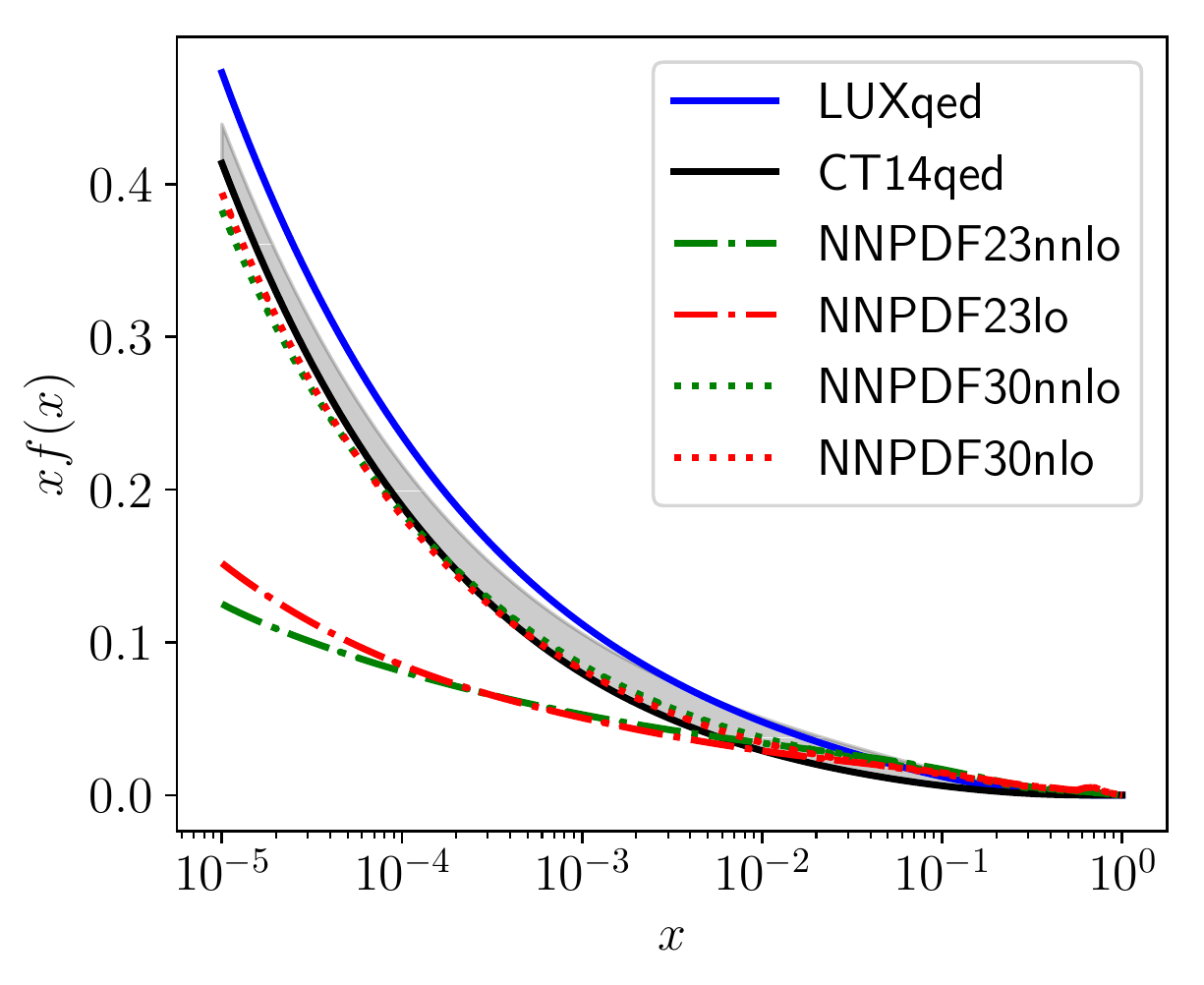}}
\caption{\label{fig:uncband1_CT14} We compare the uncertainty band of the photon CT14qed set shown, at a scale of 100 GeV, in linear (left) and logarithmic (right) scales.}
\end{figure}

\begin{figure}[h]
\centerline{\includegraphics[width=0.47\textwidth]{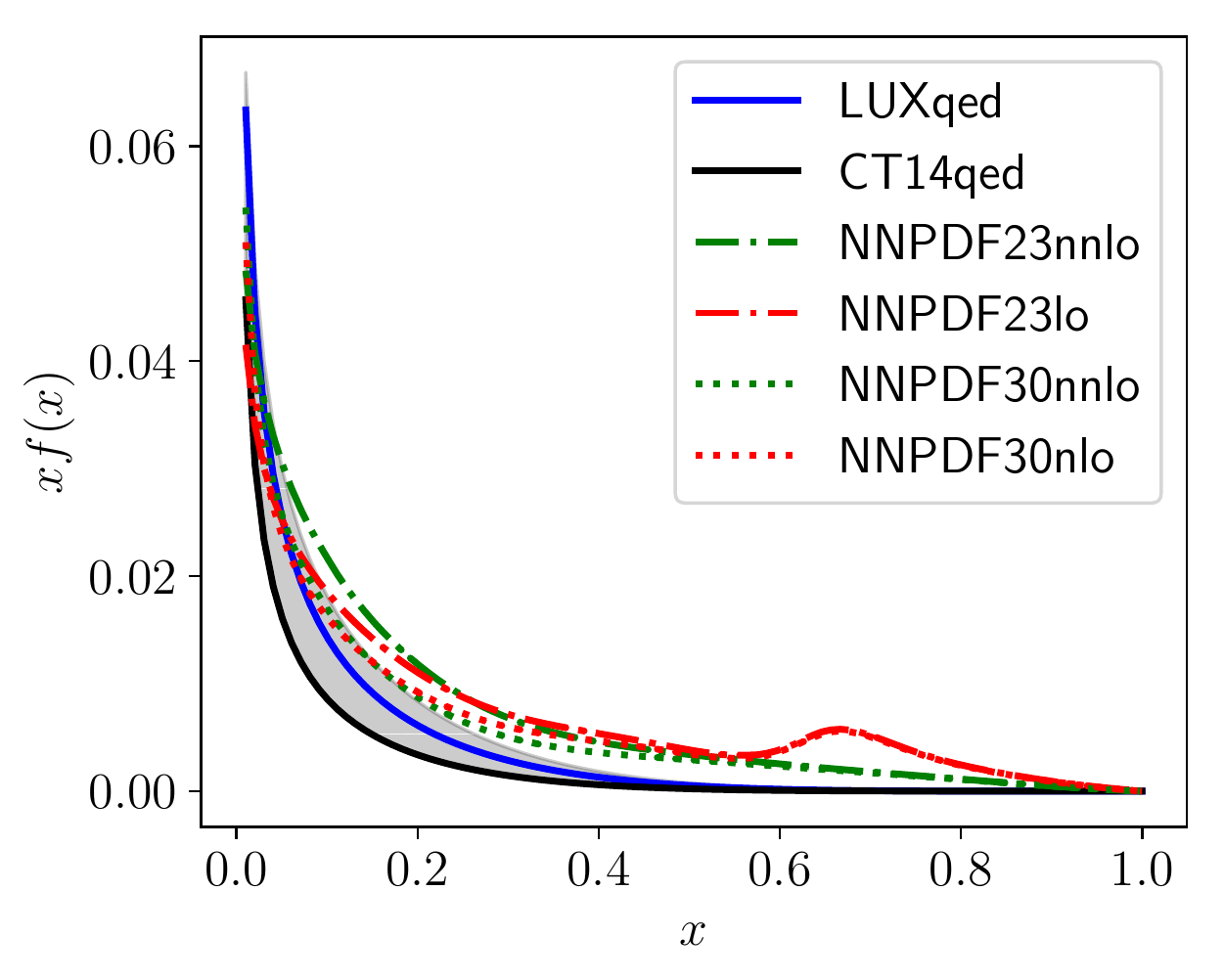}\ \
\includegraphics[width=0.47\textwidth]{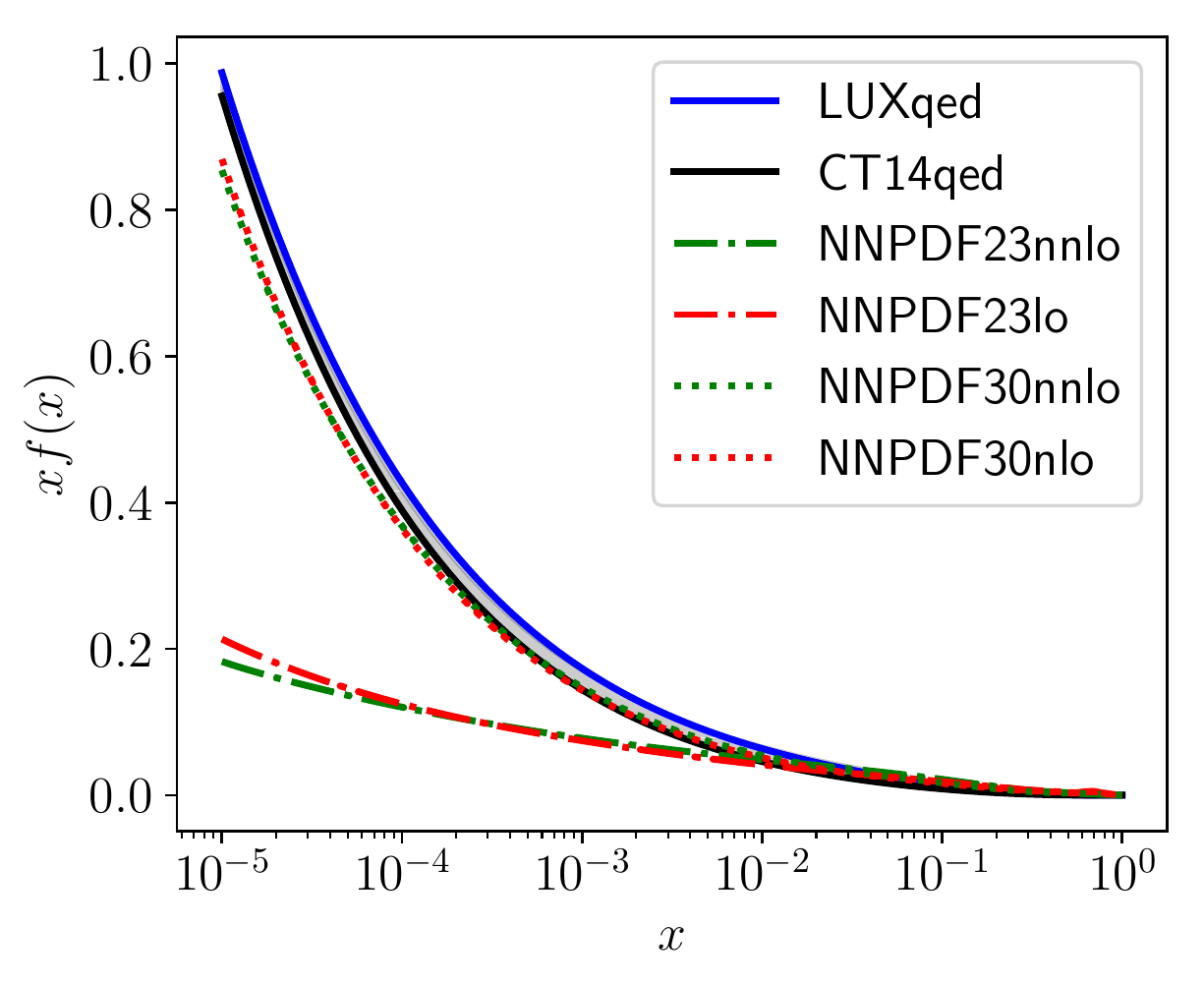}}
\caption{\label{fig:uncband2_CT14} Same as in figure~\ref{fig:uncband1_CT14} but for a scale of 1 TeV.}
\end{figure}

\begin{figure}[h]
\centerline{\includegraphics[width=0.8\textwidth]{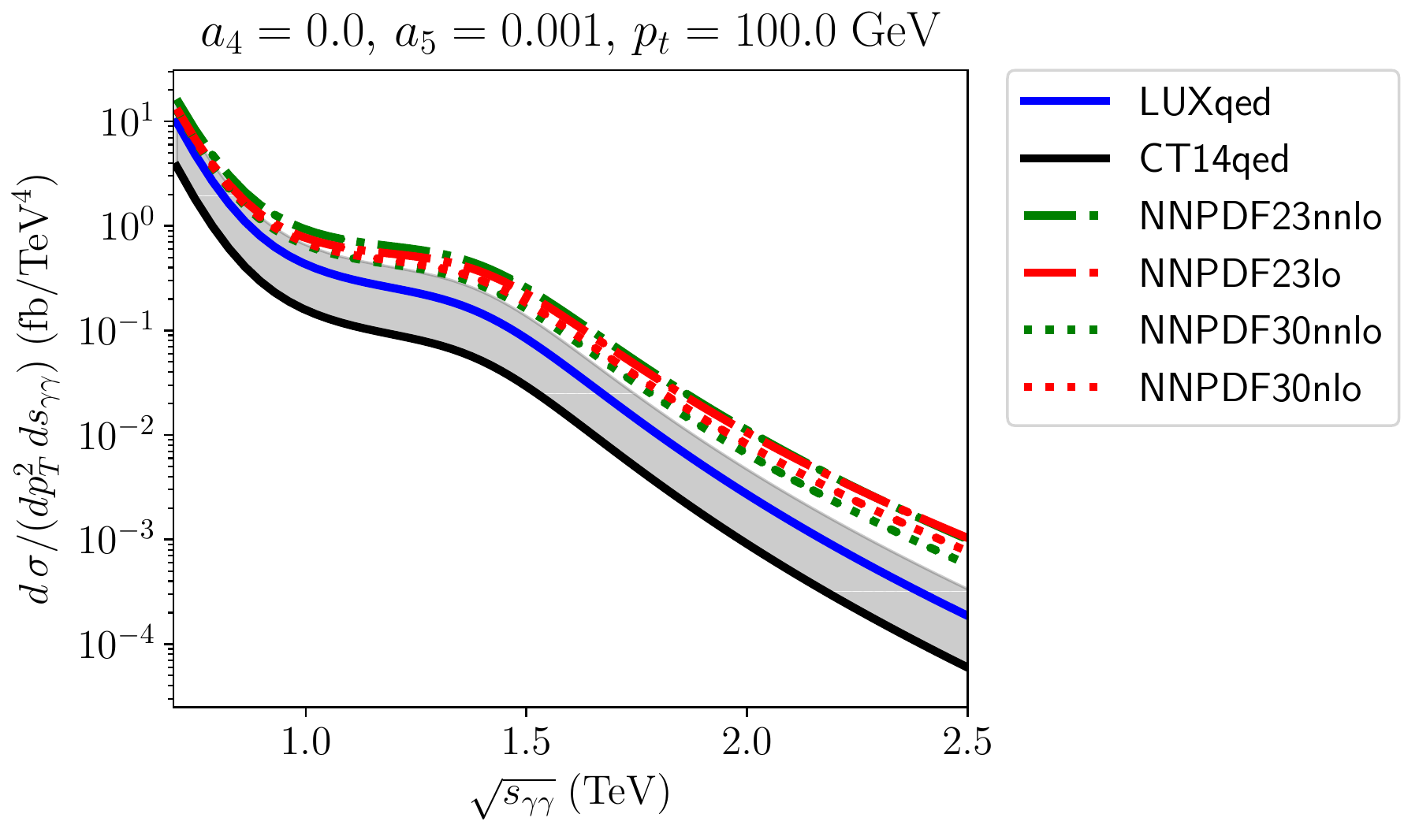}}
\centerline{\includegraphics[width=0.8\textwidth]{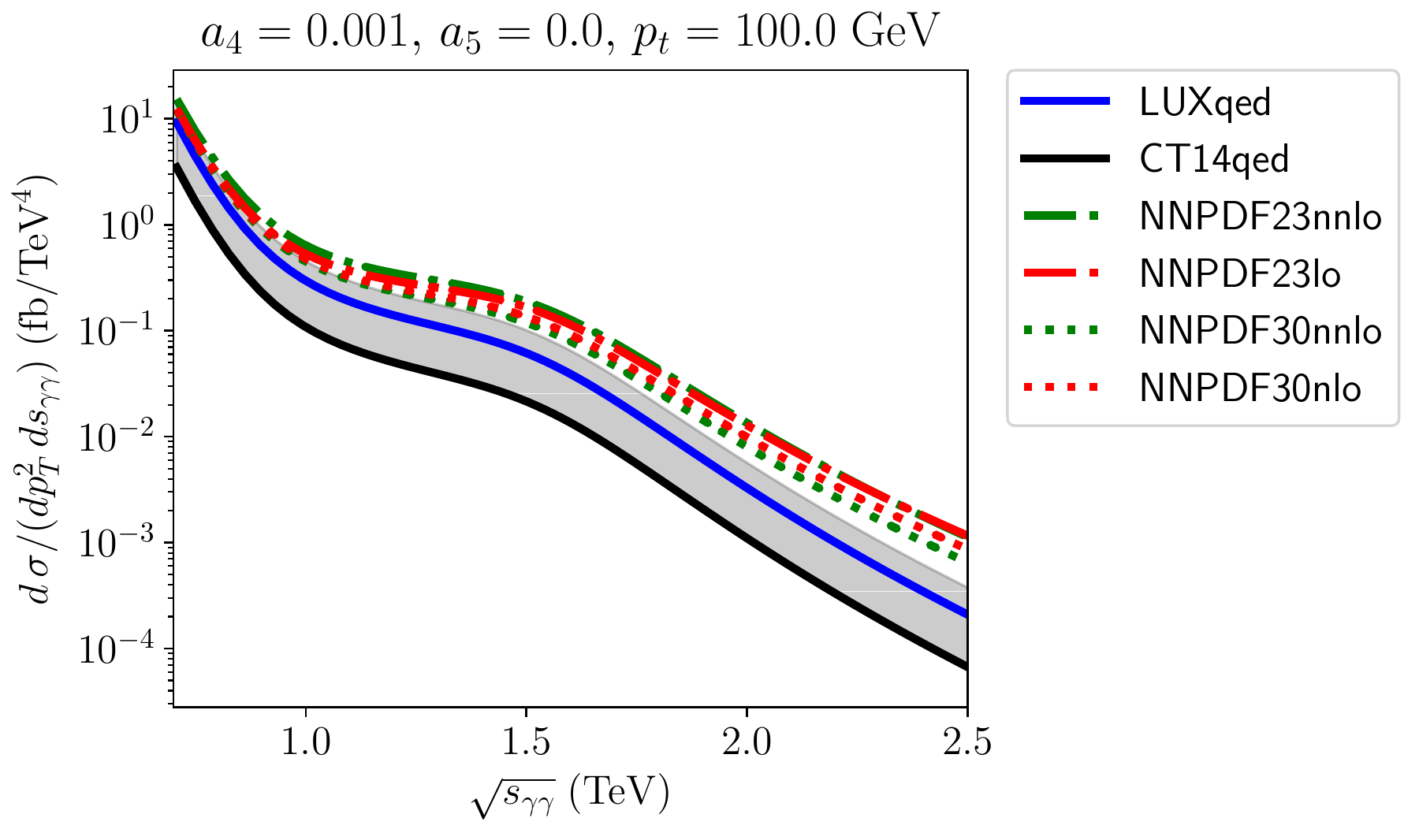}}
\centerline{\includegraphics[width=0.8\textwidth]{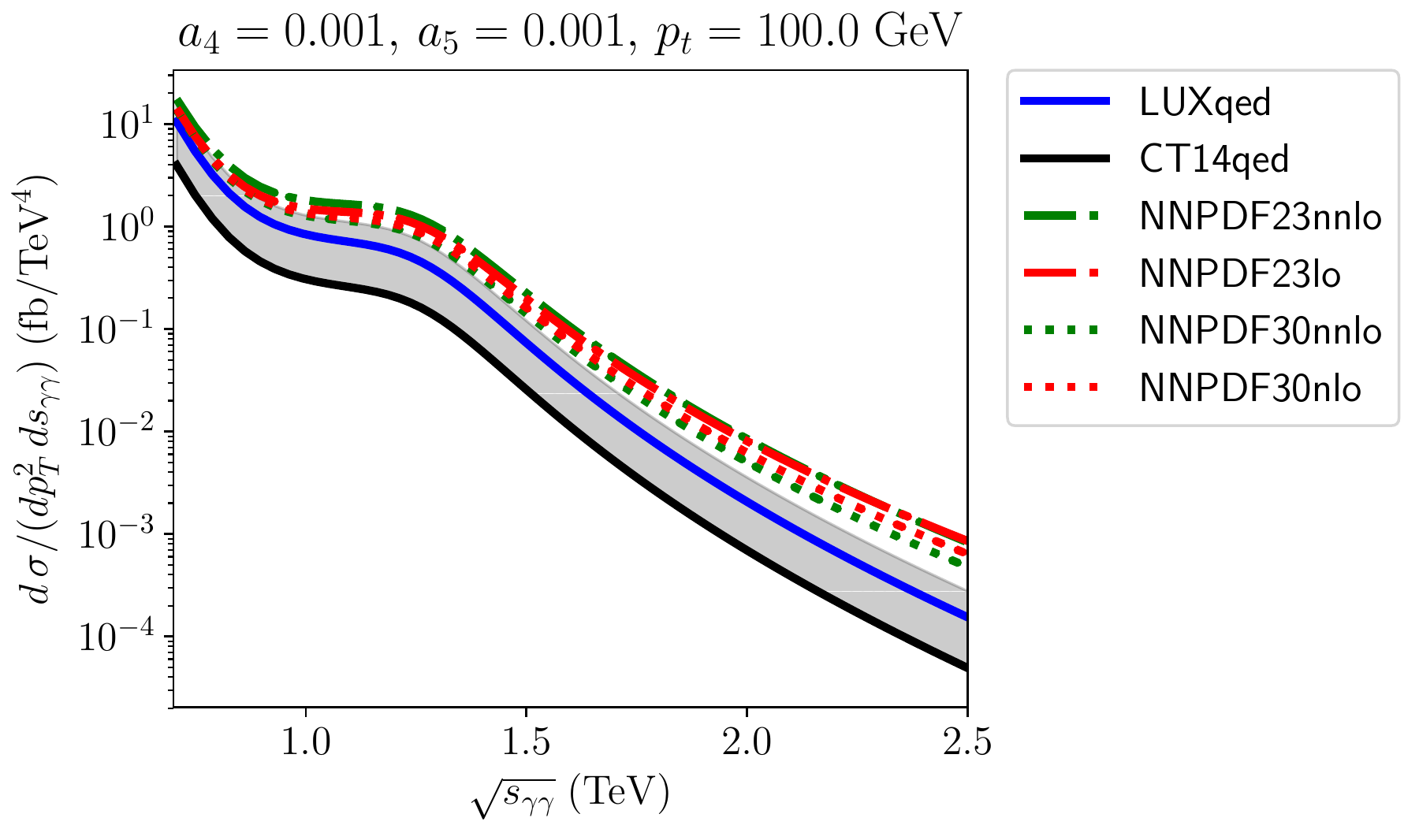}}
\caption{\label{fig:uncband_CT14_DISproduction} Same as Fig.~\ref{inelastic}, with the CT14qed error bands. Both protons dissociate in the deeply inelastic regime. (Effective) PDF energy scale $\mu^2=s_{\gamma\gamma}$.}
\end{figure}

Here, we plot the uncertainty bands for the CT14qed sets extracting a photon from the proton. The uncertainty of LUXQED has been found to be smaller than the size of the line. The CT14qed band is computed according to~\cite{Schmidt:2015zda} at 90\% CL. That is, the error band includes initial inelastic momentum fraction of the electron up to 0.30\%. The PDF line uses 0\% initial inelastic momentum fraction. Error bands for LUXqed happen to be too small for representing. Figures~\ref{fig:uncband1_CT14} and~\ref{fig:uncband2_CT14} display all the sets. Then, figure~\ref{fig:uncband_CT14_DISproduction} propagates these uncertainties to the production cross--section.

\afterpage{\clearpage}
\newpage



\end{document}